\documentclass[runningheads,english]{llncs}

\usepackage{fixltx2e,fix-cm}
\usepackage{amssymb}
\usepackage{amsmath}
\usepackage{graphicx}
\usepackage{subfigure}
\usepackage{makeidx}
\usepackage{multicol}
\usepackage{mathptmx}
\usepackage{url}
\usepackage{enumitem}

\newcommand{\keywords}[1]{\par\addvspace\baselineskip
\noindent\keywordname\enspace\ignorespaces#1}

\begin{document}

\mainmatter  

\title{An Overview of End-to-End Verifiable Voting Systems\thanks{This is a self-archived version of a chapter due to appear in the book ``Real-World Electronic Voting: Design, Analysis and Deployment'', edited by Feng Hao and Peter Y. A. Ryan, part of the Series in Security, Privacy and Trust published by CRC Press, 2016.}}

\author{Syed Taha Ali \and Judy Murray}

\institute{School of Electrical Engineering \& Computer Science,\\ National University of Sciences and Technology,\\Pakistan\\
\email{taha.ali@seecs.edu.pk},\\
\and
School of Geography, Politics and Sociology,\\
Newcastle University, \\United Kingdom\\
\email{judy.murray@newcastle.ac.uk}
}

\maketitle

\begin{abstract}
Advances in E2E verifiable voting have the potential to fundamentally restore trust in elections and democratic processes in society. In this chapter, we provide a comprehensive introduction to the field. We trace the evolution of privacy and verifiability  properties in the research literature and describe the operations of current state-of-the-art E2E voting systems. We also discuss outstanding challenges to the deployment of E2E voting systems, including technical, legal, and usability constraints.

Our intention, in writing this chapter, has been to make the innovations in this domain accessible to a wider audience. We have therefore eschewed description of complex cryptographic mechanisms and instead attempt to communicate the fundamental intuition behind the design of E2E voting systems. We hope our work serves as a useful resource and assists in the future development of E2E voting.
\end{abstract}

\section{Introduction}
\label{sec:Introduction}

The use of cryptography to secure elections was first suggested by Chaum in 1981 in a highly influential paper on anonymous communications \cite{chaum1981untraceable}. Chaum described new cryptographic primitives, or building blocks, and proposed various applications for these, including the possibility of conducting remote electronic elections. This idea proved popular in the academic research community, and, over the next two decades, a multitude of voting systems appeared in the literature, a handful of which resulted in actual prototype solutions \cite{cranor1997sensus}  \cite{herschberg1997secure}. However, research in this domain was, for the most part, confined to academia and progressed independent of developments in real-world election technology.

In the last fifteen years however the situation has dramatically altered. Research in securing elections has received a significant boost because of two main reasons: First, the gridlock in the US presidential election of 2000 between George W. Bush and Al Gore cast a spotlight on the deficiencies of the US voting infrastructure where voters faced considerable difficulties in casting votes due to confusing ballot designs and punch card voting machines \cite{levine08hanging}. This flurry of national attention led to passage of the Help America Vote Act (HAVA) in 2002 which, among other provisions, allocated generous funding for research to improve voting technology \cite{coleman2011help}.

Second, multiple investigations in recent years have highlighted glaring reliability and security issues in electronic voting machines which render them vulnerable to hackers and compromise the integrity of elections \cite{kohno2004analysis} \cite{feldman2006security} \cite{wolchok2010security} \cite{springall2014security}. Indeed, there have been numerous documented instances of voting machines inexplicably malfunctioning during live elections, flipping candidate votes, or randomly adding and subtracting candidate votes. Since then, academic research interest in securing elections has soared, and prompted the formation of dedicated conferences and journals and research organizations \cite{accurate} \cite{iavoss}.

This renewed focus has led to the development of a new type of voting system with \textbf{end-to-end verifiable (E2E) security} properties. This highly promising advance, due to individual efforts by Chaum \cite{chaum2004secret} and Neff \cite{neff04practical} in 2004, harmonizes theoretical research innovation with the ground realities of election administration to provide strong guarantees on the integrity of elections. We describe next a typical usage scenario which encapsulates the spirit of E2E voting systems.

On election day, our voter, Alice goes to the polling station to cast her vote. She identifies herself to polling staff as a legitimate voter and is guided to a private booth with a voting machine, where she chooses her candidate on the touchscreen. The machine issues her a \emph{receipt}, which is essentially a cryptographically masked copy of her vote. In the booth, Alice can also verify, either visually, or by challenging the machine, that her vote was \textbf{cast as intended}, i.e. the cryptographic obfuscation does indeed represent her chosen candidate and not another. After polls close, the system posts copies of all receipts on a public bulletin board or website where Alice confirms that her vote has been \textbf{recorded as cast}, i.e. her receipt is included in the lot and it matches the physical copy she has in her hand. In the final step, all receipts on the website are processed in a series of cryptographic computations to yield the election result. The algorithms and parameters for these operations are specified on the website, and any technically-minded person, including Alice herself, can therefore verify that her own vote was \textbf{tallied as recorded} and that the tally is indeed correct.

This scenario is starkly different from the current state of real-world elections where Alice has to implicitly trust the voting system and its administrators for the credibility of the election. Dishonest polling staff have been known to manipulate election results. Furthermore, the internal operations of voting machines are opaque to Alice and she has no assurance that the machine is actually doing what it is supposed to.\footnote{This realization was fundamental to the decision of the Federal Constitutional Court of Germany in 2009 declaring electronic voting as `unconstitutional' and thereby marking Germany's return to paper-based voting \cite{fcc09}.} E2E voting systems, on the other hand, preclude trust in personnel and machines and make the voter herself an active participant in auditing the election at every step and certifying its result. If a voting machine loses Alice's vote or switches it to another candidate, her receipt on the website will reflect this change and Alice can file a complaint, using her own physical copy as evidence. If polling staff tamper with the final tally, any third party running the tallying or verification algorithm on their own computers will pick up on it. Furthermore, since Alice's receipt is an obfuscation of her choice, she cannot use it to convince a third party of how she voted, thereby thwarting vote-selling and coercion.

We provide here a comprehensive high-level introduction to the field of E2E voting. The writing is aimed at the layman with little knowledge of cryptography and attempts to communicate a clear and intuitive appreciation of  E2E voting systems. This chapter is organized as follows: we introduce security properties of voting systems in Sec.~\ref{sec:securitypropertiesofvotingsystems}. In Sec.~\ref{sec:cryptosystems}-\ref{sec:noncryptosystems}, we summarize the workings of some twenty of the most influential E2E voting systems, classified into four distinct categories, as per their reliance on cryptography (cryptographic and non-cryptographic systems), ballot format (physical and electronic ballots), and mode of voting (precinct-based and remote voting). This is followed by a discussion of open challenges to mainstream deployment of E2E voting systems in Sec.~\ref{sec:wayforward}. We conclude in Sec.~\ref{sec:conclusion}.

\section{Security Properties of Voting Systems}
\label{sec:securitypropertiesofvotingsystems}

Here we describe key security properties characterizing E2E voting systems. Many of these properties are intimately related whereas others are in direct conflict. The merits of a system are typically assessed on the basis of what properties it provides and how successfully their inherent conflicts are harmonized within the system.

\subsection{Vote Privacy}

The privacy of the vote is widely recognized as a fundamental human right and is enshrined in Article 21 of the Universal Declaration of Human Rights \cite{assembly48universal}. The rationale behind this, dating back to ancient Greece and Rome \cite{hall1990greeks}, is that if outside parties become privy to a voter's choice, it opens the door to bribery and intimidation, ultimately corrupting the electoral process. Vote-buying was not uncommon in the Western world up until the last century \cite{stokes2012killed} and still persists in some developing countries \cite{jensen2014poverty} \cite{gonzalez2012vote}. Likewise, voters may be intimidated by criminals, local politicians, or even family members, to vote a certain way. These fears directly motivated the notion of the \emph{secret ballot}, which is typically implemented nowadays by providing the voter a private voting booth at the polling station. Her ballot bears no distinguishing marks and her vote is cast in the ballot box where it mixes with all the other votes, making it very difficult, if not impossible, to ascertain her choice.

In contrast, voting systems in the early research literature (such as \cite{cohen1985robust} \cite{benaloh1986distributing} \cite{benaloh87verifiable} \cite{chaum1988elections}) explicitly maintained the voter-ballot linkage by issuing the voter a receipt enabling her to track her vote on a public bulletin board. In 1994, Benaloh and Tuinstra \cite{benaloh1994receipt} (and independently, Niemi and Renvall \cite{niemi1995prevent}) pointed out that this strategy allowed the voter to prove her vote to third party after the election, thereby facilitating vote-buying and coercion. The authors introduced the notion of \emph{receipt-freeness} which proved influential and several voting systems that followed (such as \cite{benaloh1994receipt} \cite{niemi1995prevent} \cite{sako1995receipt} \cite{okamoto1996electronic} \cite{okamoto1998receipt} \cite{hirt2000efficient}) dispensed with receipts entirely, focussing instead on ensuring transparency and integrity of the tallying operations. However, in 2004, Chaum \cite{chaum2004secret} re-introduced the use of receipts, with the important distinction that the contents of the receipt are cryptographically masked, thereby maintaining the voter's privacy.

Juels, Catalano and Jakobsson \cite{juels2005coercion} in 2005 argued that a coercer may yet influence a voter's choice without explicit knowledge of her vote, and described three such modes of coercion: the coercer may prevent the voter from voting, he may appropriate her voting credentials, or force her to vote for a candidate at random. The exact difference between receipt-freeness and coercion resistance is a subtle one \cite{delaune2006coercion}: in receipt-freeness, the coercer is assumed to be restricted to observing the election and using evidence provided by a cooperating voter. In the case of coercion resistance, the coercer is more powerful, and may craft specific votes for the voter to cast or even interact with her in some way while she is voting. Juels et al. proposed a solution to defend against these threats which we describe in Sec.~\ref{sec:jcjandcivitas}.

Researchers have therefore progressively refined notions of vote privacy over the years into the following key properties:

\begin{itemize}[leftmargin=*]
\item\textbf{Ballot Secrecy}: the voting system must not reveal who the voter voted for.

\item\textbf{Receipt-freeness}: the voting system should not give the voter any evidence to prove to a third party how she voted.

\item\textbf{Coercion-Resistance}: the voter should be able to cast a vote for her intended choice even while appearing to cooperate with a coercer.
\end{itemize}

Intuitively, we see that each property encapsulates the previous one: coercion-resistance implies receipt-freeness which in turn implies ballot secrecy.

\subsection{Vote Verifiability}

In real-world elections, the voter has to implicitly trust voting machines and polling staff for the integrity of the elections. Typically there are ancillary processes in place to improve voter confidence in results, such as exit polls, random audits, and opening the tallying process to the public. On the other hand, voting systems in the research literature attempt to minimize the voter's dependence on personnel and machines, and use cryptography to provide ironclad guarantees on election integrity. Sako and Kilian \cite{sako1995receipt} distinguished between two forms of verifiability:

\begin{itemize}[leftmargin=*]
\item\textbf{Individual Verifiability}: a voter can verify that her vote is included in the set of all cast votes.

\item\textbf{Universal Verifiability}: an observer can verify that the tally has been correctly computed from the set of all cast votes.
\end{itemize}

E2E voting systems recast the notion of verifiability in terms of three core steps:

\begin{itemize}[leftmargin=*]
\item\textbf{Cast-as-intended}: the voter can verify the voting system correctly marked her candidate choice on the ballot.

\item\textbf{Recorded-as-cast}: the voter can verify that her vote was correctly recorded by the voting system.

\item\textbf{Tallied-as-recorded}: the voter can verify that her vote was counted as recorded.
\end{itemize}

This translates as follows: the voter first confirms the system has correctly encrypted her vote. She then tracks her vote on the bulletin board using her receipt and confirms that it is correctly recorded. Integrity of the result is ensured by rigorously auditing the tallying process and requiring the system to publish cryptographic proofs of correct operation. This three-step verification therefore covers the entire life cycle of the vote, and the voter can be confident that if there is any tampering or breakdown in the system it will be discovered in one of the checks.

These two conceptions of verifiability are also linked: in most E2E voting systems, cast-as-intended and recorded-as-cast checks are verified by the voter, i.e. together they provide individual verifiability, whereas the tallied-as-recorded step may be undertaken by voters and observers alike, i.e. it provides universal verifiability.

\subsection{Other Properties}

We list here certain additional properties also vitally important in voting systems:

\begin{itemize}[leftmargin=*]
\item\textbf{Eligibility verifiability}: an observer can verify that each vote in the set of all cast votes was cast by an eligible voter.

\item\textbf{Accountability}: in case vote verification fails at some stage, possibly due to error or fraud, the voter should be able to conclusively prove that it failed to the relevant authorities, without compromising the secrecy of her ballot.

\item\textbf{Robustness}: the system should be robust to a certain degree of malfunction or corruption and still deliver correct results. A small number of misbehaving voters or system failures should not disrupt the election.

\item\textbf{Usability}: the system should enable voters to cast their votes easily and effectively.

\item\textbf{Accessability}: the system provides equal opportunity for access and participation (including guarantees on vote privacy and verifiability) to voters with disabilities.
\end{itemize}

\subsection{Conflicts and Challenges}
\label{sec:conflictsandchallenges}

Several of the key properties described thus far conflict with each other in subtle ways, resulting in a variety of technical and legal challenges. For instance, as noted in the discussion on receipt-freeness, vote privacy clashes with vote verifiability. If a voter can successfully prove her vote to a third party outside the polling station, she could easily sell her vote or be coerced into voting for a certain candidate. A detailed treatment of this tension between verifiability and privacy in voting systems may be found in \cite{jonker2013privacy} and \cite{neumann2014analysis}.

Similarly, there is a tension between vote verifiability and usability. Requiring voters to verify their vote negatively impacts usability by adding extra steps to the process, which may prove confusing for the voter \cite{karayumak2011usability} \cite{acemyan2014usability} \cite{murray2015usability}. Furthermore, experimental trials have noted that only a very small percentage of voters actually verify their vote \cite{carback2010scantegrity} \cite{moher1diffusion}, which may critically undermine the security guarantees of these voting systems. We discuss these issues in greater detail in Sec.~\ref{sec:usability}.

Accessability also conflicts with vote privacy. Adapting voting systems to provide audio/visual aids or human assistance for voters with disabilities may create situations where the voter's candidate choice is revealed to a third party. Likewise, deploying voting systems for remote scenarios, such as postal or Internet voting, while significantly more convenient compared to in-person voting at polling stations, results in a marked deterioration in vote privacy \cite{rivest2001electronic}. In her home the voter is not guaranteed privacy and is also vulnerable to coercion.

Faced with such situations, designers of voting systems typically incorporate technological remedies or procedural safeguards into the system, prioritise certain security requirements over others, or perhaps even focus on satisfying certain properties in a weaker form. We encounter several such examples in the following sections.

\section{Cryptographic E2E Voting Systems}
\label{sec:cryptosystems}

In this section, we describe notable E2E voting systems which rely on cryptography to guarantee E2E verifiability. These systems are further categorized depending on the ballot format they support: \emph{physical ballots} (i.e. paper ballots) as opposed to \emph{electronic ballots}, and their mode of deployment, for \emph{precinct-based voting} (i.e. in-person voting at polling stations) versus \emph{remote voting}.

There is considerable cross-fertilization across categories. Many systems incorporate common cryptographic primitives and procedural innovations. Furthermore, several systems share a common lineage and design philosophy which we have attempted to highlight in our presentation. We have also included certain systems which provide only a subset of E2E security properties, i.e. they are not strictly E2E verifiable, but nevertheless make an important contribution to the field.

\subsection{Precinct-based Voting with Physical Ballots}
\label{sec:precinctbasedvotingwithphysicalballots}

This category comprises precinct-based systems in which the tally is computed from receipts of paper ballots which voters mark by hand or using a machine. These systems are directly descended from Votegrity, and prominent examples include Scantegrity and Pr\^{e}t \`{a} Voter, which have been deployed in politically binding elections.

\subsubsection{Votegrity}
\label{sec:votegrity}

Votegrity \cite{chaum2004secret} was invented by Chaum in 2004 and is one of the very first and most influential E2E voting systems.\footnote{Several systems in the literature predating Votegrity offer what are arguably E2E security properties but these are primarily theoretical protocols based on abstract and idealized assumptions. Votegrity is the first E2E verifiable voting system to address the human complexities and practical realities of elections.} Votegrity uses visual cryptography to render the verification process more intelligible to the layman.

\begin{figure}[t]
\centering
\subfigure[Construction of \emph{white/black} pixels]{
  \hspace{2em}
  \includegraphics[width=0.4\textwidth]{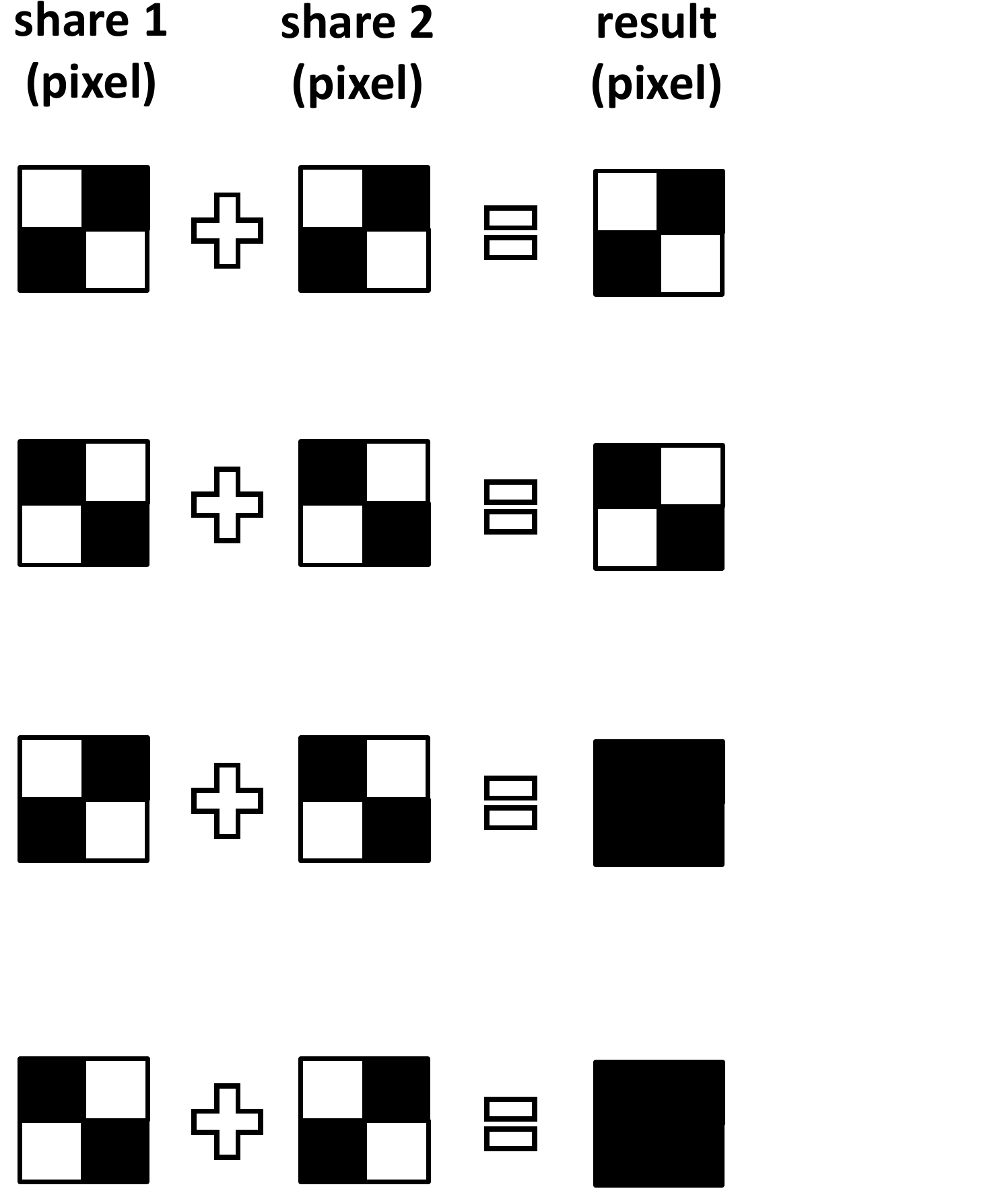}
  \label{fig:visualcrypto}}
  \hspace{2em}
\subfigure[Reconstruction of vote from shares]{
  \includegraphics[width=0.4\textwidth]{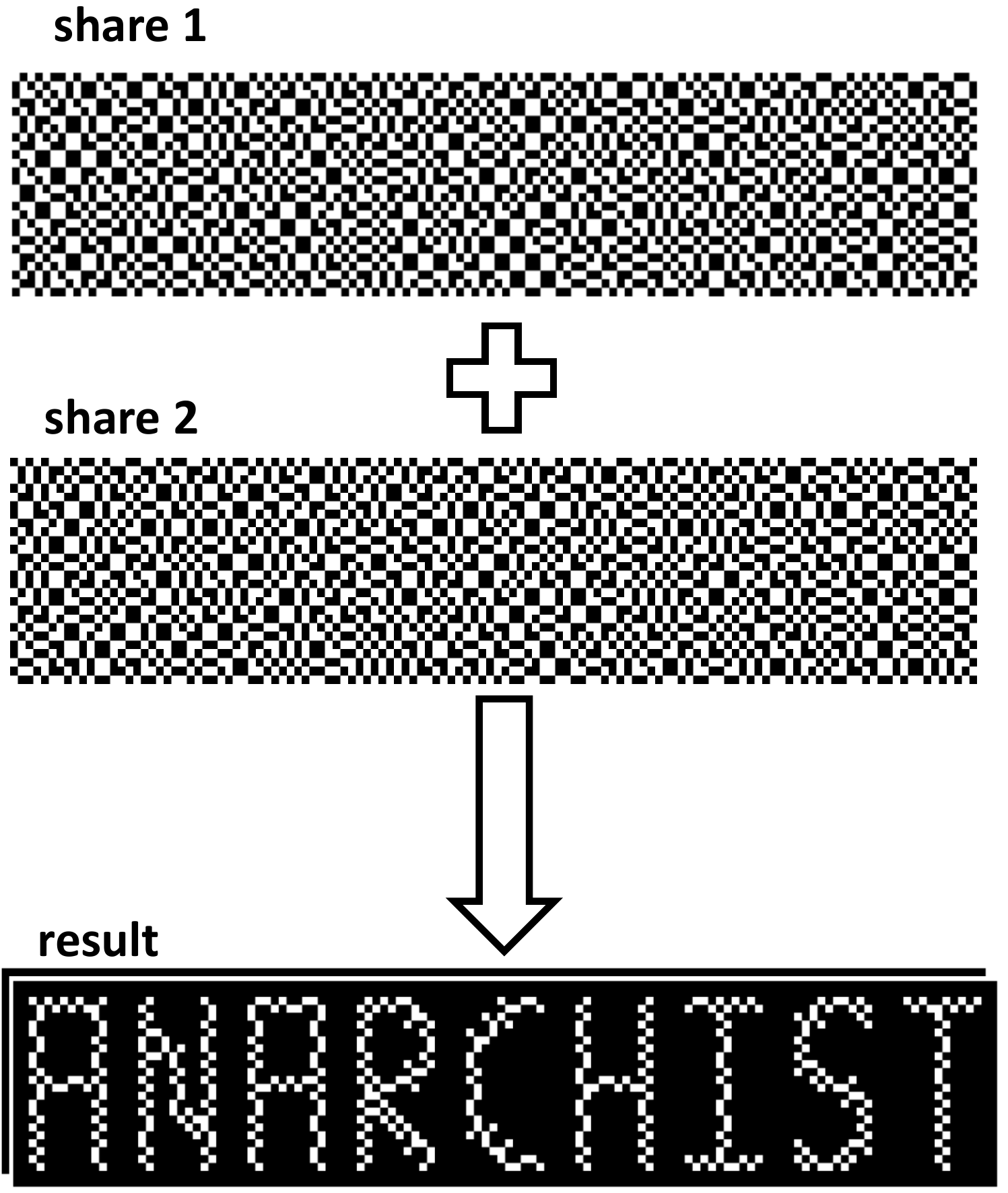}
  \label{fig:visualcryptovote}}
  \caption{Generating a Votegrity receipt (with a vote for \emph{Anarchist})}
\label{fig:votingwithvisualcrypto}
\end{figure}

In \textbf{visual cryptography} \cite{naor1995visual}, an image is split into multiple shares, such that individual shares do not yield any meaningful information about the original image and appear to contain random information. However, when the shares are overlaid the original image is reconstructed. As depicted in Fig.~\ref{fig:visualcrypto} for the case of two shares, each pixel symbol is subdivided into four sub-pixels. When the shares are superimposed, identical pixels on both layers result in a semi-transparent or \emph{white} pixel, whereas, if the two pixels are different, the resulting pixel will be opaque or \emph{black}. Individual pixels on a layer reveal no information about the final result.

Vote processing in Votegrity is undertaken by a group of election trustees.\footnote{The term \emph{election trustee} in this chapter is taken to refer specifically to election personnel who handle cryptographic keys used in the voting systems.}  \textbf{Public-key cryptography} is employed to maintain voter privacy and integrity of the tally. In this paradigm, each trustee possesses two keys, a \emph{public} key he uses for data encryption and which is publicly posted, for example, on the election website or in a directory, and a \emph{private} key used for decryption, which he keeps strictly private. These two keys are mutually related such that any entity can use the trustee's public key to encrypt data but only the trustee himself can decrypt it since he alone owns the corresponding private key. Many security protocols, including voting systems, specify multiple trustees to diffuse responsibility and trust. In this case, a system's security, may only be compromised if all trustees secretly collude. Trustees are therefore typically chosen such that they are mutually distrusting, i.e. they may belong to competing political parties and include activists and members of citizen groups.

On election day, Alice casts her vote at her local polling station. She enters her candidate choice on a voting machine, and presses a button, and a printer attached to the machine prints patterns on two strips of paper using visual cryptography. These strips are presented superimposed under a custom viewfinder, and the voter's choice is clearly visible. The example in Fig.~\ref{fig:visualcryptovote} depicts a vote for \emph{Anarchist}.

Pixels on both strips are generated using a pseudo-random function, i.e. the patterns on the individual images appear random to an observer but are actually generated in a deterministic manner.\footnote{Details on construction of the patterns can be found in \cite{chaum2004secret} and \cite{chaum2007secret}.} The voting machine also imprints  a serial number and some validating information on both strips. Alice randomly chooses one of the strips as a receipt to take home. The machine shreds the other strip, issues the chosen strip to Alice, and saves a digital image of it to memory.

After polls close, the system publishes all saved voter receipts on a public bulletin board or website. Alice can locate her receipt using the serial number and verify that the digital image and validating information matches with her physical receipt. If there is any discrepancy, i.e. if the online receipt is missing or if the two versions do not match, she can initiate an inquiry with election authorities.

Votegrity's key breakthrough is this new role of the receipt. The printed information includes character strings which appear random to an observer but are actually a digital encoding of data which enables election trustees to fully reconstruct both original strips, and thereby the vote itself. When the voting machine generates the strips, it encrypts a digital copy of the visual patterns on both using the trustees' public keys, and this encryption, or \emph{ciphertext}, is then printed on the receipt. The encryption may be understood with the analogy of an onion, where the vote is successively encapsulated in multiple skins or layers of encryption, each corresponding to a particular trustee. In the public-key paradigm only trustees can decrypt a vote by applying their private keys individually to remove corresponding layers of encryption.

\begin{figure}[t]
\begin{center}
  \includegraphics[width=1\textwidth]{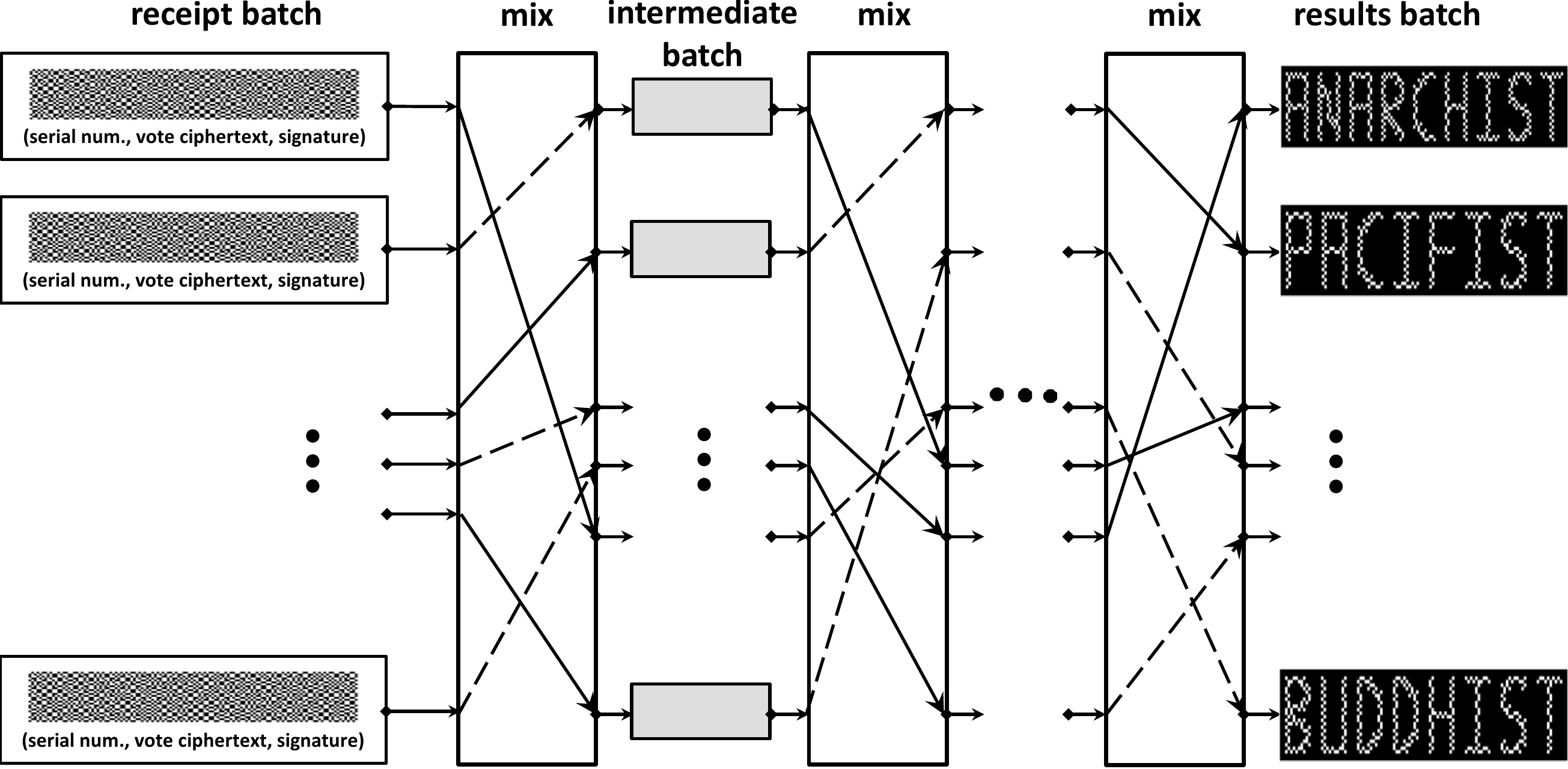}
   \caption{Votegrity decryption, mixing, and audit process}\label{fig:chaummixnet}
\end{center}
\end{figure}

The system decrypts cast votes in a privacy-preserving and verifiable manner using \textbf{decryption mixes}. A mix, invented by Chaum \cite{chaum1981untraceable}, is essentially a \emph{permutation box} which accepts multiple data items, removes a layer of encryption, shuffles the items, and outputs them. A \textbf{mixnet} consists of multiple such mixes connected together in cascading order such that the outputs of one feed into the next, as depicted in Fig.~\ref{fig:chaummixnet}. A mixnet effectively obscures the link between an input and an output and is a key ingredient in various anonymous communications protocols. Each mix is operated by an individual election trustee.

All receipts on the bulletin board are fed into the mixnet. The first mix strips the serial numbers from the receipts in addition to decryption and shuffling. Each mix removes a layer of encryption from the ciphertexts, figuratively peeling the onion, using the private key of the trustee operating the mix. The final result is a set of fully decrypted, or \emph{plaintext}, votes stripped of any identifying information regarding the voter. These votes are then tallied in a straightforward manner.

Each mix processes its batch privately. Alice can be assured her vote is secret as revealing it would require all the trustees collude to track her vote across each mix. However, there is still the issue of integrity: Alice cannot be sure the mixes are correctly processing the receipts and not altering them in transit. Chaum suggests election authorities publicly audit the mixes using a technique known as \textbf{randomized partial checking} \cite{jakobsson2002making}. The key idea is very simple: after tallying is concluded, each mix is forced to reveal a randomly chosen selection of half its inputs and outputs. These are posted on the bulletin board so that any observer can verify that the mix decrypted and mixed the items correctly. If a mix tampers with a single vote, there is a 50\% chance it will get caught in a random check. The odds of being detected increase with every additional vote the mix manipulates, and would be close to a 100\% if a mix manipulates enough votes to influence a large-scale election.

However, while it is essential that mixes be subjected to this check, it is also vital that no receipt on the bulletin board be traced through the mixes to a plaintext vote. This is accomplished by selecting input-output links in an exclusory manner such that no \emph{end-to-end} path across the mixnet is revealed. For example, considering the first two mixes in Fig.~\ref{fig:chaummixnet}, using dashed lines to denote unmasked links, we see that paths revealed in one mix are kept strictly private in the adjacent mix.

Now that the entire system has been described, we discuss the security properties it offers: when Alice casts her vote, she can visually verify that the printer correctly printed her candidate choice. Furthermore, each strip is digitally signed by the voting machine. \textbf{Digital signatures} are cryptographic equivalents of real signatures and have legal standing in several countries. In this case, the voting machine itself has a public/private key pair and their role is reversed: the machine's private key is used to mathematically compute a`signature' on the information on the receipt. This is a string of data which any observer can easily verify using the voting machine's public key (available via the bulletin board or in a public directory). Successful verification is undeniable proof of origin, as only the machine possesses the private key responsible for the signature. In case of a dispute, a digital signature is thus non-repudiable proof that the receipt was issued by a legitimate voting machine.

By digitally signing each strip, the machine may therefore be considered to have \emph{committed} itself to both strips. Alice then chooses one layer at random to take home. If the machine were to cheat and put fraudulent information on single strips for a significant number of voters, it would be detected with very high probability.

Outside the polling station, Alice can verify, using the public keys of the trustees, that the ciphertext on the receipt is indeed a correct encryption of the visual pattern on her receipt. This gives her high assurance that her vote was cast as she intended. She can ensure the system correctly recorded her vote by verifying her physical receipt on the bulletin board. When mixing and tallying are undertaken, the inputs and outputs of every mix are publicly posted on the bulletin board. Any observer, including Alice, can monitor the random checks on the mixnets and confirm the results. Randomized partial checking is well-suited to audit elections as it does not rely on complex cryptography and is therefore efficient and relatively easier for the voter to appreciate. Any observer can verify the final tally for themselves by adding up all the decrypted votes. Furthermore, none of the information on the receipt or the bulletin board leaks any knowledge of the contents of Alice's vote which a third party may exploit to bribe or intimidate her. The visual image on her receipt is incomprehensible without the second strip which the voting machine shredded during the vote-casting phase, and theoretically it could represent a vote for \emph{any} candidate.

In short, Alice can now audit the election at every key stage herself and need not trust election authorities for election integrity. Furthermore, election verifiability is independent of underlying infrastructure such as voting machines or vendor software. Technically-minded voters can create (and distribute) software to perform the necessary verifications using the information on their receipts and the bulletin board.

Votegrity has proved immensely influential and inaugurated new directions for research in secure elections. Several systems the followed borrow the basic receipt-and-mixnet template and improve on practical aspects. We consider these next.

\subsubsection{Pr\^{e}t \`{a} Voter}

\begin{figure}[t]
\centering
\subfigure[Ballot]{\raisebox{8mm}{
  \includegraphics[width=0.2\textwidth]{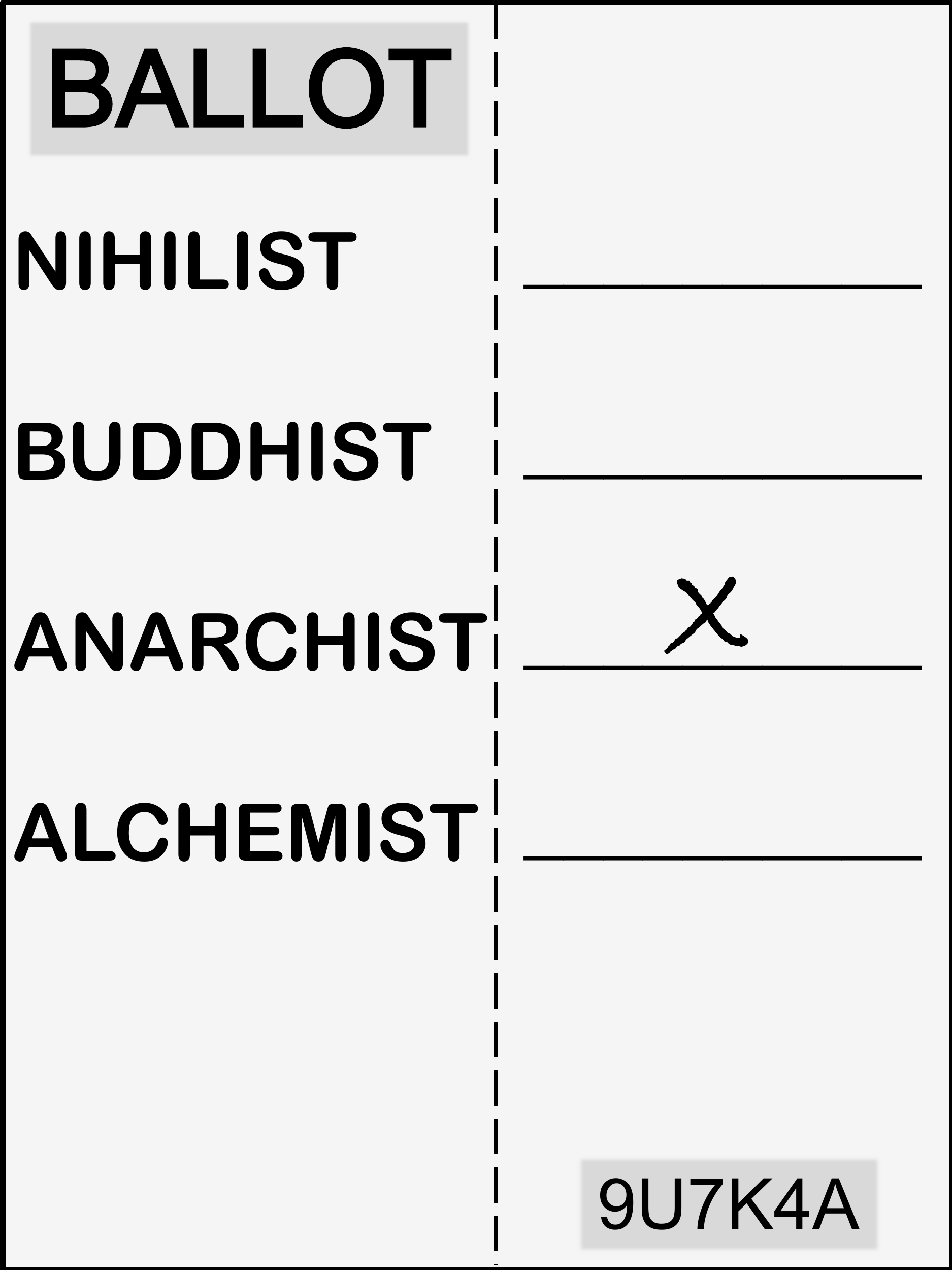}
  \label{fig:pretballot}}}
\subfigure[Vote casting and tallying]{
  \includegraphics[width=0.75\textwidth]{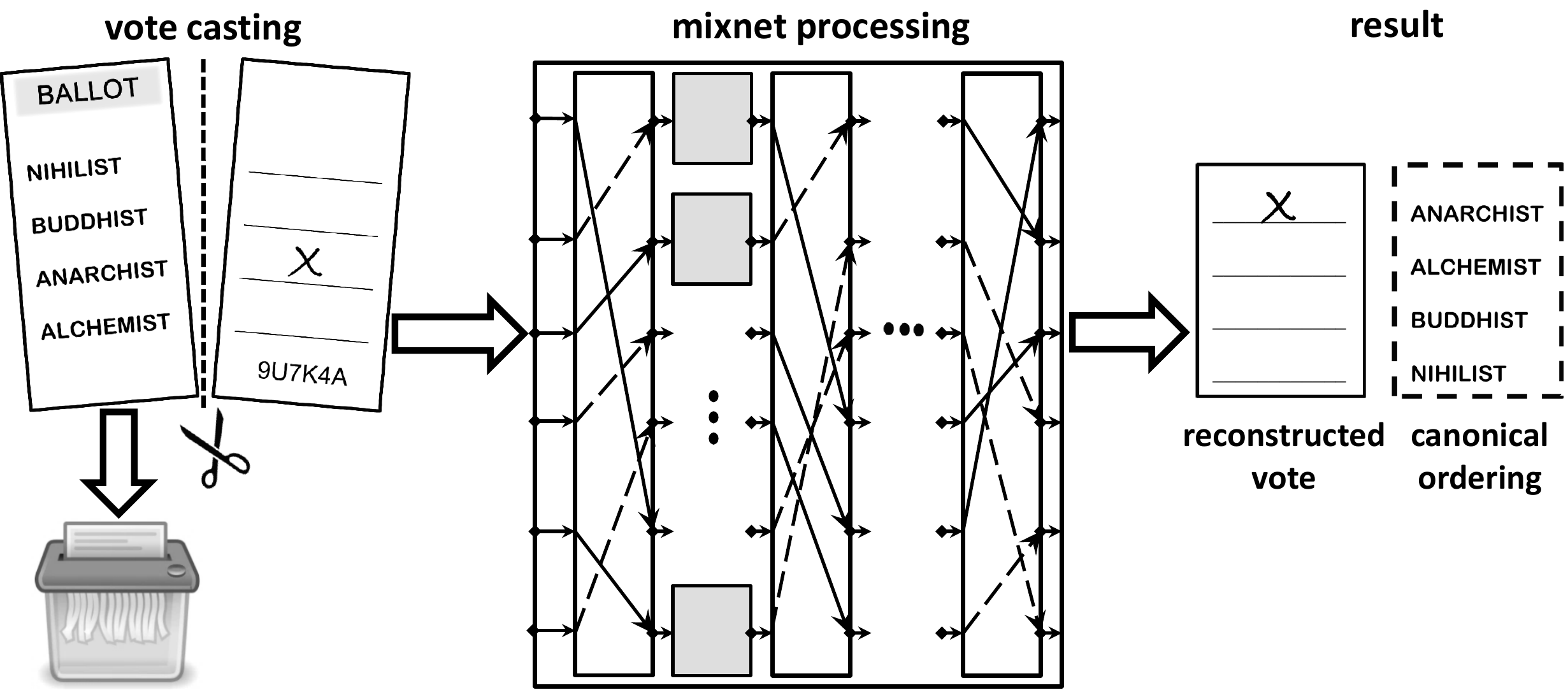}
  \label{fig:pretprocess}}
  \caption{Voting with Pr\^{e}t \`{a} Voter (casting a vote for \emph{Anarchist})}
\label{fig:pretavoter}
\end{figure}

Pr\^{e}t \`{a} Voter \cite{ryan2009pret}, initially proposed by Ryan in 2005 as a variant of Votegrity \cite{ryan2005variant}, was later developed into an independent system by Chaum, Ryan, and Schneider \cite{chaum2005practical}. We provide a brief overview.

Pr\^{e}t \`{a} Voter replaces Votegrity's visual cryptography element with a paper ballot and randomized candidate ordering. The ballot, depicted in Fig.~\ref{fig:pretballot}, is detachable into two halves along the perforated line in the middle. The left half lists candidates in randomized order. The right half has corresponding marking spaces and a random-looking string of alphanumeric characters, referred to as the \emph{onion}, printed at the bottom. The onion encodes the permutation of candidate names on the specific ballot relative to a standard \emph{canonical} ordering, encased in multiple layers of encryption using the public keys of the election trustees, as in Votegrity.

The workings of the system are presented in Fig.~\ref{fig:pretprocess}. To vote, Alice marks her candidate choice on the ballot, detaches the two halves and shreds the left side with the candidate ordering. She then scans the right-hand side and takes it home as her receipt. After polls close, all scanned receipts are posted on the bulletin board where Alice can verify her vote is correctly recorded. However, she cannot prove her choice to a third party with just the right half of the ballot since her mark could potentially correspond to any candidate.

Vote processing is analogous to Votegrity. All receipts on the bulletin board are passed through a series of mixes operated by election trustees who use their private keys to successively peel the onion, thereby reconstructing the voter's choice as per the canonical candidate ordering. The mixing process is audited using randomized partial checking. The reconstructed votes are tallied in a straightforward manner.

Verification needs to be performed on another front as well: voters need assurance that ballots are well-formed, i.e. the onions in the lower right half of the ballot correctly encrypt the randomization of candidates on the ballots. Pr\^{e}t \`{a} Voter includes a step allowing Alice to audit her ballot in the polling booth by requesting the system to decrypt the onions and prove that they correspond to the candidate ordering. Audited ballots do not count towards the tally and are discarded and Alice may audit as many ballots as she likes until she is satisfied the system is not cheating and then cast her actual vote. The authors also recommend that election authorities publicly audit a set of ballots prior to the election to create public confidence. A sufficiently sized random sampling of ballots should be picked, making it very hard for a malicious party to distribute large numbers of malformed ballots without detection. After polls close, any leftover ballots should be similarly audited for greater confidence.

Pr\^{e}t \`{a} Voter is one of the most prominent E2E voting systems and extensive work has been done in maintaining and extending it. Notable contributions include adapting  Pr\^{e}t \`{a} Voter for different electoral systems \cite{heather2007implementing} \cite{xia2010versatile} and voting scenarios \cite{popoveniuc2007simple}, enhancing its usability \cite{lundin2008human}  \cite{ryan2011preta} \cite{culnane2013faster}, and implementing the system \cite{bismark2010experiences}.

A variant of Pr\^{e}t \`{a} Voter was used in binding state elections in the Australian state of Victoria in 2014 \cite{burton2012using} \cite{culnane2014vvote}.

\subsubsection{Punchscan}

Punchscan \cite{fisher2006punchscan}, invented by Chaum in 2006, is especially suited for elections where rules mandate a uniform candidate ordering on ballots. Punchscan takes inspiration from Votegrity's notion of layers and randomizes voter choices on the ballot. The latter innovation is in contrast to Pr\^{e}t \`{a} Voter's approach where the candidate ordering is randomized on ballots.

\begin{figure}[t]
\begin{center}
  \includegraphics[width=0.7\textwidth]{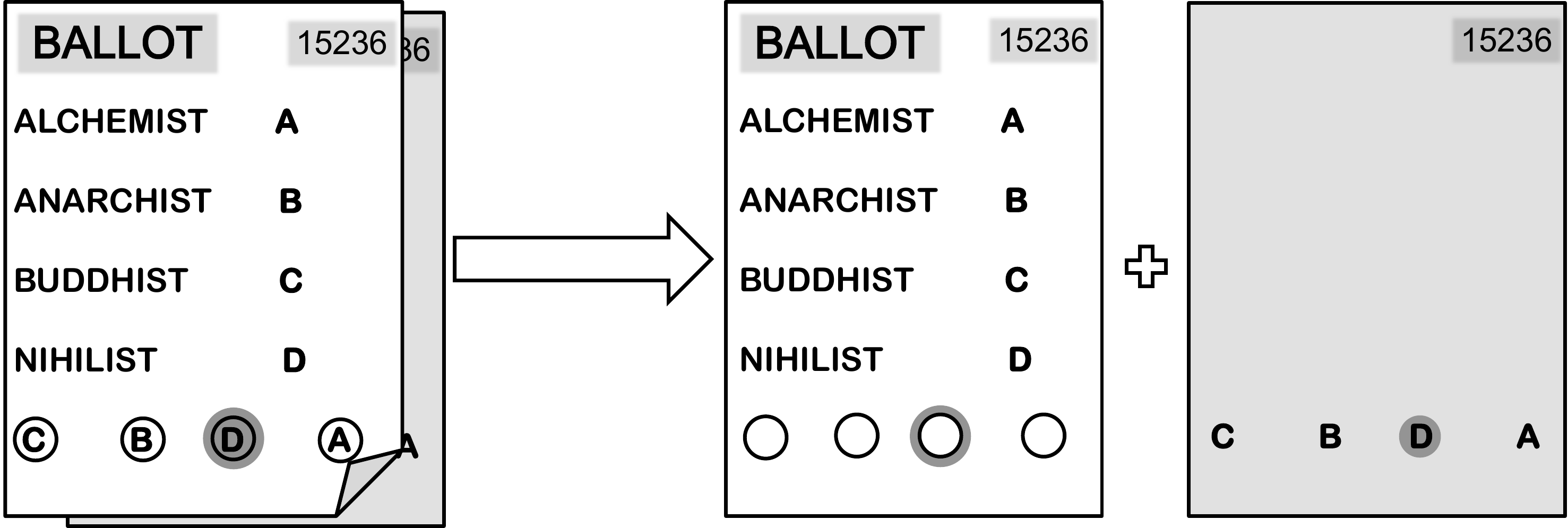}
   \caption{The Punchscan ballot (with a vote for \emph{Nihilist})}\label{fig:punchscanballot}
\end{center}
\end{figure}

The Punchscan ballot, shown in Fig.~\ref{fig:punchscanballot}, consists of two layers. The top layer has the candidate names ordered using numbers or letters (\emph{a, b, c, d}, etc.) and a set of holes at the bottom. The lower layer contains a random ordering of the same letters. When the layers are overlaid, the letters are visible through the holes. In the polling booth, Alice marks her choice, using a marker to daub the appropriate hole, thereby marking both layers at the same time. The layers are then separated. When considered individually, the top layer only indicates which hole she chose, whereas the lower indicates the letter she selected, and both layers are needed by a viewer to discern her candidate choice. Alice randomly picks one layer to cast and shreds the other. She retains a copy of the cast layer as her receipt.

Vote processing is similar to the systems described earlier. When polls close, Alice can check the bulletin board to verify that her vote was recorded as cast. The ballot serial number, printed in the top right corner on both layers of the ballot, encodes encrypted information which enables election trustees to determine the voter's choice. Punchscan replaces the back-end decryption mixnet of earlier systems with an anonymizing database referred to as the \emph{Punchboard}.

The Punchboard consists of three interlinked tables. The first one lists all markable options on all created ballots whereas the last lists the candidate corresponding to each option. For instance, for Fig.~\ref{fig:punchscanballot}, the first table will list an entry for the third option which corresponds to a vote for \emph{Nihilist} in the last table. However, the direct links between entries on these these two tables are shuffled via an intermediate table and the shuffling pattern is masked using encryption. This is conceptually similar to how a mix randomizes its inputs but is far more efficient as votes do not have to be encrypted and decrypted multiple times. Instead the system transmits the marks on each receipt over the set of secret paths directly to the appropriate candidates, incrementing their vote count accordingly.

The Punchboard and ballots are generated by election trustees prior to the election and their configuration is kept strictly secret. A preliminary audit is conducted in which half the ballots are randomly picked and publicly revealed along with the corresponding paths in the Punchboard. Any party can verify that votes cast on these ballots would most certainly have been routed to the right candidates and thereby derive confidence in the system. Audited ballots are then destroyed and the remaining half used in the polls. After tallying concludes, the Punchboard is again audited using randomized partial checking, i.e. each path is partially decrypted to prove the system is operating correctly without revealing any end-to-end paths in the process.

Some attacks have been discovered against Punchscan \cite{kelsey2010attacking} and corresponding procedural safeguards have been recommended. Other notable research contributions include adapting Punchscan for remote scenarios \cite{popoveniuc2007simple}, reducing opportunities for privacy violation \cite{carback2007independent}, and merging Pr\^{e}t \`{a} Voter and Punchscan for stronger privacy \cite{van2009voting}.

Punchscan source code was released under an open-source license and the system was trialled in elections of the University of Ottawa's Graduate Students' Association in 2007 \cite{essex2007punchscan}. Punchscan was later merged into the Scantegrity voting system.

\subsubsection{Scantegrity}
\label{sec:scantegrity}

Scantegrity \cite{chaum2008scantegrity}, developed in 2008 by a team of researchers including David Chaum and Ron Rivest, enhances existing real-world voting systems with E2E verifiability. The advantage of this approach is that infrastructure, such as optical scan technology, is already widely deployed and voters are familiar with its usage.

\begin{figure}[t]
\begin{center}
  \includegraphics[width=0.7\textwidth]{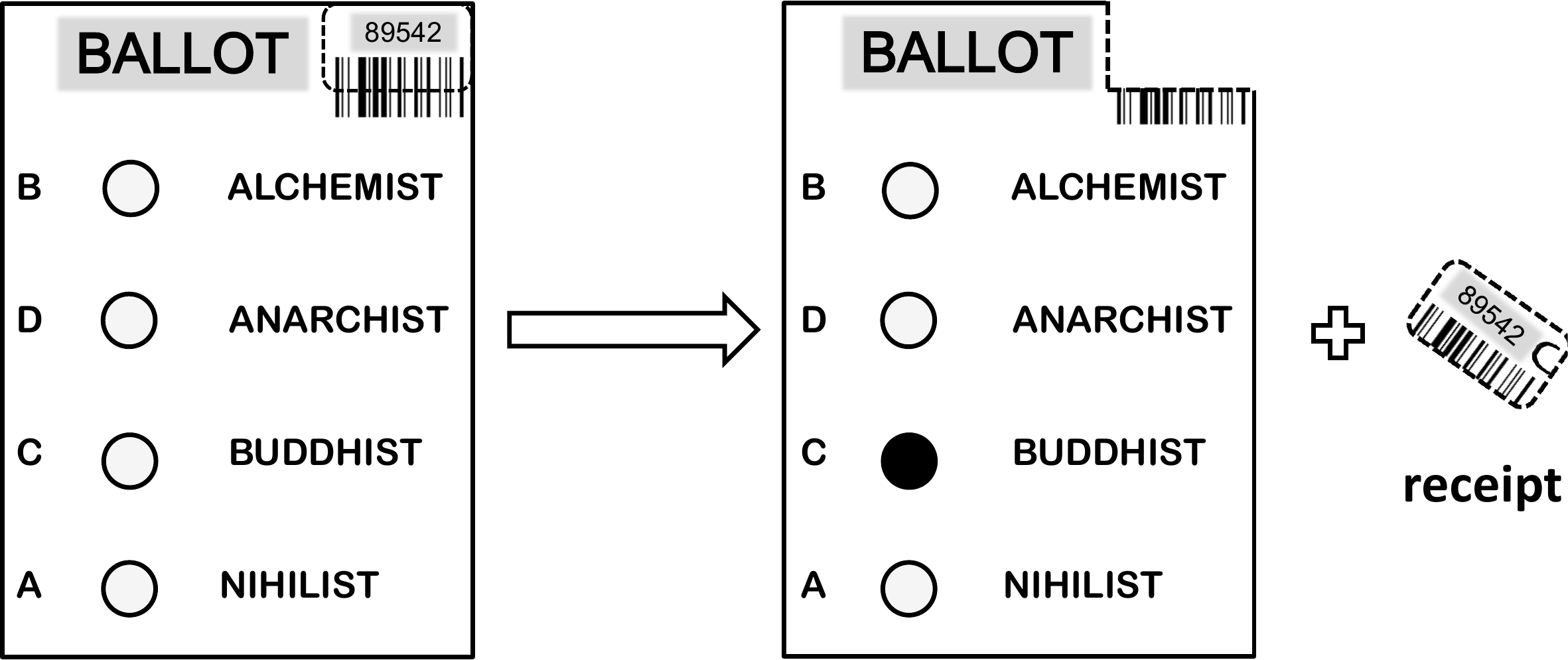}
   \caption{The Scantegrity ballot (with a vote for \emph{Buddhist})}\label{fig:scantegrityballot}
\end{center}
\end{figure}

The Scantegrity ballot, depicted in Fig.~\ref{fig:scantegrityballot}, lists the candidates with randomly picked code letters assigned to each. A ballot serial number is printed in the top right hand corner, expressed both in human-readable and barcode format. In the polling booth, Alice marks the candidate of her choice and scans the ballot into an optical scanner as she would using a regular optical scan voting system. She then detaches the perforated corner of the ballot and notes down on it the code letter for her candidate. This small stub is her Scantegrity receipt.

After polls close, election authorities tally the results of the optical scan system. Scantegrity vote processing runs independently of this count. Election authorities post all ballot serial numbers and the corresponding code letters onto the bulletin board, where voters can verify that their code letters have been correctly recorded by the system. The code letter by itself does not reveal the voter's choice of candidate to a third party as the letters are randomly assigned on ballots. Similar to Punchscan, every code letter on every ballot translates to a vote for a specific candidate in an anonymizing database, in this case referred to as the \emph{switchboard}. Votes are routed accordingly over encrypted paths in a privacy-preserving and auditable manner.

This system suffers a few disadvantages: a coercer may force a voter to mark a certain code letter regardless of which candidate it is assigned to, thereby effectively randomizing her vote. Furthermore, there is no guarantee the voter correctly records the same code on her receipt that she marks on the ballot, and dispute resolution entails a complicated mechanism which involves producing the physical ballot.

\begin{figure}[t]
\begin{center}
  \includegraphics[width=0.7\textwidth]{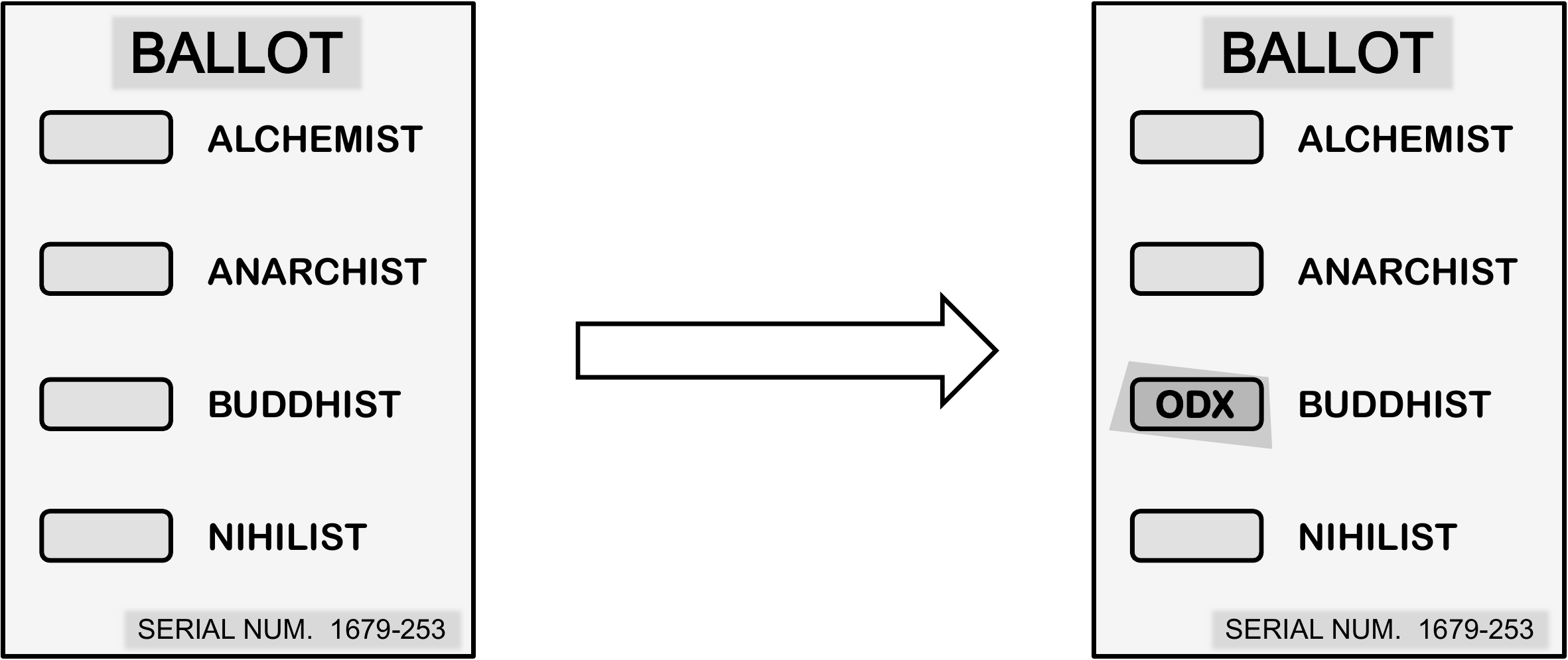}
   \caption{The Scantegrity II ballot (with a vote for \emph{Buddhist})}\label{fig:scantegrity2ballot}
\end{center}
\end{figure}

Scantegrity II \cite{chaum2008scantegrity} resolves these issues by printing candidate codes on the ballot using invisible ink which are revealed only when the voter marks her candidate choice using a special pen. These codes are randomly assigned and are of sufficient length that they cannot simply be deduced by guesswork. If a voter therefore lodges a complaint citing a valid code, her complaint is very likely genuine.  The process is depicted in Fig.~\ref{fig:scantegrity2ballot} with a vote for \emph{Buddhist}. The voter retains the candidate code \emph{ODX} and ballot serial number \emph{1679-253} as a receipt.

Scantegrity II was trialled in a binding governmental election for the offices of mayor and city council members in Takoma Park, Maryland \cite{sherman2010scantegrity}. The experience led to further modifications to the system to incorporate features requested by voters and election officials, resulting in Scantegrity III \cite{sherman2011scantegrity}.

\subsubsection{Scratch \& Vote}
\label{sec:scratchandvote}

In a paper in 2006, Adida and Rivest \cite{adida2006scratch} noted that mixnets entail complicated mixing and auditing protocols, and require the voter to have faith that election trustees will not collude to subvert the election. To address these shortcomings, they invented Scratch \& Vote  (S\&V), which combines the user-friendly voting experience of Pr\^{e}t \`{a} Voter with the backend simplicity of homomorphic encryption.

Ballots are encrypted using \textbf{homomorphic encryption}, a cryptographic primitive which enables them to be tabulated while still in encrypted form. The result of a homomorphic addition on a set of ciphertexts is equivalent to an addition operation performed on the set of plaintexts. Only the result of the addition operation is decrypted, thereby preserving the individual voter's privacy. Homomorphic encryption was first applied to electronic voting in 1985 by Josh Benaloh in his landmark work \cite{cohen1985robust} \cite{benaloh1986distributing} and is a key ingredient in several voting systems that follow.

To provide voters with further security assurances, S\&V distributes the corresponding decryption key among multiple election trustees using threshold cryptography, thereby minimizing risk of abuse of cryptographic credentials. \textbf{Threshold cryptography} \cite{pedersen1991threshold} is  conceptually similar to visual cryptography (discussed earlier in Sec.~\ref{sec:votegrity}) in that a decryption key is mathematically split into several shares which are distributed among mutually distrusting trustees. The decryption therefore requires the explicit cooperation of a threshold amount of these trustees, and at no time does any party possess the complete key.

Furthermore, vote encryption is also \textbf{probabilistic}. In public-key cryptosystems, we recall that the public key used for encryption is not secret, it is accessible to everyone. If Alice's vote were therefore encrypted in a deterministic manner, an attacker could easily mount a \textbf{brute-force attack} to deduce it, i.e. he could use the trustees' public keys to encrypt every possible voting option on Alice's ballot and compare the ciphertexts thus generated with those printed on Alice's receipt or on the bulletin board. Probabilistic encryption techniques insert random numbers into the encryption algorithm so that every encryption operation on the same plaintext yields different ciphertexts. These random numbers are large enough that they cannot be simply guessed and they are kept strictly secret. Probabilistic encryption is important in several security protocols and is used in all the voting systems described in this chapter reliant on encryption, including the ones described earlier. The reason for highlighting its use here is that, in S\&V, the probabilistic nature of the encryption plays a vital role in the vote verification process.

\begin{figure}[t]
\begin{center}
  \includegraphics[width=0.75\textwidth]{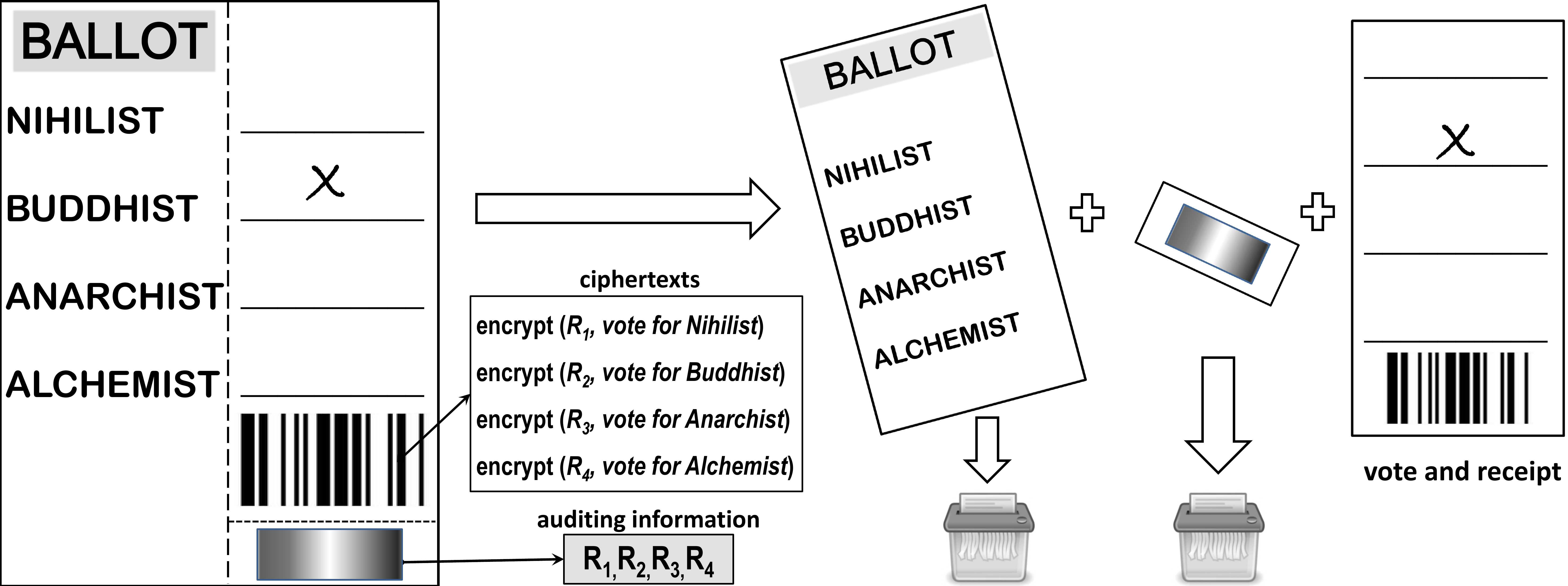}
   \caption{The S\&V ballot and vote casting process (with a vote for \emph{Buddhist})}\label{fig:scratchvoteballot}
\end{center}
\end{figure}

The S\&V ballot, depicted in Fig.~\ref{fig:scratchvoteballot}, detaches into three parts. The left half lists candidate names in randomized order. The right side bears corresponding markable spaces and a barcode. This barcode encodes multiple encrypted data, each ciphertext corresponding to a \emph{vote} for a particular candidate on the ballot. In the lower right corner, a detachable scratch surface conceals the random numbers used to encrypt the barcode data. Any party can therefore scratch off this surface to reveal the random numbers, use them to independently encrypt the candidate choices, and compare it with the barcode data, thereby confirming that the ballot is correctly formed.

At the polling station, every voter is handed two ballots, one to audit and one to cast. Alice randomly selects one, removes the scratch surface and verifies that the candidate ordering corresponds to the barcode data. The authors note that voters generally do not possess the technical expertise to perform this check and suggest that trusted helper organizations may assist them at the polls by providing equipment and guidance. Removing the scratch surface also invalidates the ballot and it can no longer be used. Alice then proceeds to cast her vote with the second ballot, as depicted in Fig.~\ref{fig:scratchvoteballot}. In a manner similar to Pr\^{e}t \`{a} Voter, she marks her chosen candidate, removes and discards the left side, and hands the ballot to a member of polling staff. He ensures the scratch surface on the ballot is intact, and detaches and destroys it without revealing the underlying content. This step is vital as knowledge of the random numbers would allow a coercer to reconstruct Alice's vote from her receipt. The ballot, now consisting only of the voter mark and the barcode, is cast. Alice is issued a scanned copy as a receipt.

After polls close, Alice can verify on the bulletin board that her vote is correctly recorded. To tabulate results, election staff examine each cast ballot and extract the particular ciphertext from the barcode corresponding to the voter's mark. This ciphertext is essentially an encrypted electronic ballot where the underlying plaintext assigns a `1' to the candidate of the voter's choice and `0' for all others. An addition operation is performed over all extracted ciphertexts such that the the ones and zeros assigned to candidates in the individual votes are directly summed up. Election trustees then engage in a protocol to decrypt the result using their individual shares of the decryption key, and post cryptographic proofs of correct decryption on the bulletin board. Any observer can easily download the contents of the bulletin board and run his own checks to verify that votes have been correctly summed and decrypted. Individual votes are never decrypted, thereby maintaining voter privacy.

Adida and Rivest also present extensions to this basic scheme, including steps to adapt S\&V to a PunchScan ballot, facilitate multiple races, and include larger numbers of candidates. They discuss procedural missteps and attacks which weaken the security of S\&V and note that these are common to systems like Pr\^{e}t \`{a} Voter and Punchscan, where ballots are filled out by hand.

\subsection{Precinct-based Voting with Electronic Ballots}

This category consists of systems where electronic votes are cast and tallied in a precinct-based setting. Systems in this category are characterized by inventive combinations of various cryptographic primitives and include influential systems such as MarkPledge and the notion of Voter Initiated Auditing.

\subsubsection{MarkPledge}
\label{sec:MarkPledge}

MarkPledge \cite{neff04practical}, invented by Neff in 2004, is, alongside Chaum's Votegrity, one of the earliest E2E voting systems. The key innovation of this system is a novel front-end protocol allowing voters to directly challenge voting machines in the polling booth to prove that they are encrypting votes correctly.

\begin{figure}[t]
\subfigure[Bit encryption and challenge]{
\centering
  \includegraphics[width=0.4\textwidth]{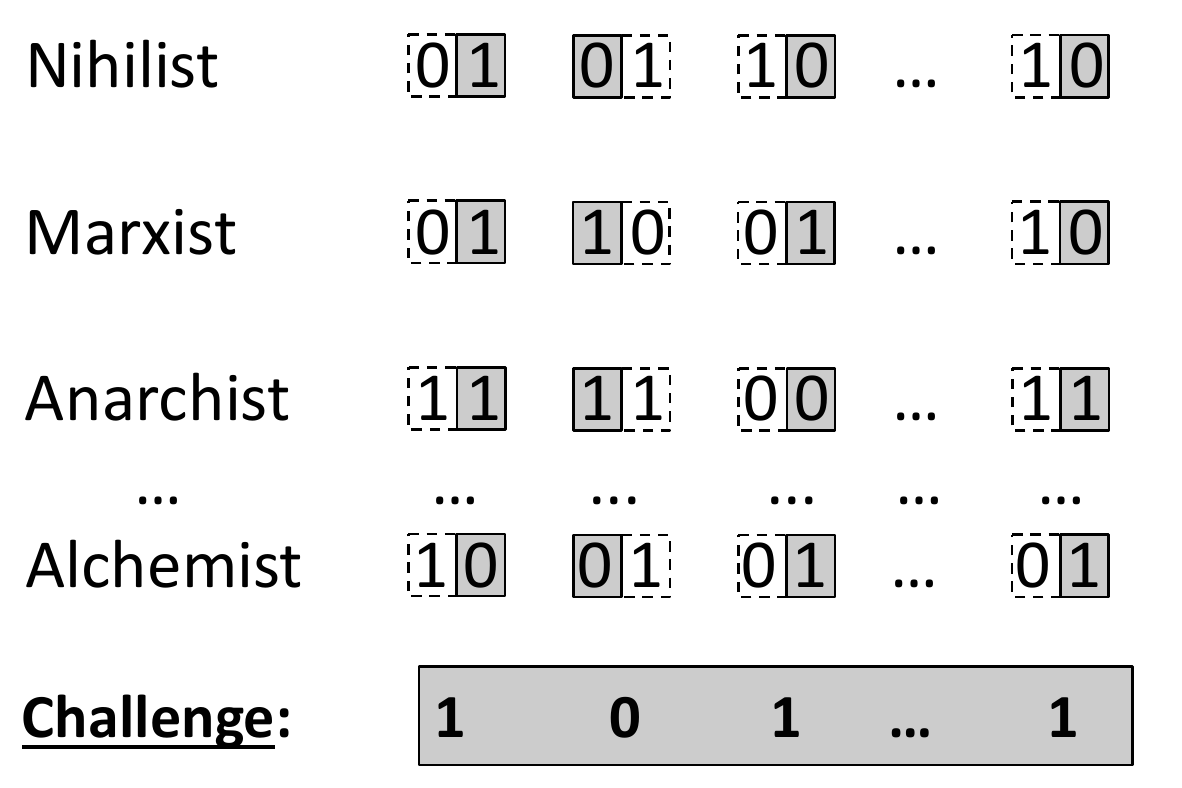}
  \label{fig:markpledge}}
  \hspace{2em}
\subfigure[Receipt]{
\centering
  \includegraphics[width=0.2\textwidth]{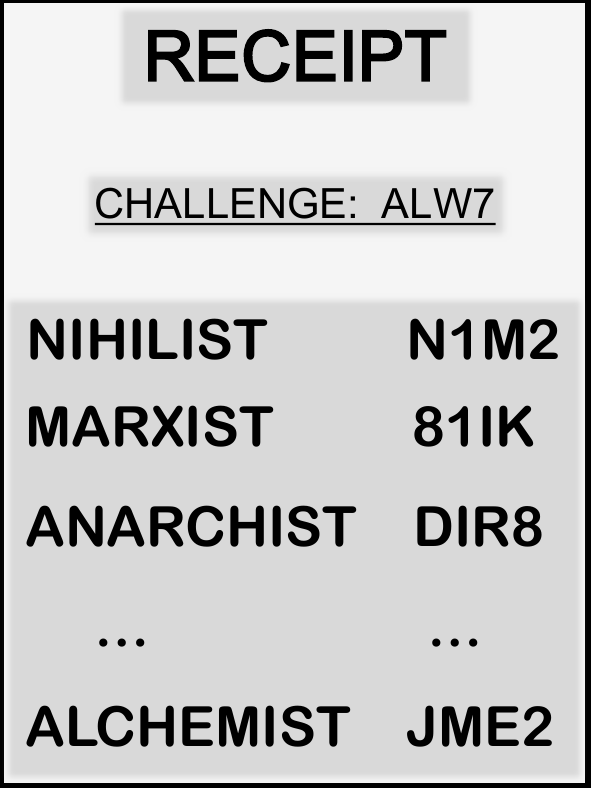}
  \label{fig:markpledgereceipt}}
  \caption{Vote verification and receipts for MarkPledge (with a vote for \emph{Anarchist})}
\label{fig:votingwithmarkpledge}
\end{figure}

On the MarkPledge ballot, each candidate is assigned a row of \emph{indexes} denoted by pairs of bits, as shown in Fig.~\ref{fig:markpledge}. A \emph{No} vote is denoted by dissimilar combinations such as (0,1) and (1,0), and a \emph{Yes} vote is denoted by a pairs of similar bits, (1,1) or (0,0). When Alice chooses a candidate, the voting machine will assign her chosen candidate a \emph{Yes} vote and \emph{No} votes to all the rest. The machine then individually encrypts every bit in each index on the ballot and displays on-screen a \emph{commitment} for the chosen candidate. This commitment is essentially a bitstring, a series of bits where each individual bit corresponds to the \emph{Yes/No} content of each index, i.e. \emph{both} ciphertext bits for the chosen candidate. For example, the commitment 1,1,0,...,1 for \emph{Anarchist} (encoded in an alphanumeric format as \emph{DIR8}) implies that the bits, as depicted, are (1,1), (1,1), (0,0),...,(1,1) which are all \emph{Yes} combinations whereas all other candidates are assigned \emph{No} combinations.

As in Scratch \& Vote, a probabilistic encryption technique is used to encrypt votes. Once the machine has encrypted her vote, Alice can challenge it by making it reveal the randomness used in encrypting either the right or the left bit of indexes (but not both) for all candidates. This process is depicted in Fig.~\ref{fig:markpledge}. Alice's challenge consists of a bitstring, where a 0 denotes bits on the left-hand side and 1 denotes bits on the right-hand side. Alice can thus confirm that each revealed bit matches with the commitment displayed earlier by the machine. The machine cannot predict beforehand which bits Alice will choose, and if it has tampered with her vote (by encrypting \emph{No} bit combinations for her candidate of choice), there is a very high probability that the deception will be revealed in the bits that Alice forces the machine to unmask. This mechanism assures Alice that the machine has encrypted her vote correctly and that her vote is cast as intended.

Once the challenge is concluded, the machine then generates \emph{No} votes for all other candidates on the ballot with the important innovation that it populates Alice's chosen right/left bit positions to make it appear that they are \emph{Yes} votes. All this information, including encrypted indexes, the machine's commitment, Alice's challenge, and the random elements are printed on Alice's receipt in the form of short strings of alphanumeric characters, as depicted in Fig.~\ref{fig:markpledgereceipt}. After polls close, all receipts are posted on the bulletin board and voters can confirm that their votes are correctly recorded. Votes are decrypted and tallied using mixnets in an auditable process similar to schemes described earlier in Section~\ref{sec:precinctbasedvotingwithphysicalballots}.

Alice only needs to ensure that the commitment string the machine originally displayed is printed next to the candidate of her choice (i.e. \emph{DIR8} for \emph{Anarchist} in this case) and she has to retain this string in memory for a short period while challenging the machine. A coercer, examining her receipt at a later time, will only see \emph{Yes} votes for every candidate, and will be unable to deduce her actual vote. Alice alone can differentiate her original \emph{Yes} vote from a fake Yes vote, due to the challenge interaction in the privacy of the polling booth, but is unable to prove it to a third party. This is also the reason that the machine only decrypts the right or left bit position, and not both, during a challenge: if Alice were to record the randomness used to encrypt both bits of an index, she could easily prove her vote to a third party.

MarkPledge is analyzed in greater detail by Adida and Neff \cite{adida2006ballot} who propose that \emph{helper organizations}, consisting of political party members or activist groups, assist the voter with more detailed verification checks. Joaquim and Ribeiro suggest a mechanism to make the Markpledge cast-as-intended verification process faster and more efficient \cite{joaquim2012efficient}. This front-end verification technique has also inspired other systems: Joaquim et al. propose two remote E2E voting systems, VeryVote \cite{joaquim2009veryvote} and EVIV (an End-to-end Verifiable Internet Voting system) \cite{joaquim2013eviv} which combine this technique with code voting (described later in Section~\ref{sec:remotevotingwithelectronicballots}) and mixnets.

\subsubsection{Bingo Voting}
\label{sec:bingovoting}

Bohli, M\"{u}ller-Quade and R\"{o}hrich invented Bingo Voting \cite{bohli2007bingo} in 2007, inspired by the concept of a bingo machine as a source of randomness to protect voter privacy.

In the election setup phase, election trustees generate strings of random numbers, such that there is one random number per voter for each candidate on the ballot (i.e. for \emph{m} voters and \emph{n} candidates, \emph{m}$\times$\emph{n} random numbers are generated, grouped such that each candidate is assigned one set of \emph{n} numbers). These numbers, referred to as \emph{dummy votes}, are kept secret, but the trustees publicly commit to them on the bulletin board before elections using a commitment scheme.

A \textbf{commitment scheme} is a cryptographic primitive which enables a party to publicly commit to a chosen secret without revealing it until a later time. This may be intuitively visualized with the analogy of a locked box which a \emph{prover} publicly entrusts to a \emph{verifier}. At a later time, the prover uses his secret key to unlock the box and reveal its secret contents. No one else can open the box. Commitment schemes are also \emph{binding} in the sense that the prover cannot alter the contents of the box without being detected by the verifier. In our case, election trustees cannot change the dummy votes once the commitment has been published. Commitment schemes are employed in various security protocols, and applications include timestamping data and trusted computation.

Alice chooses her candidate on a voting machine in the polling booth. The booth also contains a mechanical device, similar to a bingo machine, which Alice uses to generate a fresh random number. The device displays this number and transmits it to the voting machine which assigns it to Alice's candidate choice. The other candidates are assigned numbers chosen from the pool of pre-generated dummy votes. This particular configuration of candidate names and random numbers constitutes Alice's vote and the voting machine issues her a receipt. Alice simply has to confirm that the random number she has generated herself in the booth is assigned to the candidate of her choice on the screen of the voting machine and on her receipt. After polls close, all issued receipts are published on the bulletin board.

Tallying is straightforward. Every time a vote is cast for a certain candidate, all other candidates are assigned dummy votes. In this way, a dummy vote initially allocated to the chosen candidate in the pool of dummy votes is not used. Therefore, the final tally can be computed by examining this pool and the candidate with the largest number of unused dummy votes is the winner of the election.

However, election trustees need to prove that voting machines correctly allocated the dummy votes to candidates. Moreover, this needs to be accomplished without publicly differentiating between dummy votes and voter-generated random numbers on receipts, else voter privacy would be compromised. First, election trustees publish the list of unused dummy votes on the bulletin board alongside their respective commitments and the corresponding reveal information. Next, the trustees publish cryptographic proofs testifying to the integrity of the receipts, i.e. each receipt contains exactly one voter-generated random number, each dummy vote used in the elections was correctly allocated by the machines, and it was used only once. These proofs can be verified by any observer and do not reveal any information which may enable him to identify dummy votes and voter-generated random numbers.

Alice verifies her vote was cast as intended by ensuring the random number she generates in the voting booth is assigned to the candidate of her choice on her receipt. She ensures her vote is recorded correctly by verifying her receipt on the bulletin board. Her privacy is preserved and she is protected from coercion as the newly generated random number assigned to her candidate is indistinguishable from the dummy votes assigned to the other candidates on the receipt. She checks the revealed commitments and the cryptographic proofs to verify that every random number used in the election is strictly accounted for and that the tally has been correctly computed.

Bingo Voting was deployed in student parliament elections in Karlsruhe University, Germany, in 2008  as described in \cite{bar2008real}. Another study examines the ramifications of corrupted voting machines on Bingo Voting and prescribes solutions which may be generalized for other voting systems \cite{bohli2009enhancing}. Liu et al. \cite{liu2012improved} note that reliance on a random number generator is a potential security vulnerability and devise an alternative solution. Henrich \cite{henrich2012improving} analyzes the feasibility of deploying Bingo Voting for recent real-world elections in the United States, Germany, and India.

\subsubsection{Voter Initiated Auditing}
\label{sec:voterinitiatedauditing}

In most of the systems described thus far, random checks are integral to the security guarantees of the systems. When Alice casts her vote, she has no explicit guarantee that the machine has encrypted her vote correctly. For systems like Pr\^{e}t \`{a} Voter or Scantegrity which use physical ballots, she has no direct assurance that the candidate randomization or marking options on her ballot are correctly encoded. Her confidence, instead, derives from the random audits conducted on the system at every key juncture and the knowledge that these audits are overseen by multiple trustees. With these checks, Alice may be certain that any significant malfeasance will be detected.

In contrast to this approach, in 2006, Josh Benaloh introduced the notion of \textbf{voter initiated auditing} \cite{benaloh2006simple} \cite{benaloh2007ballot} \cite{benaloh2008administrative} to give the voter immediate confidence that her vote has been correctly cast. In this paradigm, when Alice chooses her candidate, the voting machine prints an encryption of the vote on her receipt but does not dispense it. Instead, the machine asks Alice whether she wishes to \emph{cast} the vote or \emph{challenge} it. If she trusts the machine, she selects the \emph{cast} option and the printer issues her the receipt. If Alice is uncertain that the machine has encrypted her vote correctly, she can challenge it to prove it is honest (this step is often referred to as a `Benaloh' challenge). The machine then prints on the receipt, alongside the encrypted vote, the contents of Alice's ballot and the randomness used to generate the encryption. This data allows Alice, or any other party, to replicate the encryption step independently using the election public key, as described earlier in Section~\ref{sec:scratchandvote}.

The novelty here is that the machine essentially \emph{commits} to the vote encryption by printing it on the receipt before Alice chooses whether to cast or challenge. If Alice initiates a challenge, she can check if the machine performed the encryption correctly. She is welcome to challenge the machine as many times as she likes until she is confident that the machine is honest before finally casting her vote. If the machine has cheated by encrypting a vote for a different candidate, the odds that it will get caught increase with every challenge.

Voter Initiated Auditing was originally proposed by Benaloh, not as a new system, but as a vote-casting technique to augment existing real-world voting systems with E2E security guarantees, a goal it shared with Scantegrity. It has since then directly inspired other well-known systems such as Helios, VoteBox, and STAR-Vote.

\subsubsection{VoteBox}
\label{sec:votebox}

VoteBox \cite{sandler2008votebox}, proposed by Sandler, Derr and Wallach in 2008, attempts to resolve myriad practical issues encountered in building secure and usable voting systems.

VoteBox follows the template set by Voter Initiated Auditing. The system uses homomorphic encryption and the decryption key is pre-distributed among multiple election trustees using threshold cryptography. On election day, Alice makes her candidate selection on a voting machine, which encrypts her vote and then offers her the option to challenge or cast it. If challenged, the machine prints extra encryption and randomness information on her receipt. Alice takes this to a special terminal in the polling station dedicated to assisting voters to verify challenged votes. Alice may mount as many challenges as she likes before casting her vote.

All receipts are posted on the bulletin board where Alice can verify her vote is correctly recorded. Tallying consists of straightforward homomorphic addition after which election trustees engage in a protocol to decrypt the final result and provide cryptographic proofs attesting to correct decryption. Voters and observers alike may download the contents of the bulletin board and verify the computations.

Apart from its E2E security qualifications, a key contribution of VoteBox is how it brings together various innovations in the research literature to address a range of security and reliability issues regarding the practical side of electronic voting. We list a few highlights: the authors go to considerable lengths to minimize and partition the amount of trusted code running on voting machines. Distinct functions of the voting machine are modularized in hardware and clearly partitioned, thereby enabling administrators to efficiently audit hardware faults. The voting experience is standardized across all voting machines by specifying a graphical user interface (GUI) consisting of pre-rendered bitmap images which correspond to the actions of voters using the machine \cite{yee2006prerendered}.

Voting machines log all \emph{machine events} (system messages, the complete record of encrypted votes, and voter challenges and machine responses) using secure logging techniques \cite{sandler2007casting}. All machines broadcast and record event logs over a local network, so that, were a voting machine to fail on election day, all its data up to the point of failure may be retrieved from the storage of other machines. This also makes things considerably more difficult for an attacker who now has to compromise all machines in the precinct to tamper with votes undetected. Furthermore, the local network is connected to the Internet using a data diode \cite{jones2006secure} which enforces strict one-way data flow i.e. logs of voting machines in the precinct are publicly broadcast over the Internet, yet remote attackers are not able to hack into the machines.

VoteBox has inspired some enhancements: RemoteBox \cite{sandler2008case} extends VoteBox to secure elections in remote voting facilities such as overseas embassies and consulates. VoteBox Nano \cite{oksuzoglu2009votebox} reimagines the VoteBox design using only low-cost hardware components, such as field programmable gate arrays. This approach obviates the need for an operating system or software on voting machine, which an attacker may potentially corrupt. Furthermore, any tampering with hardware may be visually detected by polling staff very easily on election day.

\subsubsection{Wombat}

Wombat \cite{ben2012new}, invented by Ben-Nun et al. in 2011, combines the frontend of Voter Initiated Auditing with Pr\^{e}t \`{a} Voter-style mixnet processing in the backend.

\begin{figure}[t]
\centering
\subfigure[The ballot]{
\centering
  \includegraphics[width=0.2\textwidth]{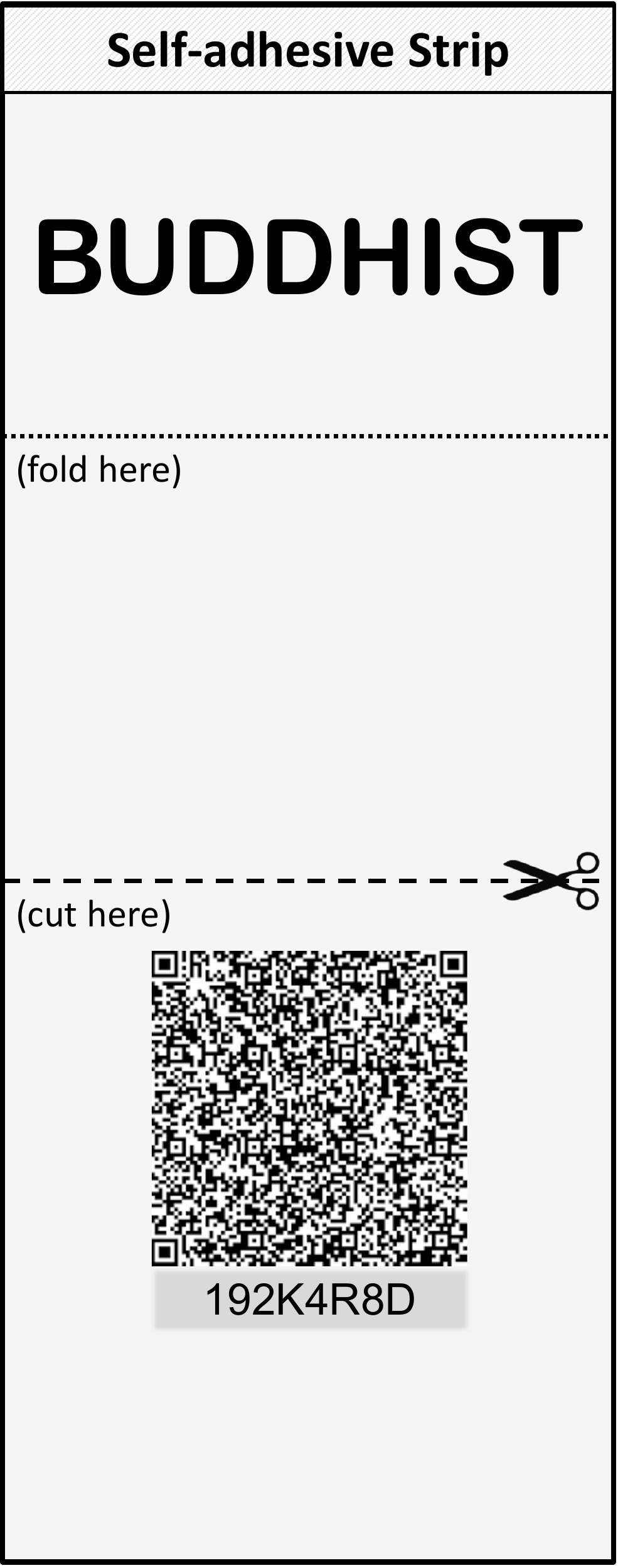}
  \label{fig:wombatballot}}
\subfigure[Auditing the ballot]{
\centering
  \includegraphics[width=0.2\textwidth]{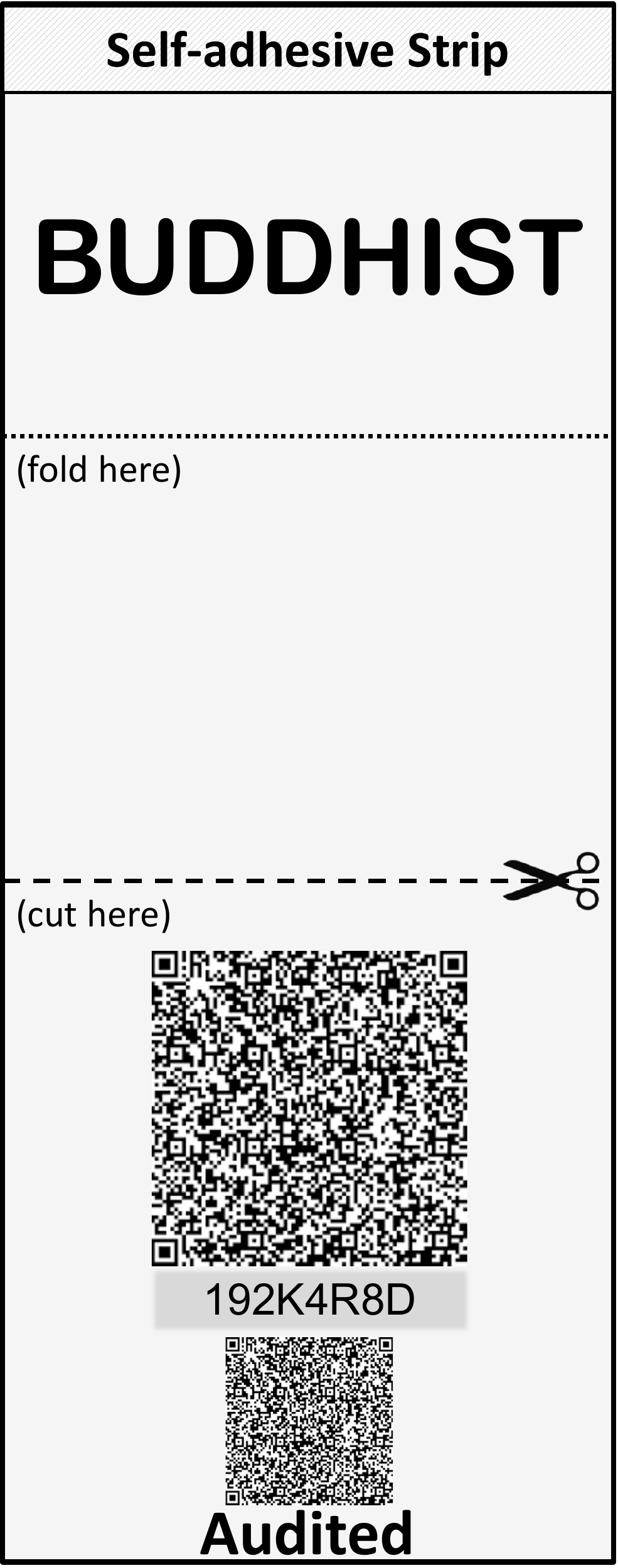}
  \label{fig:auditingwombat}}
\subfigure[Folding the ballot]{
\centering
  \includegraphics[width=0.24\textwidth]{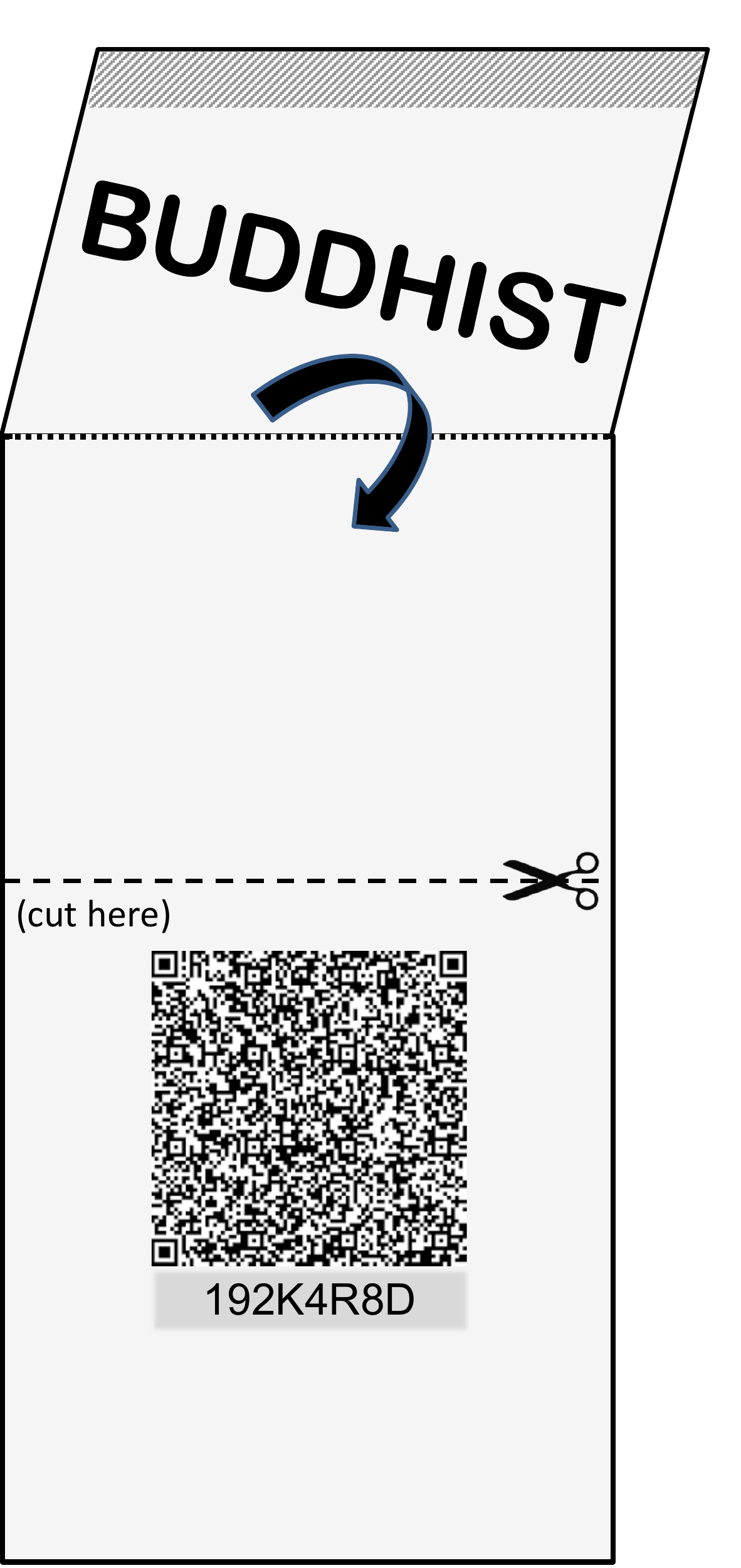}
  \label{fig:wombatfolding}}
  \caption{Verifying and casting a Wombat ballot (with a vote for \emph{Buddhist})}
\label{fig:votingwithwombat}
\end{figure}

On election day, Alice makes her candidate choice on a voting machine, which prints a \emph{dual} ballot as shown in Fig.~\ref{fig:wombatballot}. Her candidate choice is printed in plaintext followed by a detachable portion bearing her electronic vote, i.e. a barcode representing her encrypted vote. Before dispensing the ballot, the machine offers Alice the choice to challenge or cast her vote. If challenged, the machine prints a barcode with randomness information at the bottom of the ballot, as shown in Fig.~\ref{fig:auditingwombat}, which Alice can use to verify her vote. She can mount as many challenges as she likes before casting her vote.

If Alice chooses to cast her ballot, the machine prints ``For Casting" on her ballot as in Fig.~\ref{fig:wombatfolding} and dispenses it. Alice folds the ballot as shown, using the adhesive strip to conceal the plaintext. She then exits the voting booth and hands her ballot to a poll worker. He scans the encrypted vote and uploads it to the election website. He next stamps both halves of the ballot and detaches them. The folded plaintext vote is cast in a ballot box, and the barcode is issued to Alice as her receipt. Alice can later verify that her receipt is correctly recorded on the bulletin board.
Mixnets are used to shuffle and decrypt the votes which are then tallied. The mixnet protocols also provide proofs of correct operation. All decrypted votes and mixnet proofs are published on the bulletin board where they may be verified by voters and observers.

Like Scantegrity, Voter Initiated Auditing, and STAR-Vote (described next), Wombat produces both an electronic voting record and a parallel paper trail. This \textbf{dual voting} approach has considerable advantages: it is familiar to voters who may be more used to paper-based voting systems. The electronic system may fail, for instance, due to network outages or trustees may misplace cryptographic credentials, in which case the paper votes are a handy backup. Paper trails are also a good tool to audit the elections, and make an attacker's job much harder. To subvert an election, he would need to successfully compromise both the electronic and the paper record.

The team behind Wombat has placed considerable emphasis on practical aspects of the system. Guidelines to implement Wombat securely are detailed in \cite{grundland2012analysis}. At an early phase they discovered that giving voters the challenge-or-cast option proves confusing and is unnecessary as only 2-3\% of voters need exercise the challenge option to assure election integrity. This inspired the design decision to hide the challenge option, so that voters intending to audit the machine issue the challenge themselves by tapping the screen while the machine is printing the ballot.

Wombat uses Verificatum \cite{wikstrom2013user}, a free and open-source mixnet solution which has also been used in the 2013 Norwegian Parliamentary elections. The system is supplemented by an open-source Android app for vote verification. Alice can take a picture of the ciphertext and audit information on her ballot using her smartphone and the app verifies that the vote is correctly encrypted and signed. The app also confirms that cast votes are correctly posted on the election website.

Wombat has been deployed in binding elections \cite{ben2012new}. In 2011, it was used for student body elections at the Interdisciplinary Center in Israel, involving multiple races and more than two thousand voters. In 2012, the Israeli political party Merez, used Wombat for intra-party elections, involving approximately 830 voters.

\subsubsection{STAR-Vote}

In 2011, election administrators from Travis County in Texas, seeking to overhaul their election system, reached out directly to the academic research community to assist in building a secure and low-cost voting system. The result is STAR-Vote (STAR stands for Secure, Transparent, Auditable and Reliable) \cite{bell2013star}. Like Wombat, STAR-Vote brings together the convenience and usability of electronic voter machines with end-to-end cryptographic verifiability and a paper trail audit mechanism. We provide a brief overview.

At the polling station, Alice is issued a paper slip with her `voting credentials' which identifies the appropriate ballot for her. She inputs the credentials in a voting machine and makes her candidate choice. The machine electronically encrypts and records Alice's vote and issues her a receipt consisting of an encrypted commitment to her vote and related information (such as the ID of the voting machine, the time of voting, etc.). The system also generates a paper trail: the machine additionally prints out a physical copy of the marked ballot. Alice visually confirms her choice is correctly marked and then casts it by hand into a nearby ballot box. The election result is tallied from the electronic vote record. In parallel to E2E verification checks, audits are conducted on the paper ballots to further validate the results.

After casting her vote, Alice may challenge the machine to prove it correctly encrypted the vote. In this case, she informs a poll worker who assists her in verifying that the encrypted commitment on her receipt is indeed correct. Alice may repeat this process until she is satisfied and then cast a final vote. This check is in the spirit of Voter Initiated Auditing but with the key difference that here the challenge and verification steps occur \emph{after} vote casting, not before, and are initiated by voters. This is in the interests of system usability: for voters not interested in challenging the machine, the voting process is fairly similar to that of existing real-world systems.

After polls close, all encrypted votes are posted on the bulletin board where voters can compare against their receipts. Tallying consists of homomorphic addition of the cast ballots. The decryption key is pre-shared among multiple trustees who cooperate to decrypt the result in a verifiable manner. Challenged votes are posted as well, along with verifiable decryptions, and these are excluded from the final tally.

After the elections, \textbf{risk-limiting audits} \cite{lindeman2012gentle} are conducted on the paper ballots. A random selection of ballots is checked on two counts: first, verifying if data on the paper ballots corresponds to the electronic record. The second check determines if the tally of the random sample approximates the electronic results within an acceptable margin. This ``belt and suspenders approach" \cite{letzter2014new} of incorporating multiple safeguards aims at giving voters and observers strong confidence in the system.

The STAR-Vote system architecture inherits several engineering innovations from VoteBox including strict partitioning of program modules, a GUI consisting of pre-rendered graphics, and secure auditable logging of machine events.

STAR-Vote is slated for deployment in elections in Travis County in 2017 \cite{lim2014travis}.

\subsubsection{DRE-i}

Hao, Randell, and Clarke noted that voting systems are vulnerable due to their reliance on \emph{tallying authorities}, i.e. trustees tasked with tallying election results who need to be implicitly trusted not to misuse their privileges \cite{hao2012self}. Furthermore, as demonstrated in a real-world trial of the Helios system \cite{adida2009electing}, trustees may lack the technical expertise to manage cryptographic credentials securely. To address these concerns, Hao et al. proposed DRE-i (Direct Recording Electronic with Integrity) \cite{hao2014every}, an E2E voting system which dispenses with tallying authorities entirely.

The DRE-i ballot consists of pairwise strings of encrypted data, or \emph{cryptograms}, assigned to each candidate, depicting \emph{Yes} and \emph{No} votes. When Alice makes her candidate choice on a voting machine, a \emph{Yes} cryptogram is assigned to her candidate and \emph{No} cryptograms to all the others. The machine issues her a receipt listing the candidates with the assigned cryptograms. Individually the \emph{Yes/No} cryptograms do not reveal any information about Alice’s choice and resemble strings of random data. Before casting her vote, Alice has the option of challenging the voting machine to prove it is not cheating. If challenged, the machine reveals both \emph{Yes/No} cryptograms for each candidate on the ballot. Any party can now differentiate between \emph{Yes} and \emph{No} votes, and confirm that they are correctly assigned.

All receipts are published onto the bulletin board where they may be verified by voters, alongside proofs that the vote is \textbf{well-formed}. This is a privacy-preserving cryptographic proof, used in several of the voting systems described in this chapter, which allows any observer to verify that the ciphertext on the bulletin board contains a vote for only one candidate, without revealing the voter's identity or candidate choice. The votes are then combined in a homomorphic operation to yield the final tally. There is no decryption step. Any observer can therefore download the contents of the bulletin board and replicate the addition to confirm the results.

Self-tallying voting systems have been proposed earlier in the literature, for small-scale scenarios such as boardroom elections \cite{groth2004efficient} \cite{kiayias2002self} \cite{hao2010anonymous}, but Hao et al. differentiate these from DRE-i which enables self-tallying within the broader framework of E2E verifiable voting, a paradigm they refer to as \emph{self-enforcing electronic voting} \cite{hao2012self}. A verifiable classroom voting system based on DRE-i has been implemented and used \cite{hao2013verifiable}.

\subsection{Remote Voting with Electronic Ballots}
\label{sec:remotevotingwithelectronicballots}

This category comprises systems which enable voters to cast their votes over the Internet and includes prominent systems JCJ/Civitas and Helios. As we noted earlier in Sec.~\ref{sec:conflictsandchallenges}, remote voting poses unique challenges which restrict the use of some of these systems in politically binding elections.

\subsubsection{Adder}
\label{sec:Adder}

Adder \cite{kiayias2006internet} invented by Kiayias, Korman and Walluck in 2006, is a fully-functional open-source online voting system. Adder is suited to both small and large-scale elections, as well as surveys and data collection applications.

Adder includes a preliminary setup phase in which multiple election trustees engage in a cryptographic protocol to create a public key for the election which is then published on the election website. The corresponding decryption key is distributed among the trustees using threshold cryptography.

To vote, Alice navigates her web browser to the election website, logs in using voting credentials which identify her as an eligible voter, and downloads the election public key. Alice makes her candidate choice and her web browser then generates and encrypts her vote using the election public key and computes and appends a proof of well-formedness. All encrypted votes received by the system are posted on an online bulletin board, where voters can verify they have been correctly recorded.

After polls close, an addition operation is performed over all the encrypted votes on the bulletin board. Election trustees then publicly provide partial decryption information using their individual shares of the decryption key, which enable the server to decrypt the tally. All intermediate computations are posted on the bulletin board.

Adder's security properties are fairly intuitive: the voter encrypts and transmits her own vote and can later verify it is correctly recorded on the bulletin board. Privacy is ensured because individual votes cannot be decrypted unless a threshold amount of these election trustees collude. Voters and observers alike can audit the tallying and decryption process by accessing the bulletin board and verifying each step.

However, Adder is vulnerable to critical security issues common to Internet voting systems. The first is the \textbf{untrusted terminal problem}: it is trivially easy for malicious software on Alice's computer to leak knowledge of her vote or switch her vote to another candidate without her knowledge, and she cannot detect this by examining the ciphertext. Second, physical privacy cannot be guaranteed. Someone may be standing at Alice's shoulder while she votes. Furthermore, a coercer, such as a political activist or a family member, may force her to vote a particular way. This threat is also common in other remote voting scenarios, such as postal voting \cite{stewart2006banana}.

\subsubsection{JCJ and Civitas}
\label{sec:jcjandcivitas}

The system designed by Juels, Catalano and Jakobsson in 2005, commonly referred to as JCJ \cite{juels2005coercion}, is notable in that it is the first online voting scheme to offer a high level of coercion-resistance. JCJ combines various cryptographic primitives in a novel protocol that has proved very influential in the research literature.

JCJ assumes that prior to elections, there is an in-person voter registration phase, conducted in a trusted and private environment, where Alice receives a voting credential from the election registrar. This credential is her proof of eligibility and can be used to vote in several different elections. After the registration phase concludes, the registrar publishes an encrypted list of all voting credentials on the bulletin board. Alice then uses a public algorithm to generate any number of \emph{fake} credentials which are indistinguishable from her real one to a third party. She can use these to cast votes under coercion or even freely yield these to coercers to vote on her behalf. Votes cast using fake credentials will show up on the bulletin board as legitimate cast votes but will ultimately be rejected by the system prior to tallying. It is assumed the coercer is not monitoring Alice for the entire duration of the election and she gets some private moments to vote using her real credential.

Votes are cast online. Alice's vote includes her candidate choice and her voting credential and is encrypted with the public key of the election trustees. Also attached is a cryptographic proof enabling any observer to verify that the vote is well-formed and includes a usable credential (real or fake). After polls close, all vote ciphertexts received by the system are published onto the bulletin board where Alice can verify that hers was correctly recorded.

Election trustees then process votes on the bulletin board. Vote proofs are examined and malformed votes or those with unusable credentials are discarded. A \textbf{plaintext equivalence test} is used to check if multiple votes have been cast using the same credential. This cryptographic test compares encrypted votes in a pair-wise manner and simply identifies, without decryption, any ciphertexts which include a common credential. No direct information about the credentials or votes themselves is revealed in this test, i.e. voter identities and candidate choices are still secret. If multiple votes using the same credential are identified, they may be processed according to some \emph{revoting} policy such that only one vote per credential is actually counted. For instance, only the most recent vote could be retained and all others using the same credential could be eliminated.

However, this batch still contains votes cast using fake credentials which have to be identified and discarded. All votes are passed through a \textbf{re-encryption mixnet}. As opposed to decryption mixes (described earlier in Sec.~\ref{sec:votegrity}) which peel off layers of encryption from the vote, re-encryption mixes instead simply inject fresh randomness into the ciphertext. This \emph{re-randomization} generates a batch of completely new ciphertexts which are still encryptions of the original plaintext data, i.e. votes and credentials. These new ciphertexts are then shuffled and output.

The plaintext equivalence test is then again run, this time to compare the encrypted credential in each new ciphertext with the encrypted list of valid voting credentials published by the election authority prior to the election. All votes with fake credentials can then be identified and discarded by the system without revealing the credentials themselves. Leftover votes are then decrypted and tallied in a straightforward manner to yield the election result.

The equivalence tests and the mixing and tallying processes are conducted in a publicly verifiable manner, and all intermediate results are posted on the bulletin board where observers can audit them for correctness. This mix-and-compare process is what gives JCJ the property of coercion-resistance. The re-encryption mixnet effectively breaks the association between cast votes and counted votes such that a coercer has no way of determining if the credential he holds is real or fake and if the vote he cast was included in the final tally or not.

However JCJ has a key disadvantage: the processing time and computation costs of this scheme are very high because compute-intensive plaintext equivalence tests need to be performed for every vote. There have been several suggestions in the literature to optimize this process, including \cite{araujo2010practical}, \cite{spycher2012new}, and \cite{haghighat2013efficient}.

Civitas \cite{clarkson2008civitas}, proposed by Clarkson, Chong and Myers, is the first concrete implementation of JCJ along with important modifications, most notably a secure protocol for voter registration. In the original JCJ protocol, a corrupt registrar could simply issue a fake credential to Alice without her being able to detect the deception. However, in Civitas, the registrar functionality is distributed across multiple mutually distrusting trustees. Alice individually contacts these to receive portions of her credential which she then combines into a final credential. The implicit assumption is that some trustees are honest, thereby allowing her to construct the right credential.

Distributing registration functionality for JCJ has also been discussed in \cite{krivoruchko2007robust}. Koenig et al. \cite{koenig2011preventing} suggest enhancements, including protection against a potential denial-of-service attack whereby attackers flood the bulletin board with votes with fake credentials, thereby creating severe processing delays. Other proposals include formalizing the voting-process in Civitas, a user interface inspired in part by Helios (described next), and credential management using smartcards \cite{neumann2012civitas} \cite{neumann2013towards}.

The core JCJ architecture has inspired other E2E voting systems which improve upon its various usability aspects. For instance, Selections \cite{clark2012selections} provides an easier way for voters to register and manage credentials by incorporating the use of \emph{panic passwords}, i.e. voting credentials that the voter may create on the fly. Trivitas \cite{bursuc2012trivitas} adapts Civitas to further simplify the non-intuitive and complex verifiability process for the layman voter by introducing the notion of \emph{trial votes}. These are cast alongside regular votes, and are processed in much the same way, except that trial votes are decrypted and revealed at different stages, enabling the voter to derive strong confidence that her actual vote is being correctly handled. Caveat Coercitor \cite{grewal2013caveat} extends JCJ's coercion-resistance property for more general scenarios including leakage of voting credentials by malware, corrupt registrars, and impersonation attacks.

\subsubsection{Helios}
\label{sec:helios}

Helios \cite{adida2008helios}, invented by Adida in 2008, is an online voting system intended for scenarios where privacy and integrity concerns are paramount but the risk of coercion is low, i.e. in settings such as student body elections, clubs, and online communities. Helios does not introduce any novel features as such, but successfully brings together different innovations in the literature to build an efficient and highly usable voting system. We present a brief overview:

To vote, Alice uses her web browser to navigate to the election website and marks her choice on an online ballot form. Her vote is encrypted locally on her computer, a digital receipt is printed on the screen. In the spirit of Voter Initiated Auditing, Alice is then offered the opportunity to cast or challenge her vote. She can challenge the system multiple times until she is convinced it is encrypting her choice correctly. When she is ready to cast her vote, she authenticates herself to the system as an eligible voter and uploads the ciphertext. The system posts it on an online bulletin board where she can verify it at a later time against the receipt she received earlier.

After polls close, encrypted votes on the bulletin board are processed using re-encryption mixes to effectively break the association between voter identity and cast votes after which votes are decrypted and tallied. The initial version of Helios employed the Sako-Kilian mixnet protocol \cite{sako1995receipt} which published cryptographic proofs allowing third parties to confirm that the mixnet processed the data correctly. In 2009, Helios was updated to Helios 2.0 which improved efficiency and voter privacy by replacing the mixnets with a homomorphic encryption scheme and distributing the election decryption key among multiple election trustees.

Helios provides E2E guarantees but, as noted earlier, is not suited for political elections where the risk of coercion is high. Adida acknowledges this as a general limitation of remote voting systems and suggests that Helios might in fact educate voters on this threat. For this reason, the initial version of Helios included a ``Coerce Me!" button, which Alice could press when casting her ballot, thereby emailing a coercer complete details of how she voted. Adida also notes that a corrupted version of Helios may impersonate absent voters and cast ballots on their behalf, making it all the more important that voters engage in the verification process.

Helios has proved very influential and is actively being maintained. The source code is available under an open-source license. Helios has been the subject of multiple usability studies \cite{weber2009usability} \cite{karayumak2011user} \cite{acemyan2014usability}. A number of vulnerabilities have been discovered in the implementation of Helios \cite{estehghari2010exploiting} \cite{heiderich2012bug} \cite{bernhard2012not}. Examples include cases where attackers may copy and re-cast previously cast votes \cite{cortier2013attacking}, and a corrupted system may dispense identical receipts to different voters while modifying their actual votes undetected \cite{kusters2012clash}. Many of these issues have been fixed in later versions of Helios. Improvements have also been suggested: Helios has been modified to enable new properties such as self-tallying \cite{dossogne2014blinded} and stronger privacy guarantees \cite{bernhard2011adapting} \cite{demirel2012improving}.

Helios has been used in several binding elections: the Universit\'{e} Catholique de Louvain used Helios in 2009 to elect the University President in an election with about 5000 registered voters \cite{adida2009electing}. Helios has also been deployed for student elections at Princeton \cite{morell2010secret} and internal elections of the International Association for Cryptologic Research (IACR) \cite{heliosiacr} and the Association for Computing Machinery (ACM) \cite{HeliosACM}. Helios has also been adapted to accommodate different voting schemes such as single transferable vote (STV) and ranked voting schemes. Examples include Zeus, built by Louridas et al. \cite{louridas2014zeus}, and another variant developed by Bulens et al. \cite{bulens2011running}. Both of these systems have been deployed in dozens of university-level elections.


\subsubsection{Pretty Good Democracy}

Pretty Good Democracy \cite{ryan2013pretty} is an online voting system developed by Ryan and Teague in 2013. Pretty Good Democracy relies on threshold cryptography and code voting to provide a subset of E2E verifiability properties.

\begin{figure}[t]
\begin{center}
  \includegraphics[width=0.6\textwidth]{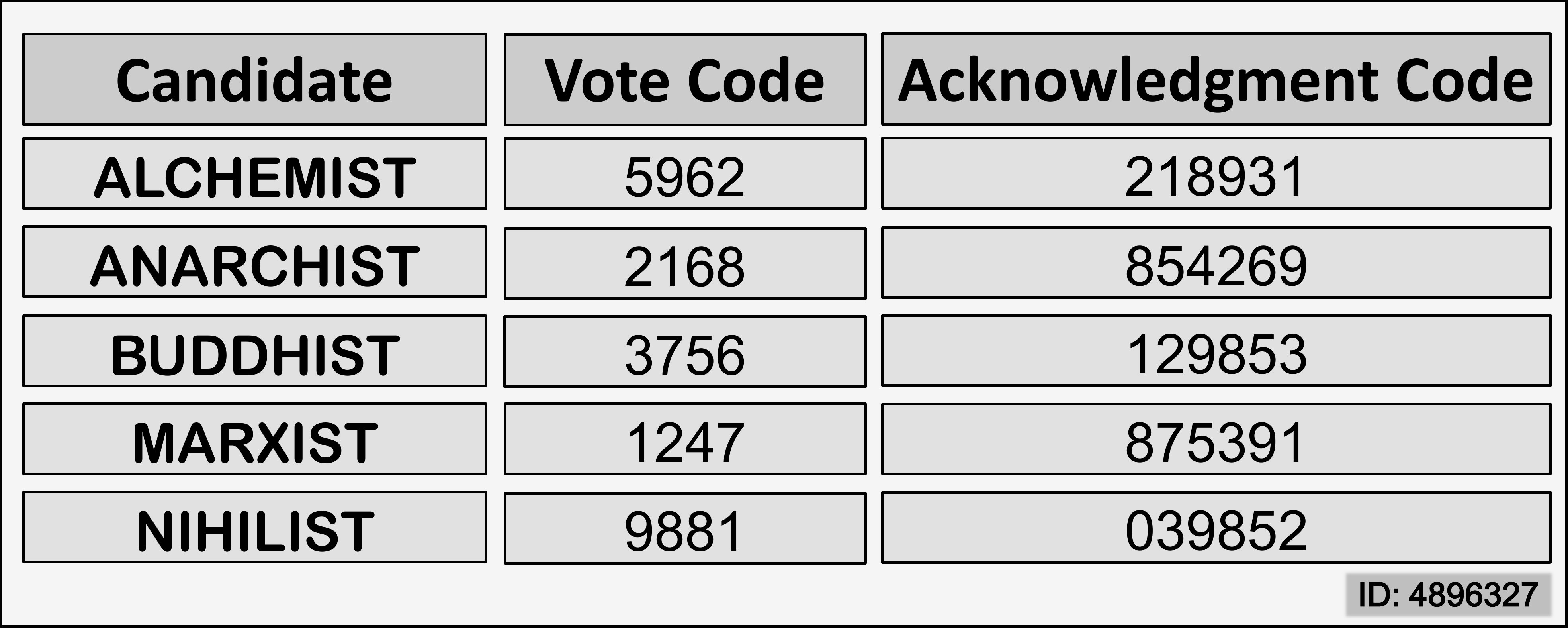}
   \caption{Sample code sheet}\label{fig:codesheet}
\end{center}
\end{figure}

\textbf{Code voting} was first suggested by Chaum as a solution to the untrusted terminal problem \cite{chaum2001surevote} and now serves as a key component in several online E2E voting systems \cite{joaquim2009veryvote} \cite{kutylowski2010scratch} \cite{zagorski2013remotegrity} \cite{joaquim2013eviv} \cite{pereira2014scroll}. The key idea is simple: codes are used to set up a private authenticated channel between voters and election trustees. Prior to polls, the trustees send each voter a physical \emph{code sheet} in the mail which assigns unique, randomly picked \emph{vote codes} and \emph{acknowledgement codes} to each candidate on the ballot as depicted in Fig.~\ref{fig:codesheet}. These codes are alphanumeric strings of a length that resists simple guessing attacks. Each code sheet bears a unique ballot ID.

To vote, Alice logs onto the system, and inputs the ID on her code sheet, followed by the vote code for the candidate of her choice. For example, she would type in \emph{2168} to vote for \emph{Anarchist}. The system responds with the corresponding acknowledgement code, in this case, \emph{854269}. Since the codes are randomly assigned and known only to Alice and the trustees, a malware on her computer or an attacker eavesdropping on network traffic cannot deduce her vote nor alter it in any meaningful way without being detected. Furthermore, Alice's vote code assures the trustees that they are communicating with Alice herself, since it should be difficult for an attacker to procure her code sheet. Likewise, if Alice receives the correct acknowledgement code, she can be certain her vote was correctly received by the trustees and not an attacker masquerading as a trustee. If either party enters a wrong code, the protocol fails.

After polls close, ballot IDs for all votes received by the system are published on the bulletin board and Alice can check that hers is included in the list. Decryption mixnets are then used to anonymize votes prior to tallying, similar to Votegrity and Pr\^{e}t \`{a} Voter, and the process is likewise audited using randomized partial checking.

The bulletin board does not display vote receipts otherwise it would be trivially easy for Alice to prove her vote to a third party using her code sheet. However, displaying only ballot IDs of cast votes opens up the possibility that a corrupt tallying authority may alter votes undetected. To address this vulnerability, Ryan and Teague employ threshold cryptography to distribute system functionality among the election trustees. When the system receives Alice's vote code, it engages in a cooperative protocol with a threshold number of trustees to generate the appropriate acknowledgement code and record Alice's vote in the system. It is therefore not possible to alter votes unless a large number of these trustees are corrupt, which should reassure Alice to a degree that her vote has been correctly recorded by the system.

However, Pretty Good Democracy only partially addresses the problems associated with remote voting. There is still the threat that someone may be standing at Alice's shoulder when she votes and may even coerce her to vote for a certain candidate. Alice may sell her vote by giving her code sheet to someone else. An online attacker may pose as a fake election server and her vote may be lost. Similar attacks have been observed in remote voting scenarios such as postal voting. The authors, therefore, explicitly recommend that Pretty Good Democracy only be used for low-risk elections such as for student bodies and professional societies and not for politically binding elections.

Pretty Good Democracy has inspired other systems: Pretty Understandable Democracy \cite{budurushi2013pretty} simplifies the way votes are processed. Virtually Perfect Democracy \cite{bella2014virtually} introduces personalized voter smartcards into the protocol to increase trust in the system and provide an element of coercion resistance. Pretty Good Democracy has also been adapted for ranking-based voting schemes such as single transferable vote (STV) and instant-runoff voting (IRV) \cite{heather2010pretty}.

\subsubsection{Remotegrity}

Remotegrity \cite{zagorski2013remotegrity} is an online E2E voting protocol, proposed by Zagorski et al. in 2013. Remotegrity is not a full-fledged system but an extension to precinct-based systems like Scantegrity and Pr\^{e}t \`{a} Voter, specifically aimed at absentee voters, and uses an innovative combination of code voting and scratch surfaces to ensure votes are securely recorded.

\begin{figure}[t]
\centering
\subfigure[Scantegrity II ballot]{
  \includegraphics[width=0.275\textwidth]{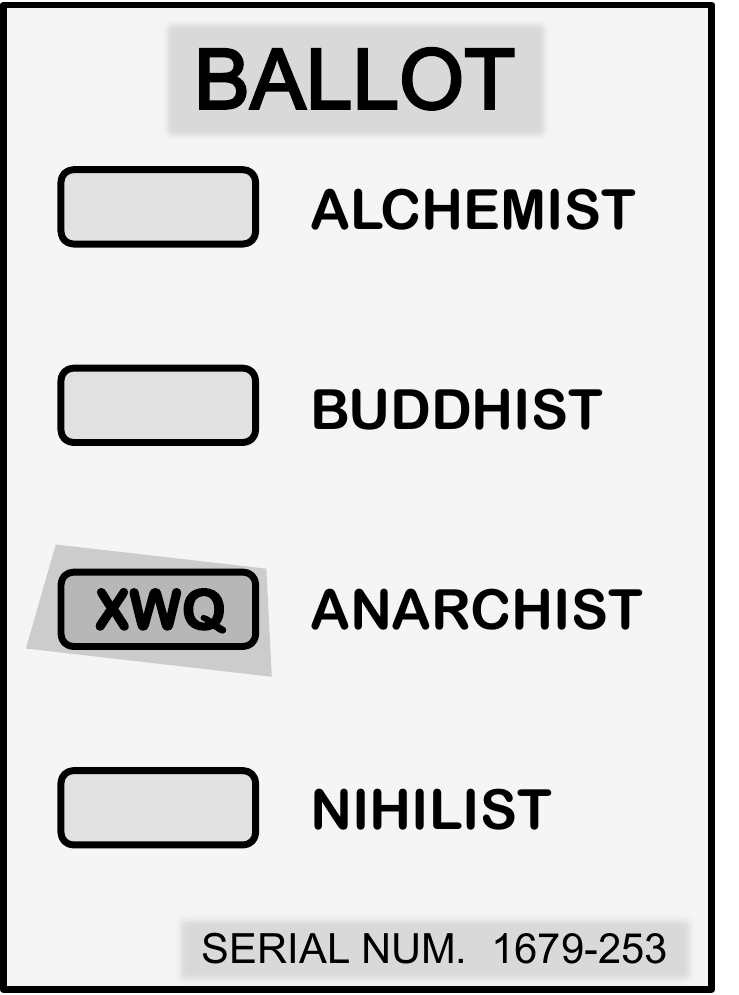}
  \label{fig:remotegrityscantegrity2ballot}}
\subfigure[Remotegrity authorization card]{
  \includegraphics[width=0.275\textwidth]{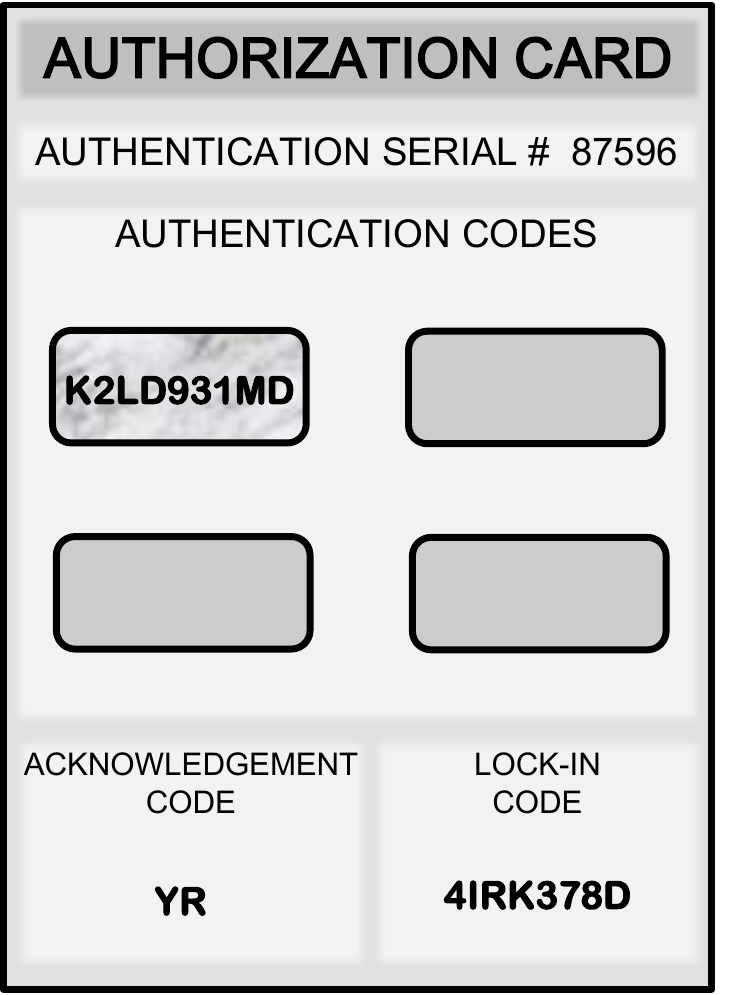}
  \label{fig:remotegrityauthorizationcard}}
  \caption{Voting with Remotegrity (casting a vote for \emph{Anarchist})}
\label{fig:remotegrity}
\end{figure}

As an absentee voter, Alice receives via post a Scantegrity ballot (described in Sec.~\ref{sec:scantegrity}) and a Remotegrity authorization card, as depicted in Fig.~\ref{fig:remotegrity}. The authorization card contains an authentication serial number, and three types of codes: multiple authentication codes, a lock-in code, and an acknowledgement code.

To vote, Alice logs on to the system and enters her ballot serial number and the authentication serial number on her authorization card. This allows the system to check that she has not already cast a vote. Alice next obtains the vote code on the ballot for the candidate of her choice (\emph{XWQ} for \emph{Anarchist}). From the authorization card, she scratches off an authentication code at random. These two codes are sent to the system as a single message and are posted on the bulletin board. The election trustees verify that the authentication code is valid. They then append an acknowledgement code to it, digitally sign the whole message, thereby certifying it as genuine, and update the entry on the bulletin board. At a later time, Alice checks the bulletin board and verifies that her vote has not been maliciously altered and confirms the acknowledgement code and the trustees' signature. She then approves the updated entry by scratching off and submitting  the lock-in code on her authorization card. The system again updates the bulletin board entry to include the lock-in code.

When polls close, the trustees check all entries to verify the lock-in codes are correct. They then input the ballot serial numbers and the corresponding vote codes for each entry into the Scantegrity tallying system. These votes are then tallied alongside precinct-cast votes using the Scantegrity Switchboard in a publicly verifiable manner.

The use of codes effectively defeats malware and hackers who may tamper with Alice's computer. The scratch surfaces act as tamper-evident seals. If the election trustees were to cast a vote on Alice's behalf, Alice could produce her authorization card with the authentication code scratch surfaces still intact as proof of malfeasance. If they were to change her vote after she submitted it, she could refuse to commit to it with her lock-in code. If the trustees were to append the lock-in code themselves, her authorization card, with the lock-in scratch surface still intact, would prove that she did not authorize that vote. The trustees may issue authorization cards with invalid codes, but this may be detected by conducting public audits in which randomly chosen authorization cards are examined for integrity. The protocol therefore ensures that Alice's vote is cast and recorded by the system as she intended. Regarding the verifiability of the tally, Remotegrity inherits Scantegrity's security guarantees. However, Remotegrity does not protect against vote coercion or vote buying.

Remotegrity was successfully trialled in municipal elections in Takoma Park, Maryland in 2008 \cite{zagorski2013remotegrity}.

\section{Non-cryptographic E2E Voting Systems}
\label{sec:noncryptosystems}

These systems represent an interesting experiment in the overall research on E2E voting. Systems such as ThreeBallot, Twin, and Aperio emulate E2E verifiability properties in a wholly precinct-based paper-ballot format without resorting to any cryptography. This approach has only produced a handful of workable systems thus far, but it has generated valuable new insights about E2E voting in general. Additionally, due to their exclusion of complex cryptography, these systems aid in communicating fundamental principles of E2E verifiability to the layman.

\setcounter{subsubsection}{0}

\subsubsection{ThreeBallot, VAV, and Twin}

\begin{figure}[t]
\centering
\subfigure[The ThreeBallot ballot]{
  \includegraphics[width=0.45\textwidth]{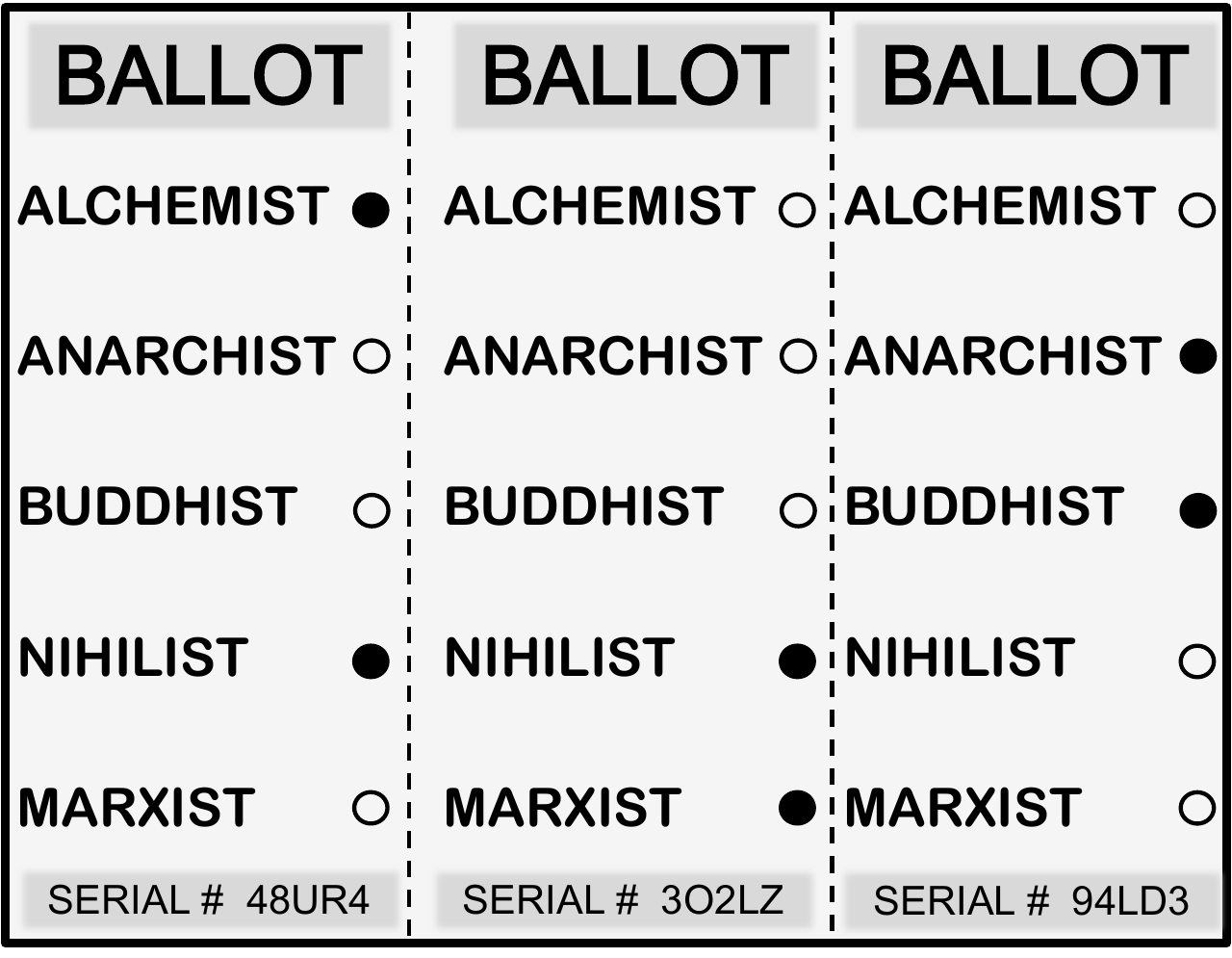}
  \label{fig:threeballotballot}}
\subfigure[The VAV ballot]{
  \includegraphics[width=0.45\textwidth]{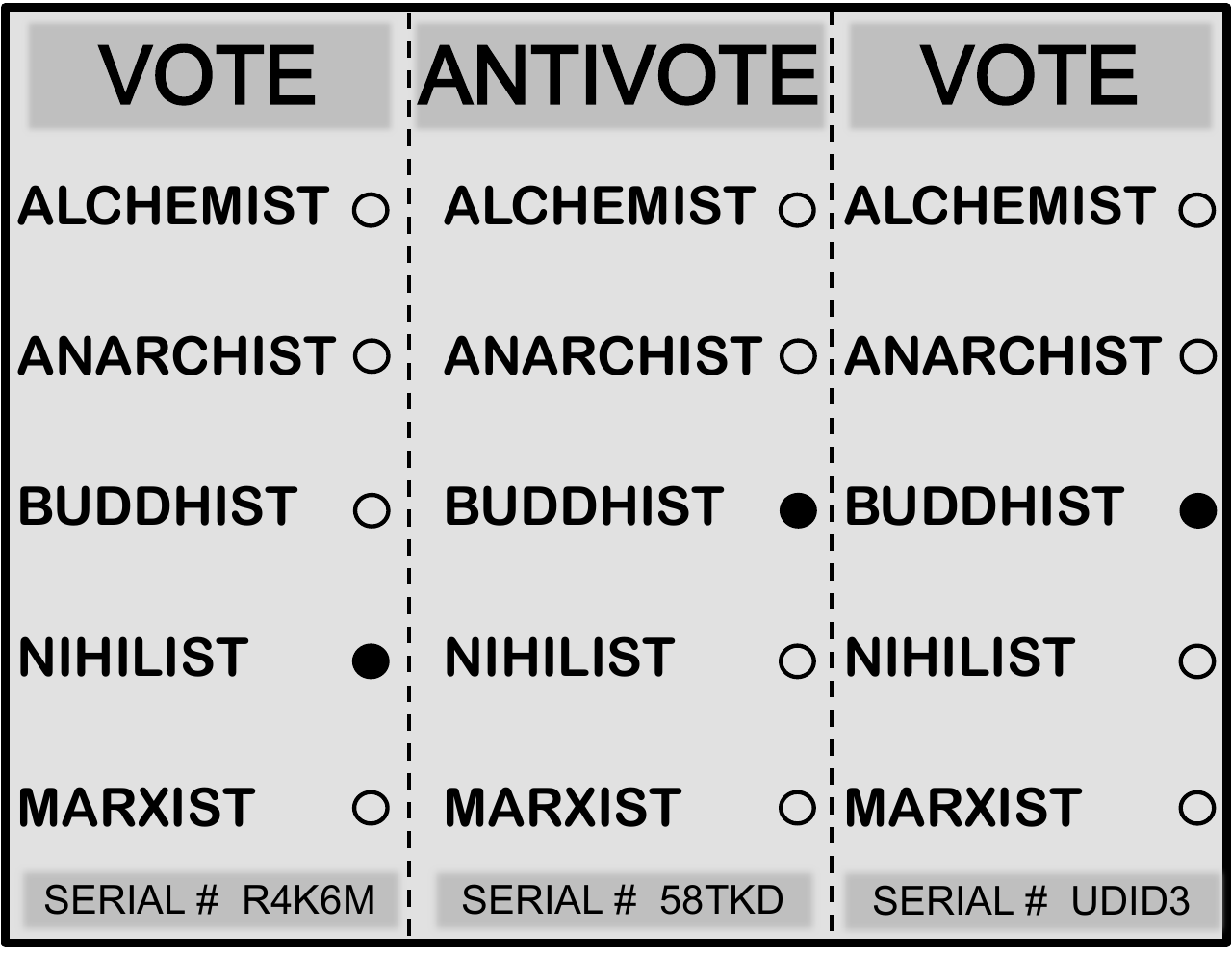}
  \label{fig:vavballot}}
  \caption{Voting with ThreeBallot and VAV (with votes for \emph{Nihilist})}
\label{fig:votingwiththreeballot}
\end{figure}

ThreeBallot \cite{rivest2007three}, introduced by Rivest and Smith in 2007, derives its name from its unique ballot design, consisting of three detachable ballots, as depicted in Fig.~\ref{fig:threeballotballot}. Each ballot is assigned a unique serial number printed at the bottom. In the polling booth, Alice marks the ballots such that each candidate is marked once whereas the candidate of her choice is marked \emph{twice} {Fig.~\ref{fig:threeballotballot} represents a vote for \emph{Nihilist}}. Alice's vote can therefore be understood as a composite of the three ballots. The ballots are then detached. Alice randomly chooses one as a receipt and scans a copy to take home. She then casts all three ballots into the ballot box.

After polls close, all cast ballots are published on the bulletin board, and Alice uses her receipt to verify the corresponding ballot is correctly recorded. The tally is computed by simply adding up all ballots on the bulletin board. Since each candidate receives one vote by default, the result is inflated by the number of voters, which is accordingly subtracted from the tally. Any observer can replicate and verify the tally.

An important distinction to be made here is that Alice's receipt is not an encryption of her vote. Instead, the receipt testifies to a key component of the vote which has been delinked from the other components. Alice's privacy is maintained as the receipt reveals nothing about her actual vote unless examined in conjunction with the two other ballots she cast, and these could be any on the bulletin board. Furthermore, if any party tampers with her ballots, Alice has a one in three chance of detecting it on the bulletin board. For large-scale tampering, the odds are much higher.

ThreeBallot's non-intuitive ballot marking scheme may impact usability: in a mock election, a large number of voters experienced difficulty using ThreeBallot and more than 30\% of initial cast votes proved invalid \cite{jones2006threeballot}. There are also security issues: a coercer may force Alice to mark her ballots in certain unique patterns which he can later identify on the bulletin board. These issues and potential solutions are documented in \cite{rivest2007three} \cite{appel2006defeat} \cite{henry2009effectiveness} \cite{cichon2008short} \cite{kelsey2010attacking}.

Rivest and Smith also presented two systems that are variations on ThreeBallot, VAV (Vote/AntiVote/Vote) and Twin \cite{jones2006threeballot}. The VAV ballot suite, depicted in Fig.~\ref{fig:vavballot}, consists of two \emph{Vote} ballots, and one \emph{AntiVote} ballot. The marking process differs from ThreeBallot in that only one candidate is marked on each ballot. The \emph{AntiVote} ballot essentially \emph{cancels} out one \emph{Vote} ballot, and both have to be marked identically. The leftover \emph{Vote} determines Alice's actual choice (Fig.~\ref{fig:vavballot} depicts a vote for \emph{Nihilist}). Alice scans one of the three ballots at random to use as a receipt. Tallying and verification processes for VAV are very similar to ThreeBallot.

Twin is based on traditional voting systems in that the voter casts only one ballot, but with the innovation that Alice does not take home her own receipt but that of another voter. In this scenario, a big bin in the polling station is periodically filled with valid receipts. After casting her vote, Alice makes her way to the bin, draws out a receipt at random, and takes a copy home, which she later verifies on the bulletin board. These receipts are therefore referred to as \emph{floating receipts}. This system is very easy to use, and though individual vote verifiability is sacrificed, voters can still verify the election tally and coercion resistance is ensured.

\subsubsection{Randell \& Ryan's Scratch Card Voting System}

In 2006, Randell and Ryan proposed a voting system \cite{randell2006voting} that augments the Pr\^{e}t \`{a} Voter ballot with scratch surfaces and dispenses with cryptography and mixnets.

\begin{figure}[t]
\begin{center}
  \includegraphics[width=0.7\textwidth]{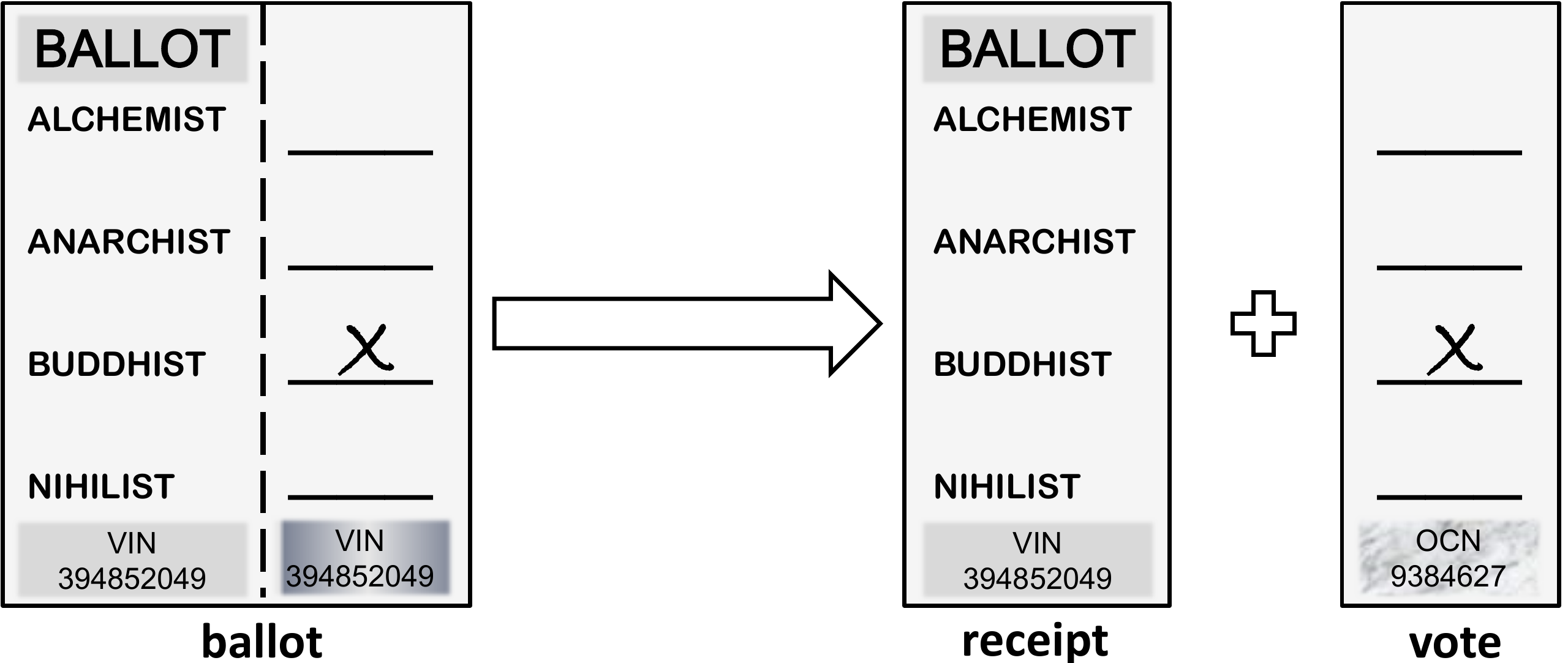}
   \caption{The Randell \& Ryan Scratch Card Voting ballot}\label{fig:randellryanballot}
\end{center}
\end{figure}

The ballot, depicted in Fig.~\ref{fig:randellryanballot}, is detachable in two halves along the middle. Candidate names are listed in randomized order on the left side with corresponding markable spaces on the right. The Vote Identification Number (VIN), functioning as a unique serial number for the ballot, is duplicated on both halves, except on the right it is printed atop a scratch surface. The scratch surface conceals an encoding of the candidate ordering on the ballot, referred to as the Order of Candidate Names (OCN).

On election day, Alice marks her candidate choice on the ballot, detaches the right side and casts it in the ballot box. The left side serves as her receipt. Election staff will accept her vote only if the scratch surface is still intact. After polls close, VINs of all cast votes are published on a bulletin board and Alice can ensure the system received her ballot. To process votes, election officials first remove the scratch surface of each cast vote. This reveals the OCNs, enabling them to reconstruct the original candidate ordering on the ballots, and thereby deduce the voters' choices and compute the tally. More importantly, removing the scratch surface also destroys the VIN on top, protecting voter privacy. Once the VIN is removed, Alice's vote can no longer be differentiated from other cast votes.

The OCNs are simple codes, each of which maps to a different candidate ordering. Prior to casting her vote, Alice can audit her ballot by simply removing the scratch surface, decoding the OCN, and verifying if it matches the candidate ordering on the ballot. However, removing the scratch surface invalidates the ballot, and she will need a new one to cast her vote. Alice can audit as many ballots as she likes until she is convinced the ballots are correctly formed, and then cast her vote.


This system provides a subset of E2E properties. Only received VINs are published on the bulletin board, providing Alice assurance that her vote has been received, but not that it has been recorded and tallied exactly as she cast it. Cast ballots are not published as it would make it trivially easy for any party to deduce voters' choices from their receipts. Furthermore, removing the scratch surface irretrievably destroys the link between cast votes and counted votes, and Alice has no means of verifying that election officials correctly included her vote in the tally and did not replace it with a fake one. To increase transparency, Randell and Ryan suggest instituting random audits, and incorporating paper trails into the system.

\subsubsection{Aperio}

Aperio \cite{essex2010aperio}, developed by Essex, Clark, and Adams in 2010, is intended for low-tech minimal environments, such as elections in developing countries. Aperio derives inspiration from PunchScan in its use of ballots with stacked sheets which enable audit trails to defend against ballot box stuffing and ballot tampering.

\begin{figure}[t]
\begin{center}
  \includegraphics[width=1\textwidth]{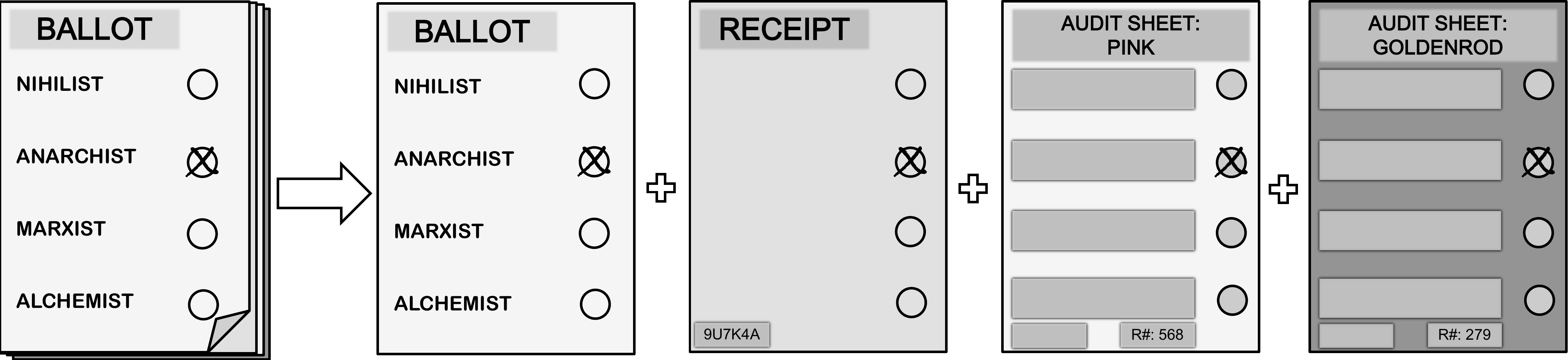}
   \caption{The Aperio ballot assembly with a vote for \emph{Anarchist}}\label{fig:aperioballot}
\end{center}
\end{figure}

The Aperio ballot suite, depicted in Fig.~\ref{fig:aperioballot}, consists of a ballot with randomized candidate names, a receipt layer and multiple distinctly coloured audit layers (we assume two audit layers in this example, coloured pink and goldenrod). The sheets are backed with carbon paper so that marks made on the ballot are copied onto the lower layers. Receipt and audit sheets each bear a unique reference number, the receipt serial number and the audit reference number, respectively.

\begin{figure}[t]
\begin{center}
  \includegraphics[width=0.75\textwidth]{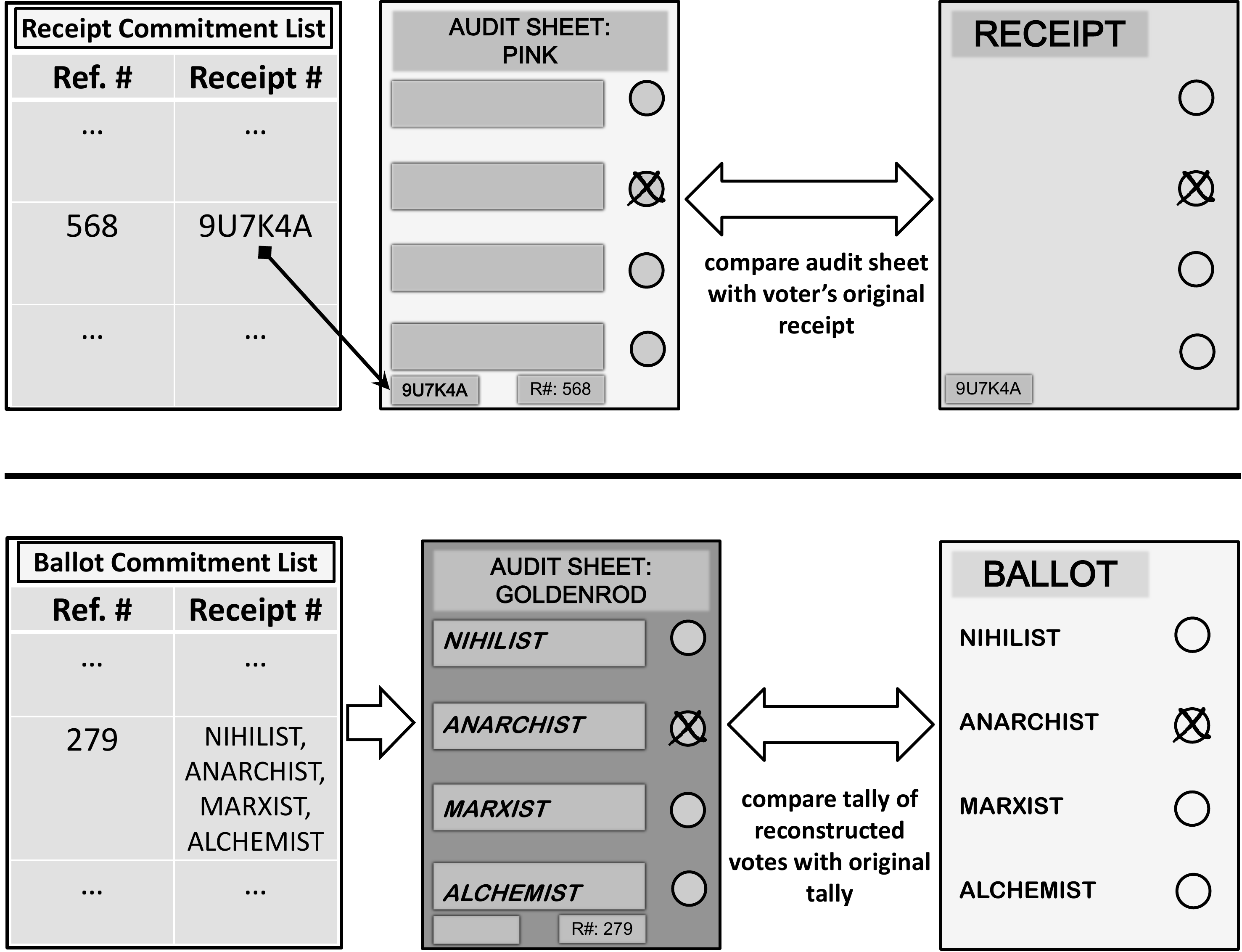}
   \caption{Auditing receipts and ballots in Aperio}\label{fig:aperioaudit}
\end{center}
\end{figure}

In the election setup phase, election trustees assemble the ballot suites by superimposing the individual sheets and binding them together. They also generate two lists for each individual audit layer: a \emph{receipt commitment list} linking each receipt serial number to an audit reference number, and a \emph{ballot commitment list} linking the audit reference number to ballot candidate ordering. Fig.~\ref{fig:aperioaudit} includes sample entries for these lists for the ballot suite in Fig.~\ref{fig:aperioballot}. These commitment lists are secured in separate tamper-evident envelopes and placed in safe custody prior to polls.

To cast her vote, Alice marks her candidate choice on the ballot suite and hands it to a member of the polling staff who ensures that the counterfoil is still intact. If not, the vote is rejected. If intact, the staff member then separates the four sheets. The ballot is cast in the ballot box, the receipt is issued to Alice to take home, and the audit sheets are cast in corresponding pink and goldenrod-coloured audit boxes. After polls close, results are tallied by simply counting the votes in the ballot box.

To audit the election, the trustees retrieve the receipt and ballot commitment lists. They check the envelope to ensure there has been no tampering. A coin is publicly tossed, and, depending on the result, one of the two coloured audit boxes is used to generate a \emph{receipt audit trail} and the other to generate a \emph{ballot audit trail}. For example, if the pink box is chosen to audit receipts, the envelopes containing the pink receipt commitment list and the goldenrod ballot commitment list are opened. The envelopes containing the pink ballot commitment list and the goldenrod receipt commitment list are destroyed.

The commitment lists and audit sheets enable individual reconstruction of the vote and the receipt, as depicted in Fig.~\ref{fig:aperioaudit}. The pink audit sheet already bears the mark for Alice's candidate choice. Election staff search for the sheet's audit reference number (\emph{568} in our example) in the receipt commitment list and note the corresponding receipt serial number (\emph{9U7K4A}) by hand on the sheet. This step effectively reconstructs Alice's receipt. Reconstructed receipts are posted on the bulletin board where voters can verify them against the receipts issued to them earlier.

Likewise, the reference number on the goldenrod audit sheet (\emph{279} in this case) identifies a specific candidate ordering in the ballot commitment list. Trustees write this ordering on the sheet, thereby reconstructing the vote. These votes are then counted independently and results are compared with the official tally.

Aperio's key novelty is that receipts and ballots are audited separately. The act of destroying the unused commitment lists effectively decouples the ballots from the receipts, thereby preserving voter privacy. Alice can verify her vote was correctly recorded by comparing her receipt against the reconstructed receipts. Election trustees and observers ensure the votes have been correctly counted if the tally of the reconstructed votes matches the original results. However, as in the case of paper-based E2E systems, certain coercion attacks, such as randomized voting, still apply.

Aperio has inspired an electronic counterpart, Eperio \cite{essex2012eperio}, which employs cryptographic primitives to similarly verify votes against commitment lists. Another system, Hash-Only Verification (\emph{Hover}) \cite{essex2012hover}, uses this verification approach in conjunction with the Scantegrity optical-scan system to reduce reliance on election trustees and make the voting system intelligible to non-technical voters.

\section{The Way Forward for E2E Voting Systems}
\label{sec:wayforward}

As described thus far, considerable work has been done on developing E2E voting systems and reconciling theoretical notions with practical realities. However, to fully transition these systems into the real world, a host of pressing technical issues, usability concerns, and legal dilemmas still need to be resolved. In this section, we briefly overview progress and outstanding challenges in these domains.

\subsection{Technical Issues}

As often happens when prototyping theoretical systems, the implementation itself may introduce security flaws. Karlof et al. \cite{karlof2005cryptographic} analyzed two systems, Votegrity and MarkPledge, and described several scenarios where voter privacy and election integrity may be compromised by exploiting flaws in how the systems were built. For instance, malicious software on voting machines could encrypt votes in a manner that surreptitiously leaks information regarding the voter's candidate choice. Malicious voting machines may sabotage an election by issuing large numbers of forged or invalid receipts. Large scale attacks would be detected but may negatively impacting public confidence in the system. Human factors also need to be considered. For instance, voters who do not intend to verify their votes may discard their receipts on polling premises where a poll worker could collect them and later manipulate the corresponding votes without fear of detection.

Researchers have presented various practical guidelines to implementing voting systems:  Karlof et al. advocate a \textbf{systems approach} \cite{karlof2005cryptographic} which essentially states that security analyses of such systems should not examine system components in isolation but consider the entire system and how the individual parts interact with each other. Bruck et al. strongly caution against ``all-at-once" approaches \cite{bruck2010modular} to building voting equipment where diverse functionality is bundled together into the same hardware components. They propose a highly \textbf{modular architecture}, dubbed \emph{Frog voting}, which clearly partitions processes such as voter registration, vote recording, casting, tallying, and auditing, thereby reducing complexity, simplifying certification, and encouraging incremental innovation. Rivest and Wack have proposed the principle of \textbf{software independence} \cite{rivest2008notion} urging that the integrity of election results should not be dependent on the software used by the system. Popoveniuc et al. \cite{popoveniuc2010performance} also detail a series of security checks which ensure E2E verifiability properties of a voting system.

A second major direction of research has been to improve individual components of voting systems. There is a significant body of work on \textbf{securing electronic voting machines}. For instance, researchers have proposed the use of trusted hardware, i.e. specialized tamper-proof chips which protect cryptographic credentials and guarantee that voting machines run trusted software \cite{paul2009trustworthy} \cite{fink2009tpm} \cite{smart2011true}. Researchers have also identified critical security properties that voting machines must ensure, and proposed corresponding solutions \cite{molnar2006tamper} \cite{bethencourt2007cryptographic}. Another key component in E2E voting systems is the \textbf{public bulletin board} and various solutions have been put forward to implement it \cite{heather2009append} \cite{culnane2014peered} and make it robust against attacks \cite{koenig2011preventing} \cite{haenni2013generic}.

\textbf{Online voting} has also received considerable attention. Proposed methods to defeat hackers and malware include the use of visual cryptography techniques \cite{paul2003authentication}, CAPTCHAs \cite{popoveniuc2008remote}, voice recognition \cite{popoveniuc2010speakup}, and deploying assistive handheld devices \cite{yasinsac2013independent}. However, voter privacy and coercion concerns still persist.

Unfortunately, most of these solutions have been proposed piecemeal and it is unclear if and how they will fit together. There have been very few attempts, like VoteBox (described earlier in Sec.~\ref{sec:votebox}), to harmonize these various innovations in the context of a complete E2E voting system. Substantial work is still needed in this domain before E2E voting systems are deployed for mainstream use.

\subsection{Usability}
\label{sec:usability}

The importance of usability for a voting system cannot be emphasized enough and usability oftentimes conflicts with security properties. Ensuring system usability is also complicated by the fact that elections occur only rarely and voters must be able to vote with near 100\% success while having little or no experience or training on that voting system. This problem also extends to poll workers and election officials who may have limited technical expertise and experience regarding certain systems. For these reasons, the National Institute of Standards and Technology (NIST) has advocated a broad all-encompassing perspective to study usability within the context of the complete voting system, including the physical environment, the voting product, the ballot, the voter, and all personnel involved in the process \cite{laskowski2004improving}.

At this particular stage in the development of E2E voting systems, the central usability concerns are whether voters can successfully cast their votes using these systems, and more importantly, whether or not they are able to undertake the verification process. Regarding vote casting, evidence suggests that certain design features of these systems are problematic. In one study, the ballot design of Pr{\^e}t {\`a} Voter, requiring voters to shred half of their ballot to ensure vote privacy caused confusion \cite{schneider2011focus}. A preliminary study involving Helios, Pr{\^e}t {\`a} Voter, and Scantegrity II found that a significant number of voters failed to cast a vote with each system \cite{acemyan2014usability}. Troublingly, many of those voters thought they had in fact successfully cast a vote. It also took almost twice as long to cast a ballot as with a traditional paper-based system.

Individual vote verification is a new and novel concept for voters and adds a layer of complexity. Critically, voters are active participants in auditing the election and certifying its results, and if, for whatever reason they are unable to verify their votes, the system's security guarantees become mute and the system is not auditable. Recent testing and `live' applications of E2E systems have resulted not just in consistently low rates of voter verification but even lower rates for those who actually report discrepancies \cite{delis2014pressing} \cite{moher1diffusion}. This is likely due to two reasons.

First, voters do not see the need to verify their vote. One possible explanation for this is that \textbf{trust-transference} from election authorities to voter does not happen. As Olembo et al. \cite{olembo2014voter} discovered, voters may initially verify their vote out of curiosity, but after continued use of the system, they establish trust in the system and verification is deemed unnecessary. In line with this conclusion, recent usability testing involving Helios and a Galois-designed prototype based on STAR-Vote, found that participants had a tacit level of trust in any voting system provided it was officially branded e.g. \emph{Jurisdiction X Official Election Website} \cite{murray2015usability}. The participants considered verification unnecessary if a voting system met this basic criterion.

The second concern is complexity. The verification process of Helios has repeatedly been found to be difficult \cite{karayumak2011usability} \cite{murray2015usability}. The reasons may be technical or even that the voters may not understand what to do since the verification language is not related to prior voting experience on their part. Research suggests that terms like ``audit'', ``verifiability'', or ``ballot fingerprint'' lack clarity for voters, cause confusion, and do not engender trust in the voting system \cite{olembo2013mental}. This is critical as voters need to understand the necessity of verification to be motivated to invest the extra effort in verifying their vote. The intuition behind verification also needs to be effectively communicated. As Schneider et al. discovered in their study concerning Pr{\^e}t {\`a} Voter, simply confirming that an encrypted vote on a bulletin board corresponded to a receipt did not provide sufficient security guarantees for voters themselves \cite{schneider2011focus}.

Several technical solutions have been proposed in the literature, including simplifying security elements or enhancing voting systems to facilitate individual vote verification \cite{nandi2010stamp} \cite{ryan2011preta}, bundling multiple receipts to enable mass verification \cite{bohli2009enhancing}, or involving third parties in the process, such as activist groups or helper organizations, \cite{adida2006scratch} \cite{adida2006ballot}, but these require further study from a usability perspective.

Overall, research in the usability of E2E voting systems is still in an early stage and significant work is needed to demonstrate that these systems are accessible to non-expert users while simultaneously maintaining the desired security properties.

\subsection{Legal Framework}
\label{sec:legalandethicalissues}

It is a common perception that numerous attempts to introduce new voting technology, for example in Britain, Finland, Ireland, and Quebec, have failed in large part due to the absence of a comprehensive supporting legal framework \cite{alvarez2010electronic}. Indeed, many of the issues regarding new voting technologies are not due to the technology itself, but rather the lack of ancillary processes that establish trust in the voting system. A new paradigm, like E2E voting, requires legislators to craft an appropriate legal framework that embeds election norms and details administrative procedures to address risks, problems, and threats. This includes aspects such as testing requirements, implementation instructions, and processes for problem resolution.

A useful guiding principle in constructing a legal framework is that of \textbf{functional equivalence}. Any new legal framework must essentially function in an equivalent way to the legal framework for existing voting systems, i.e. the new framework must be at least as good, if not better, than the old one. Towards this end, legislators need to identify the distinct purposes served by existing voting systems and ensure that new rules also address the same needs. Applying this criterion to E2E voting systems requires consideration that current electoral values and community expectations are preserved, while making specific legislative changes to ensure that voting experiences using E2E voting systems are as good as or better than existing procedures.

Another overarching concept is that of \textbf{nondiscrimination} \cite{loncke2004online}. A new legal framework should not have the unintended consequence of discriminating against any voter. For instance, in the United States, any electronic technology, including voting equipment, needs to be Section 508 compliant \cite{vocrehab}, i.e. it must attempt to fully accommodate disability groups. If E2E voting systems cannot successfully cater to certain voter groups, legislation must set out clear alternatives to avoid discriminatory legal challenges. A good example is the case of absentee voting, where eligibility requirements are defined, permitting certain groups to cast their votes remotely. Furthermore, legislation should not unduly limit technological choices to certain vendors or technology in cases where a better technology or solution exists. Excessive detail in legislation can inhibit innovation and create legal \emph{technology locks} \cite{alvarez2010electronic}.

An additional function of a legal framework is to \textbf{minimize and mediate error and risk}. No electoral mechanism, be it electronic or paper, is ever absolutely secure from every possible error or risk, and the greater question is whether the regulatory framework provides voters and political stakeholders confidence that risk is minimized and contingencies are in place to cover the range of possible circumstances.

Regarding specific properties to be included in legislation, Schwartz and Grice detail the following \textbf{normative values} addressing electronic voting systems \cite{schwartz2012establishing}:

\begin{itemize}[noitemsep]
\item Accessibility and facilitate reasonable accommodation
\item Voter anonymity
\item Fairness
\item Accurate and prompt results
\item Comprehensible and transparent processes
\item System security and risk assessment
\item Detection of problems and remedial contingencies
\item Legislative certainty and finality
\item Effective and independent oversight
\item Cost justification and efficiency
\end{itemize}

Many of these values could be embedded into a legal framework in the form of technological choices within the specifications of an E2E voting system. For instance, to deter coercion, voters may be permitted to change or update their vote once it has been cast, as is the case of elections in Estonia \cite{maaten2004towards}. Similarly, if multiple modes of voting are available, legislation may specify that an electronic vote can be revoked if the voter casts a paper ballot. Other technological specifications that may be considered election norms include the capability to detect undervotes and overvotes on marked ballots prior to casting and the ability to cast a blank ballot as a protest vote without it being considered technically invalid.

It should be recognized that constructing a legal framework involves trade-offs between competing norms. An illustrative example is the case of Dutch voters and the Rijnland Internet Election System (RIES) \cite{osce2006netherlands}, an online voting system used for the Rijnland District Water Board Elections in 2004. In the interest of transparency, most of the RIES technology was made open-source. RIES provides a degree of verifiability but weak privacy. Upon voting, the system generated a technical code, enabling voters to verify their vote was included in the tally. However, if a voter disclosed her code, any third party could determine her actual vote. In terms of norms, therefore, transparency trumped voter privacy. Whereas this particular example may not directly apply to most E2E voting systems, where cryptography obfuscates the contents of the voter receipt, it is indicative of the conflicts legislators will have to wrestle with in preparing a legal framework for these new systems.

\subsection{Uptake of E2E Voting Systems}

Broadly speaking, uptake of a new voting system depends on how successfully it incorporates existing election norms, and equally importantly, how effectively this perception is communicated to voters. Research on Internet voting lists four factors which may serve as a good starting point to discuss the uptake of E2E voting systems \cite{oostveen2004internet}. These are security, privacy, accountability, and economic feasibility.

Regarding security and privacy, in Sec.~\ref{sec:securitypropertiesofvotingsystems} we have already motivated the fundamental requirement for voting systems to ensure vote privacy and election integrity. However, security also needs to be considered in terms of \textbf{reliability}, i.e. how a system isolates and reacts to failure.  While a comprehensive legal framework will considerably ameliorate failure reaction, a voting system should be designed so that there is no single point of failure. As a risk assessment of large-scale events notes \cite{hole2010toward}: `If failure in one part of an information system can cause failure in other interconnected parts, then the system is susceptible to cascade failure'. Furthermore, in the words of Pieters and Becker \cite{pieters2005ethics}: ``It is not only important that a system is reliable, it is also important that people believe that the system is reliable.''

Fostering a common understanding of the risks associated with a voting system may actually advance acceptance. Risk exists even in paper-based elections but these are generally known, accepted, and tolerated risks. Transparency and accountability are key to minimizing risk, and risk perception, by detailing and perhaps even publicly demonstrating potential risks and corresponding remedies. The legal framework should require the electoral authority to establish a set of procedures and tests to be completed before the voting system goes live. Additionally, procedures should be legislated that provide solutions for potential risks and these should be updated before each election to ensure protection against new threats or vulnerabilities.

Moreover, comprehensive and high quality voter instruction is critical to uptake of a new system and typically falls under the auspices of the legal framework ensuring equal access. Voter education is particularly important for a new paradigm like E2E voting. These systems are radically different from existing systems that voters are accustomed to, the intuition behind certain procedural steps may not be immediately clear, and they have been known to confuse voters \cite{schneider2011focus} \cite{acemyan2014usability} \cite{murray2015usability}. Furthermore, these systems task voters with the extra responsibility of vote verification, the necessity of which needs to be adequately explained and motivated. Dissemination of this instruction can take many forms, including programs to educate the general public about E2E voting systems via various media, supplying user-friendly instructions, public demonstrations of the technology, and maintaining phone help-lines to walk voters through the voting and verification processes.

\section{Conclusion}
\label{sec:conclusion}

In this chapter we have undertaken a comprehensive introduction to the burgeoning field of end-to-end verifiable voting. We have traced the development of privacy and verifiability properties in the literature and described the workings of current state-of-the-art E2E voting systems. In our classification, we have attempted to track the intellectual evolution of these systems and highlight the key cryptographic and procedural techniques employed by their designers to harmonize security and usability concerns. We have also discussed current challenges to the deployment of these systems, consisting of outstanding technical, legal, and usability issues.

Advances in E2E verifiable voting have the potential to restore trust in elections and improve democratic processes in society and, for that reason, we believe that the importance of these developments must not be underestimated. Our intention, in writing this chapter, has been to make the important innovations in this field accessible to a wider audience. We hope our work serves as a useful resource in this regard and assists in the future development of E2E voting.

\section{Acknowledgements}
\label{sec:Acknowledgements}

The authors wish to thank Peter Hyun-Jeen Lee and Feng Hao for constructive discussions and Siamak F. Shahandashti and Jeremy Clark for helpful comments on the manuscript.
This work is supported by ERC Starting Grant No. 306994.


\begin{thebibliography}{100}

\bibitem{chaum1981untraceable}
David~L Chaum.
\newblock Untraceable electronic mail, return addresses, and digital
  pseudonyms.
\newblock {\em Communications of the ACM}, 24(2):84--90, 1981.

\bibitem{cranor1997sensus}
Lorrie~Faith Cranor and Ron~K Cytron.
\newblock Sensus: A security-conscious electronic polling system for the
  internet.
\newblock In {\em Thirtieth Hawaii International Conference on System
  Sciences}, volume~3, pages 561--570. IEEE, 1997.

\bibitem{herschberg1997secure}
Mark~A Herschberg.
\newblock {\em Secure electronic voting over the world wide web}.
\newblock PhD thesis, Massachusetts Institute of Technology, 1997.

\bibitem{levine08hanging}
Samantha Levine.
\newblock {Hanging Chads: As the Florida Recount Implodes, the Supreme Court
  Decides Bush v. Gore}, Jan. 17 2008.
\newblock
  \url{http://www.usnews.com/news/articles/2008/01/17/the-legacy-of-hanging-chads}.

\bibitem{coleman2011help}
Kevin~J Coleman and Eric~A Fischer.
\newblock The help america vote act and elections reform: Overview and issues,
  2011.

\bibitem{kohno2004analysis}
Tadayoshi Kohno, Adam Stubblefield, Aviel~D Rubin, and Dan~S Wallach.
\newblock Analysis of an electronic voting system.
\newblock In {\em IEEE Symposium on Security and Privacy, 2004. Proceedings},
  pages 27--40. IEEE, 2004.

\bibitem{feldman2006security}
Ariel~J Feldman, J~Alex Halderman, and Edward~W Felten.
\newblock Security analysis of the diebold accuvote-ts voting machine.
\newblock 2006.

\bibitem{wolchok2010security}
Scott Wolchok, Eric Wustrow, J~Alex Halderman, Hari~K Prasad, Arun Kankipati,
  Sai~Krishna Sakhamuri, Vasavya Yagati, and Rop Gonggrijp.
\newblock Security analysis of {I}ndia's electronic voting machines.
\newblock In {\em ACM conference on Computer and Communications Security},
  CCS'10, pages 1--14. ACM, 2010.

\bibitem{springall2014security}
Drew Springall, Travis Finkenauer, Zakir Durumeric, Jason Kitcat, Harri Hursti,
  Margaret MacAlpine, and J~Alex Halderman.
\newblock Security analysis of the {E}stonian {I}nternet voting system.
\newblock In {\em ACM Conference on Computer and Communications Security},
  CCS'14, pages 703--715. ACM, 2014.

\bibitem{accurate}
{ACCURATE: A Center for Correct, Usable, Reliable, Auditable, and Transparent
  Elections}.
\newblock \url{http://accurate-voting.org/}.

\bibitem{iavoss}
{International Association for Voting Systems Sciences}.
\newblock \url{http://www.iavoss.org/}.

\bibitem{chaum2004secret}
David Chaum.
\newblock Secret-ballot receipts: True voter-verifiable elections.
\newblock {\em IEEE security \& privacy}, 2(1):38--47, 2004.

\bibitem{neff04practical}
C~Andrew NNeff.
\newblock Practical high certainty intent verification for encrypted votes.
  vote-here (2004), 2004.
\newblock \url{http://www.votehere.net/vhti/documentation}.

\bibitem{fcc09}
The Federal Constitutional Court Press Release~No. 19/2009.
\newblock Use of voting computers in 2005 bundestag election unconstitutional,
  March 3 2009.
\newblock
  \url{http://www.bundesverfassungsgericht.de/SharedDocs/Pressemitteilungen/EN/2009/bvg09-019.html}.

\bibitem{assembly48universal}
UN~General Assembly.
\newblock Universal declaration of human rights.
\newblock {\em Resolution adopted by the General Assembly}, 10(12), 1948.

\bibitem{hall1990greeks}
Ursula Hall.
\newblock Greeks and romans and the secret ballot.
\newblock {\em Owls to Athens: Essays on Classical Subjects Presented to Sir
  Kenneth Dover, ed. EM Craik, Oxford}, 191:199, 1990.

\bibitem{stokes2012killed}
Susan Stokes, Thad Dunning, Marcelo Nazareno, and Valeria Brusco.
\newblock What killed vote-buying in britain and the united states?
\newblock 2012.
\newblock
  \url{http://www.princeton.edu/csdp/online-community/historical-theoretical-pe/What-Killed-Vote-Buying-in-Britain-and-the-US.pdf}.

\bibitem{jensen2014poverty}
Peter~Sandholt Jensen and Mogens~K Justesen.
\newblock Poverty and vote buying: Survey-based evidence from africa.
\newblock {\em Electoral Studies}, 33:220--232, 2014.

\bibitem{gonzalez2012vote}
Ezequiel Gonzalez-Ocantos, Chad~Kiewiet De~Jonge, Carlos Mel{\'e}ndez, Javier
  Osorio, and David~W Nickerson.
\newblock Vote buying and social desirability bias: Experimental evidence from
  nicaragua.
\newblock {\em American Journal of Political Science}, 56(1):202--217, 2012.

\bibitem{cohen1985robust}
Josh~D Cohen and Michael~J Fischer.
\newblock A robust and verifiable cryptographically secure election scheme.
\newblock In {\em 26th Annual Symposium on Foundations of Computer Science},
  pages 372--382. IEEE, 1985.

\bibitem{benaloh1986distributing}
Josh~C Benaloh and Moti Yung.
\newblock Distributing the power of a government to enhance the privacy of
  voters.
\newblock In {\em ACM symposium on Principles of distributed computing}, pages
  52--62. ACM, 1986.

\bibitem{benaloh87verifiable}
Josh Benaloh.
\newblock {\em Verifiable Secret-Ballot Elections}.
\newblock PhD thesis. Yale University, Department of Computer Science
  Department, 1987.

\bibitem{chaum1988elections}
David Chaum.
\newblock Elections with unconditionally-secret ballots and disruption
  equivalent to breaking {RSA}.
\newblock In {\em Advances in Cryptology--EUROCRYPT'88}, pages 177--182.
  Springer, 1988.

\bibitem{benaloh1994receipt}
Josh Benaloh and Dwight Tuinstra.
\newblock Receipt-free secret-ballot elections.
\newblock In {\em ACM symposium on Theory of computing}, pages 544--553. ACM,
  1994.

\bibitem{niemi1995prevent}
Valtteri Niemi and Ari Renvall.
\newblock How to prevent buying of votes in computer elections.
\newblock In {\em Advances in Cryptology--ASIACRYPT'94}, pages 164--170.
  Springer, 1995.

\bibitem{sako1995receipt}
Kazue Sako and Joe Kilian.
\newblock Receipt-free mix-type voting scheme.
\newblock In {\em Advances in Cryptology--EUROCRYPT'95}, pages 393--403.
  Springer, 1995.

\bibitem{okamoto1996electronic}
Tatsuaki Okamoto.
\newblock An electronic voting scheme.
\newblock In {\em Advanced IT Tools}, pages 21--30. Springer, 1996.

\bibitem{okamoto1998receipt}
Tatsuaki Okamoto.
\newblock Receipt-free electronic voting schemes for large scale elections.
\newblock In {\em Security Protocols Workshop}, pages 25--35. Springer, 1998.

\bibitem{hirt2000efficient}
Martin Hirt and Kazue Sako.
\newblock Efficient receipt-free voting based on homomorphic encryption.
\newblock In {\em Advances in Cryptology--EUROCRYPT 2000}, pages 539--556.
  Springer, 2000.

\bibitem{juels2005coercion}
Ari Juels, Dario Catalano, and Markus Jakobsson.
\newblock Coercion-resistant electronic elections.
\newblock In {\em ACM workshop on Privacy in the electronic society}, pages
  61--70. ACM, 2005.

\bibitem{delaune2006coercion}
Stephanie Delaune, Steve Kremer, and Mark Ryan.
\newblock Coercion-resistance and receipt-freeness in electronic voting.
\newblock In {\em Computer Security Foundations Workshop, 2006. 19th IEEE},
  pages 12--pp. IEEE, 2006.

\bibitem{jonker2013privacy}
Hugo Jonker, Sjouke Mauw, and Jun Pang.
\newblock Privacy and verifiability in voting systems: Methods, developments
  and trends.
\newblock {\em Computer Science Review}, 10:1--30, 2013.

\bibitem{neumann2014analysis}
Stephan Neumann, Jurlind Budurushi, and Melanie Volkamer.
\newblock Analysis of security and cryptographic approaches to provide secret
  and verifiable electronic voting.
\newblock In {\em Design, Development, and Use of Secure Electronic Voting
  Systems}. IGI Global, 2014.

\bibitem{karayumak2011usability}
Fatih Karayumak, Maina~M Olembo, Michaela Kauer, and Melanie Volkamer.
\newblock Usability analysis of {H}elios -- an open source verifiable remote
  electronic voting system.
\newblock In {\em USENIX Electronic Voting Technology Workshop/Workshop on
  Trustworthy Elections}, 2011.

\bibitem{acemyan2014usability}
Claudia~Z Acemyan, Philip Kortum, Michael~D Byrne, and Dan~S Wallach.
\newblock Usability of voter verifiable, end-to-end voting systems: Baseline
  data for {Helios}, {Pr{\^e}t {\`a} Voter}, and {Scantegrity II}.
\newblock {\em The USENIX Journal of Election Technology and Systems}, page~26,
  2014.

\bibitem{murray2015usability}
Judith Murray.
\newblock Usability testing for end-to-end verifiable internet voting project
  feasibility study.
\newblock {\em Publication Pending}, 2015.

\bibitem{carback2010scantegrity}
Richard Carback, David Chaum, Jeremy Clark, John Conway, Aleksander Essex,
  Paul~S Herrnson, Travis Mayberry, Stefan Popoveniuc, Ronald~L Rivest, Emily
  Shen, et~al.
\newblock Scantegrity ii municipal election at takoma park: the first e2e
  binding governmental election with ballot privacy.
\newblock In {\em USENIX Security}, pages 19--19. USENIX Association, 2010.

\bibitem{moher1diffusion}
Ester Moher, Jeremy Clark, and Aleksander Essex.
\newblock Diffusion of voter responsibility: Potential failings in {E2E} voter
  receipt checking.
\newblock {\em USENIX Journal of Election Technology and Systems (JETS)}, 1,
  2014.

\bibitem{rivest2001electronic}
Ronald~L Rivest.
\newblock Electronic voting.
\newblock In {\em Financial Cryptography}, volume~1, pages 243--268, 2001.

\bibitem{naor1995visual}
Moni Naor and Adi Shamir.
\newblock Visual cryptography.
\newblock In {\em Advances in Cryptology--EUROCRYPT'94}, pages 1--12. Springer,
  1995.

\bibitem{chaum2007secret}
David Chaum, Jeroen Van De~Graaf, Peter Y~A Ryan, and Poorvi~L Vora.
\newblock {\em Secret ballot elections with unconditional integrity}.
\newblock University of Newcastle upon Tyne, Computing Science, 2007.

\bibitem{jakobsson2002making}
Markus Jakobsson, Ari Juels, and Ronald~L Rivest.
\newblock Making mix nets robust for electronic voting by randomized partial
  checking.
\newblock In {\em USENIX security symposium}, pages 339--353, 2002.

\bibitem{ryan2009pret}
Peter Y~A Ryan, David Bismark, James Heather, Steve Schneider, and Zhe Xia.
\newblock Pr{\^e}t {\`a} voter: a voter-verifiable voting system.
\newblock {\em IEEE Transactions on Information Forensics and Security},
  4(4):662--673, 2009.

\bibitem{chaum2005practical}
David Chaum, Peter Y~A Ryan, and Steve Schneider.
\newblock A practical voter-verifiable election scheme.
\newblock In {\em Computer Security -- ESORICS 2005}, volume 3679 of {\em
  Lecture Notes in Computer Science}, pages 118--139. Springer Berlin
  Heidelberg, 2005.

\bibitem{heather2007implementing}
James Heather.
\newblock Implementing stv securely in pr{\^e}t {\`a} voter.
\newblock In {\em Computer Security Foundations Symposium (CSF'07)}, pages
  157--169. IEEE, 2007.

\bibitem{xia2010versatile}
Zhe Xia, Chris Culnane, James Heather, Hugo Jonker, Peter Y~A Ryan, Steve
  Schneider, and Sriramkrishnan Srinivasan.
\newblock Versatile pr{\^e}t {\`a} voter: Handling multiple election methods
  with a unified interface.
\newblock In {\em Progress in Cryptology-INDOCRYPT 2010}, pages 98--114.
  Springer, 2010.

\bibitem{popoveniuc2007simple}
Stefan Popoveniuc and David Lundin.
\newblock A simple technique for safely using {P}unchscan and pr{\^e}t {\`a}
  voter in mail-in elections.
\newblock In {\em E-Voting and Identity}, pages 150--155. Springer, 2007.

\bibitem{lundin2008human}
David Lundin and Peter Y~A Ryan.
\newblock Human readable paper verification of {P}r\^{e}t \`{a} {V}oter.
\newblock In {\em European Symposium on Research in Computer Security
  (ESORICS'08)}, pages 379--395. Springer-Verlag, 2008.

\bibitem{ryan2011preta}
Peter Y~A Ryan.
\newblock Pr{\^e}t {\`a} voter with confirmation codes.
\newblock In {\em USENIX Electronic Voting Technology Workshop}, 2011.

\bibitem{culnane2013faster}
Chris Culnane, James Heather, Rui Joaquim, Peter Y~A Ryan, Steve Schneider, and
  Vanessa Teague.
\newblock Faster print on demand for pr{\^e}t {\`a} voter.
\newblock {\em USENIX Journal of Election Technology and Systems}, 2(1), 2013.

\bibitem{bismark2010experiences}
David Bismark, James Heather, Roger Peel, Steve Schneider, Zhe Xia, and Peter
  Y~A Ryan.
\newblock Experiences gained from the first pret a voter implementation.
\newblock In {\em International Workshop on Requirements Engineering for
  e-Voting Systems (RE-VOTE)}, pages 19--28. IEEE, 2010.

\bibitem{burton2012using}
Craig Burton, Chris Culnane, James Heather, Thea Peacock, Peter Y~A Ryan, Steve
  Schneider, Sriramkrishnan Srinivasan, Vanessa Teague, Roland Wen, and Zhe
  Xia.
\newblock Using pr{\^e}t \'{a} {V}oter in the {V}ictorian state elections.
\newblock In {\em Electronic Voting Technology Workshop/Workshop on Trustworthy
  Elections (EVT/WOTE'12)}, 2012.

\bibitem{culnane2014vvote}
Chris Culnane, Peter Y~A Ryan, Steve Schneider, and Vanessa Teague.
\newblock v{V}ote: a verifiable voting system.
\newblock {\em arXiv preprint arXiv:1404.6822}, 2014.

\bibitem{fisher2006punchscan}
Kevin Fisher, Richard Carback, and Alan~T Sherman.
\newblock Punchscan: Introduction and system definition of a high-integrity
  election system.
\newblock In {\em Workshop on Trustworthy Elections (WOTE)}, 2006.

\bibitem{kelsey2010attacking}
John Kelsey, Andrew Regenscheid, Tal Moran, and David Chaum.
\newblock Attacking paper-based e2e voting systems.
\newblock In {\em Towards Trustworthy Elections}, pages 370--387. Springer,
  2010.

\bibitem{carback2007independent}
Richard~T Carback~III, Jeremy Clark, Aleks Essex, and Stefan Popoveniuc.
\newblock On the independent verification of a {P}unchscan election.
\newblock {\em Proc. VoComp}, 42, 2007.

\bibitem{van2009voting}
Jeroen Van De~Graaf.
\newblock Voting with unconditional privacy by merging pr{\^e}t {\`a} voter and
  {P}unchscan.
\newblock {\em IEEE Transactions on Information Forensics and Security},
  4(4):674--684, 2009.

\bibitem{essex2007punchscan}
Aleks Essex, Jeremy Clark, Richard Carback, and Stefan Popoveniuc.
\newblock Punchscan in practice: an e2e election case study.
\newblock In {\em Workshop on Trustworthy Elections}, 2007.

\bibitem{chaum2008scantegrity}
David Chaum, Aleks Essex, Richard Carback, Jeremy Clark, Stefan Popoveniuc,
  Alan Sherman, and Poorvi Vora.
\newblock Scantegrity: End-to-end voter-verifiable optical-scan voting.
\newblock {\em Security \& Privacy, IEEE}, 6(3):40--46, 2008.

\bibitem{sherman2010scantegrity}
Alan~T Sherman, Richard~T Carback, David Chaum, Jeremy Clark, Aleksander Essex,
  Paul~S Hernson, Travis Mayberry, Stefan Popoveniuc, Ronald~L Rivest, Emily
  Shen, Bimal Sinha, and Poorvi~L Vora.
\newblock Scantegrity mock election at {T}akoma {P}ark.
\newblock In {\em EVOTE}, 2010.

\bibitem{sherman2011scantegrity}
Alan~T Sherman, Russell~A Fink, Richard Carback, and David Chaum.
\newblock Scantegrity iii: automatic trustworthy receipts, highlighting
  over/under votes, and full voter verifiability.
\newblock In {\em Conference on Electronic voting technology/workshop on
  trustworthy elections}, pages 7--7. USENIX Association, 2011.

\bibitem{adida2006scratch}
Ben Adida and Ronald~L Rivest.
\newblock Scratch \& vote: self-contained paper-based cryptographic voting.
\newblock In {\em the 5th ACM workshop on Privacy in electronic society}, pages
  29--40. ACM, 2006.

\bibitem{pedersen1991threshold}
Torben~Pryds Pedersen.
\newblock A threshold cryptosystem without a trusted party.
\newblock In {\em Advances in Cryptology--EUROCRYPT'91}, pages 522--526.
  Springer, 1991.

\bibitem{adida2006ballot}
Ben Adida and C~Andrew Neff.
\newblock Ballot casting assurance.
\newblock In {\em USENIX/Accurate Electronic Voting Technology Workshop}, 2006.

\bibitem{joaquim2012efficient}
Rui Joaquim and Carlos Ribeiro.
\newblock An efficient and highly sound voter verification technique and its
  implementation.
\newblock In {\em E-voting and identity}, pages 104--121. Springer, 2012.

\bibitem{joaquim2009veryvote}
Rui Joaquim, Carlos Ribeiro, and Paulo Ferreira.
\newblock Veryvote: A voter verifiable code voting system.
\newblock In {\em E-voting and identity}, pages 106--121. Springer, 2009.

\bibitem{joaquim2013eviv}
Rui Joaquim, Paulo Ferreira, and Carlos Ribeiro.
\newblock Eviv: An end-to-end verifiable internet voting system.
\newblock {\em computers \& security}, 32:170--191, 2013.

\bibitem{bohli2007bingo}
Jens-Matthias Bohli, J{\"o}rn M{\"u}ller-Quade, and Stefan R{\"o}hrich.
\newblock Bingo voting: Secure and coercion-free voting using a trusted random
  number generator.
\newblock In {\em E-Voting and Identity}, pages 111--124. Springer, 2007.

\bibitem{bar2008real}
Michael B{\"a}r, Christian Henrich, J{\"o}rn M{\"u}ller-Quade, Stefan
  R{\"o}hrich, and Carmen St{\"u}ber.
\newblock Real world experiences with {B}ingo {V}oting and a comparison of
  usability.
\newblock In {\em Workshop On Trustworthy Elections (WOTE)}, volume 2008, 2008.

\bibitem{bohli2009enhancing}
J-M Bohli, Christian Henrich, Carmen Kempka, J~Muller-Quade, and S~Rohrich.
\newblock Enhancing electronic voting machines on the example of {B}ingo
  voting.
\newblock {\em IEEE Transactions on Information Forensics and Security},
  4(4):745--750, 2009.

\bibitem{liu2012improved}
Yining Liu, Peiyong Sun, Jihong Yan, Yajun Li, and Jianyu Cao.
\newblock An improved electronic voting scheme without a trusted random number
  generator.
\newblock In {\em Information Security and Cryptology}, pages 93--101.
  Springer, 2012.

\bibitem{henrich2012improving}
Christian Henrich.
\newblock {\em Improving and Analysing Bingo Voting}.
\newblock PhD thesis, Karlsruhe, Karlsruher Institut f{\"u}r Technologie (KIT),
  Diss., 2012, 2012.

\bibitem{benaloh2006simple}
Josh Benaloh.
\newblock Simple verifiable elections.
\newblock In {\em USENIX/Accurate Electronic Voting Technology Workshop}, pages
  5--5. USENIX Association, 2006.

\bibitem{benaloh2007ballot}
Josh Benaloh.
\newblock Ballot casting assurance via voter-initiated poll station auditing.
\newblock In {\em USENIX Workshop on Accurate Electronic Voting Technology},
  2007.

\bibitem{benaloh2008administrative}
Josh Benaloh.
\newblock Administrative and public verifiability: can we have both?
\newblock {\em EVT}, 8:1--10, 2008.

\bibitem{sandler2008votebox}
Daniel Sandler, Kyle Derr, and Dan~S Wallach.
\newblock Votebox: A tamper-evident, verifiable electronic voting system.
\newblock In {\em USENIX Security Symposium}, volume~4, page~87, 2008.

\bibitem{yee2006prerendered}
Ka-Ping Yee, David Wagner, Marti Hearst, and Steven~M Bellovin.
\newblock Prerendered user interfaces for higher-assurance electronic voting.
\newblock In {\em USENIX/ACCURATE Electronic Voting Technology Workshop}, 2006.

\bibitem{sandler2007casting}
Daniel Sandler and Dan~S Wallach.
\newblock Casting votes in the auditorium.
\newblock In {\em USENIX/ACCURATE Electronic Voting Technology Workshop
  (EVT'07)}, 2007.

\bibitem{jones2006secure}
Douglas~W Jones and Tom~C Bowersox.
\newblock Secure data export and auditing using data diodes.
\newblock {\em Technology}, 6:7, 2006.

\bibitem{sandler2008case}
Daniel Sandler and Dan~S Wallach.
\newblock The case for networked remote voting precincts.
\newblock {\em EVT}, 8:1--7, 2008.

\bibitem{oksuzoglu2009votebox}
Ersin {\"O}ks{\"u}zoglu and Dan~S Wallach.
\newblock Votebox nano: A smaller, stronger fpga-based voting machine (short
  paper).
\newblock In {\em USENIX/Accurate Electronic Voting Technology
  Workshop/Workshop on Trustworthy Elections}, 2009.

\bibitem{ben2012new}
Jonathan Ben-Nun, Niko Fahri, Morgan Llewellyn, Ben Riva, Alon Rosen, Amnon
  Ta-Shma, and Douglas Wikstr{\"o}m.
\newblock A new implementation of a dual (paper and cryptographic) voting
  system.
\newblock In {\em Electronic Voting}, pages 315--329, 2012.

\bibitem{grundland2012analysis}
Eitan Grundland.
\newblock An analysis of the wombat voting system model, 2012.
\newblock
  \url{http://www.grundland.org/An\%20Analysis\%20of\%20the\%20Wombat\%20Voting\%20System\%20Model.pdf}.

\bibitem{wikstrom2013user}
Douglas Wikstr{\"o}m.
\newblock User manual for the verificatum mix-net version 1.4. 0.
\newblock {\em Verificatum AB, Stockholm, Sweden}, 2013.

\bibitem{bell2013star}
Susan Bell, Josh Benaloh, Michael~D Byrne, Dana DeBeauvoir, Bryce Eakin, Gail
  Fisher, Philip Kortum, Neal McBurnett, Julian Montoya, Michelle Parker,
  Olivier Pereira, Philip~B Stark, Dan~S Wallach, and Michael Winn.
\newblock Star-vote: A secure, transparent, auditable, and reliable voting
  system.
\newblock {\em The USENIX Journal of Election Technology Systems, 1 (1)}, pages
  18--37, 2013.

\bibitem{lindeman2012gentle}
Mark Lindeman and Philip~B Stark.
\newblock A gentle introduction to risk-limiting audits.
\newblock {\em IEEE Security \& Privacy}, (5):42--49, 2012.

\bibitem{letzter2014new}
Rafi Letzter.
\newblock A new voting machine could make sure every vote really counts, 2014.
\newblock
  \url{http://www.popsci.com/article/technology/new-voting-machine-could-make-sure-every-vote-really-counts}.

\bibitem{lim2014travis}
Andra Lim.
\newblock Travis county, tx developing electronic voting system with a paper
  trail, July 15 2014.
\newblock
  \url{http://www.govtech.com/products/Travis-County-TX-Developing-Electronic-Voting-System-With-a-Paper-Trail.html}.

\bibitem{hao2012self}
Feng Hao, Brian Randell, and Dylan Clarke.
\newblock Self-enforcing electronic voting.
\newblock In {\em Security Protocols Workshop}. Springer, 2012.

\bibitem{adida2009electing}
Ben Adida, Olivier De~Marneffe, Olivier Pereira, and Jean-Jacques Quisquater.
\newblock Electing a university president using open-audit voting: analysis of
  real-world use of {Helios}.
\newblock In {\em International Conference on Electronic Voting
  Technology/Workshop on Trustworthy Elections}, 2009.

\bibitem{hao2014every}
Feng Hao, Matthew Kreeger, Brian Randell, Dylan Clarke, Siamak~F Shahandashti,
  and Peter Hyun-Jeen Lee.
\newblock Every vote counts: ensuring integrity in large-scale electronic
  voting.
\newblock {\em The USENIX Journal of Election Technology and Systems}, pages
  1--25, 2014.

\bibitem{groth2004efficient}
Jens Groth.
\newblock Efficient maximal privacy in boardroom voting and anonymous
  broadcast.
\newblock In {\em Financial Cryptography}, pages 90--104. Springer, 2004.

\bibitem{kiayias2002self}
Aggelos Kiayias and Moti Yung.
\newblock Self-tallying elections and perfect ballot secrecy.
\newblock In {\em Public Key Cryptography}, pages 141--158. Springer, 2002.

\bibitem{hao2010anonymous}
Feng Hao, Peter Y~A Ryan, and Piotr Zieli{\'n}ski.
\newblock Anonymous voting by two-round public discussion.
\newblock {\em IET Information Security}, 4(2):62--67, 2010.

\bibitem{hao2013verifiable}
Feng Hao, Dylan Clarke, and Carlton Shepherd.
\newblock Verifiable classroom voting: Where cryptography meets pedagogy.
\newblock In {\em Security Protocols Workshop}. Springer, 2013.

\bibitem{kiayias2006internet}
Aggelos Kiayias, Michael Korman, and David Walluck.
\newblock An internet voting system supporting user privacy.
\newblock In {\em Annual Computer Security Applications Conference (ACSAC'06),
  2006}, pages 165--174. IEEE, 2006.

\bibitem{stewart2006banana}
John Stewart.
\newblock A banana republic? the investigation into electoral fraud by the
  birmingham election court.
\newblock {\em Parliamentary Affairs}, 59(4):654--667, 2006.

\bibitem{araujo2010practical}
Roberto Araujo, S{\'e}bastien Foulle, and Jacques Traor{\'e}.
\newblock A practical and secure coercion-resistant scheme for internet voting.
\newblock In {\em Towards Trustworthy Elections}, pages 330--342. Springer,
  2010.

\bibitem{spycher2012new}
Oliver Spycher, Reto Koenig, Rolf Haenni, and Michael Schl{\"a}pfer.
\newblock A new approach towards coercion-resistant remote e-voting in linear
  time.
\newblock In {\em Financial Cryptography and Data Security}, volume 7035, page
  182. Springer Science \& Business Media, 2012.

\bibitem{haghighat2013efficient}
Alireza~Toroghi Haghighat, Mohammad~Sadeq Dousti, and Rasool Jalili.
\newblock An efficient and provably-secure coercion-resistant e-voting
  protocol.
\newblock In {\em Annual International Conference on Privacy, Security and
  Trust (PST)}, pages 161--168. IEEE, 2013.

\bibitem{clarkson2008civitas}
Michael~R Clarkson, Stephen Chong, and Andrew~C Myers.
\newblock Civitas: Toward a secure voting system.
\newblock In {\em IEEE Symposium on Security and Privacy}, pages 354--368. IEEE
  Computer Society, 2008.

\bibitem{krivoruchko2007robust}
Taisya Krivoruchko.
\newblock Robust coercion-resistant registration for remote e-voting.
\newblock In {\em Workshop on Trustworthy Elections (WOTE'07)}, 2007.

\bibitem{koenig2011preventing}
Reto Koenig, Rolf Haenni, and Stephan Fischli.
\newblock Preventing board flooding attacks in coercion-resistant electronic
  voting schemes.
\newblock In {\em Future Challenges in Security and Privacy for Academia and
  Industry}, pages 116--127. Springer, 2011.

\bibitem{neumann2012civitas}
Stephan Neumann and Melanie Volkamer.
\newblock Civitas and the real world: Problems and solutions from a practical
  point of view.
\newblock In {\em International Conference on Availability, Reliability and
  Security (ARES)}, pages 180--185. IEEE, 2012.

\bibitem{neumann2013towards}
Stephan Neumann, Christian Feier, Melanie Volkamer, and Reto~E Koenig.
\newblock Towards a practical jcj/civitas implementation.
\newblock {\em IACR Cryptology ePrint Archive}, 2013:464, 2013.

\bibitem{clark2012selections}
Jeremy Clark and Urs Hengartner.
\newblock Selections: Internet voting with over-the-shoulder
  coercion-resistance.
\newblock In {\em Financial Cryptography and Data Security}, pages 47--61.
  Springer, 2012.

\bibitem{bursuc2012trivitas}
Sergiu Bursuc, Gurchetan~S Grewal, and Mark Ryan.
\newblock Trivitas: Voters directly verifying votes.
\newblock In {\em E-Voting and Identity}, pages 190--207. Springer, 2012.

\bibitem{grewal2013caveat}
Gurchetan~S Grewal, Mark Ryan, Sergiu Bursuc, and Peter Y~A Ryan.
\newblock Caveat coercitor: Coercion-evidence in electronic voting.
\newblock In {\em IEEE Symposium on Security and Privacy (SP)}, pages 367--381.
  IEEE, 2013.

\bibitem{adida2008helios}
Ben Adida.
\newblock Helios: Web-based open-audit voting.
\newblock In {\em USENIX Security Symposium}, volume~17, pages 335--348, 2008.

\bibitem{weber2009usability}
Janna-Lynn Weber and Urs Hengartner.
\newblock Usability study of the open audit voting system {H}elios.
\newblock INCLUDE, 2009.

\bibitem{karayumak2011user}
Fatih Karayumak, Michaela Kauer, Maina~M Olembo, Tobias Volk, and Melanie
  Volkamer.
\newblock User study of the improved {H}elios voting system interfaces.
\newblock In {\em Workshop on Socio-Technical Aspects in Security and Trust
  (STAST)}, pages 37--44. IEEE, 2011.

\bibitem{estehghari2010exploiting}
Saghar Estehghari and Yvo Desmedt.
\newblock Exploiting the client vulnerabilities in internet e-voting systems:
  Hacking {H}elios 2.0 as an example.
\newblock In {\em Electronic Voting Technology Workship/Workshop on Trustworthy
  Elections (EVT/WOTE)}, 2010.

\bibitem{heiderich2012bug}
Mario Heiderich, Tilman Frosch, Marcus Niemietz, and J{\"o}rg Schwenk.
\newblock The bug that made me president a browser-and web-security case study
  on {H}elios voting.
\newblock In {\em E-voting and identity}, pages 89--103. Springer, 2012.

\bibitem{bernhard2012not}
David Bernhard, Olivier Pereira, and Bogdan Warinschi.
\newblock How not to prove yourself: Pitfalls of the {F}iat-{S}hamir heuristic
  and applications to {H}elios.
\newblock In {\em Advances in Cryptology--ASIACRYPT 2012}, pages 626--643.
  Springer, 2012.

\bibitem{cortier2013attacking}
V{\'e}ronique Cortier and Ben Smyth.
\newblock Attacking and fixing {H}elios: An analysis of ballot secrecy.
\newblock {\em Journal of Computer Security}, 21(1):89--148, 2013.

\bibitem{kusters2012clash}
Ralf Kusters, Tomasz Truderung, and Andreas Vogt.
\newblock Clash attacks on the verifiability of e-voting systems.
\newblock In {\em IEEE Symposium on Security and Privacy (SP)}, pages 395--409.
  IEEE, 2012.

\bibitem{dossogne2014blinded}
J{\'e}r{\^o}me Dossogne and Fr{\'e}d{\'e}ric Lafitte.
\newblock Blinded additively homomorphic encryption schemes for self-tallying
  voting.
\newblock {\em Journal of Information Security and Applications}, 2014.

\bibitem{bernhard2011adapting}
David Bernhard, V{\'e}ronique Cortier, Olivier Pereira, Ben Smyth, and Bogdan
  Warinschi.
\newblock Adapting {H}elios for provable ballot privacy.
\newblock In {\em Computer Security--ESORICS 2011}, pages 335--354. Springer,
  2011.

\bibitem{demirel2012improving}
Denise Demirel, Jeroen Van De~Graaf, and Roberto Ara{\'u}jo.
\newblock Improving {H}elios with everlasting privacy towards the public.
\newblock In J~Alex Halderman and Olivier Pereira, editors, {\em Electronic
  Voting Technology Workshop / Workshop on Trustworthy Elections,
  {EVT/WOTE}'12}. {USENIX} Association, 2012.

\bibitem{morell2010secret}
Dan Morrell.
\newblock {Secret Ballots, Verifiable Votes}, May-June 2010.
\newblock
  \url{http://harvardmagazine.com/2010/05/secret-ballots-verifiable-votes}.

\bibitem{heliosiacr}
Should the {IACR} use e-voting for its elections?
\newblock \url{http://www.iacr.org/elections/eVoting/}.

\bibitem{HeliosACM}
{ACM Council Election}.
\newblock \url{http://www.acm.org/acmelections}.

\bibitem{louridas2014zeus}
Panos Louridas, Georgios Tsoukalas, Kostas Papadimitriou, and Panayiotis
  Tsanakas.
\newblock Zeus: Bringing internet voting to greece.
\newblock In {\em E-Democracy, Security, Privacy and Trust in a Digital World},
  pages 213--223. Springer, 2014.

\bibitem{bulens2011running}
Philippe Bulens, Damien Giry, Olivier Pereira, et~al.
\newblock Running mixnet-based elections with {Helios}.
\newblock In {\em Electronic Voting Technology Workshop/Workshop on Trustworthy
  Elections. Usenix}, 2011.

\bibitem{ryan2013pretty}
Peter Y~A Ryan and Vanessa Teague.
\newblock Pretty good democracy.
\newblock In {\em Security Protocols Workshop XVII}, pages 111--130. Springer,
  2013.

\bibitem{chaum2001surevote}
David Chaum.
\newblock Surevote: technical overview.
\newblock In {\em Workshop on trustworthy elections (WOTE'01)}, 2001.

\bibitem{kutylowski2010scratch}
Miros{\l}aw Kuty{\l}owski and Filip Zag{\'o}rski.
\newblock Scratch, click \& vote: E2e voting over the internet.
\newblock In {\em Towards trustworthy elections}, pages 343--356. Springer,
  2010.

\bibitem{zagorski2013remotegrity}
Filip Zag{\'o}rski, Richard~T Carback, David Chaum, Jeremy Clark, Aleksander
  Essex, and Poorvi~L Vora.
\newblock Remotegrity: Design and use of an end-to-end verifiable remote voting
  system.
\newblock In {\em Applied Cryptography and Network Security}, pages 441--457.
  Springer, 2013.

\bibitem{pereira2014scroll}
Gon{\c{c}}alo David Martins~Tourais Pereira.
\newblock Scroll, match \& vote: An e2e coercion resistant mobile voting
  system.
\newblock In {\em EVOTE}, pages 149--152. IEEE, 2014.

\bibitem{budurushi2013pretty}
Jurlind Budurushi, Stephan Neumann, Maina~M Olembo, and Melanie Volkamer.
\newblock Pretty understandable democracy: A secure and understandable
  {I}nternet voting scheme.
\newblock In {\em International Conference on Availability, Reliability and
  Security (ARES)}, pages 198--207. IEEE, 2013.

\bibitem{bella2014virtually}
Giampaolo Bella, Peter Y~A Ryan, and Vanessa Teague.
\newblock Virtually perfect democracy.
\newblock In {\em Security Protocols XVIII}, pages 161--166. Springer, 2014.

\bibitem{heather2010pretty}
James Heather, Peter Y~A Ryan, and Vanessa Teague.
\newblock Pretty good democracy for more expressive voting schemes.
\newblock In {\em Computer Security--ESORICS 2010}, pages 405--423. Springer,
  2010.

\bibitem{rivest2007three}
Ronald~L Rivest and Warren~D Smith.
\newblock Three voting protocols: Threeballot, vav, and twin.
\newblock In {\em USENIX Workshop on Accurate Electronic Voting Technology},
  volume~16. USENIX Association, 2007.

\bibitem{jones2006threeballot}
Harvey Jones, Jason Juang, and Greg Belote.
\newblock Threeballot in the field.
\newblock {\em Term paper for MIT course}, 6, 2006.

\bibitem{appel2006defeat}
Andrew~W Appel.
\newblock How to defeat {R}ivest's {T}hree{B}allot voting system.
\newblock {\em Manuskrypt, pazdziernik}, 2006.

\bibitem{henry2009effectiveness}
Kevin Henry, Douglas~R Stinson, and Jiayuan Sui.
\newblock The effectiveness of receipt-based attacks on threeballot.
\newblock {\em Information Forensics and Security, IEEE Transactions on},
  4(4):699--707, 2009.

\bibitem{cichon2008short}
Jacek Cicho{\'n}, Miros{\l}aw Kuty{\l}owski, and Bogdan Weglorz.
\newblock Short ballot assumption and threeballot voting protocol.
\newblock In {\em SOFSEM 2008: Theory and Practice of Computer Science}, pages
  585--598. Springer, 2008.

\bibitem{randell2006voting}
Brian Randell and Peter Y~A Ryan.
\newblock Voting technologies and trust.
\newblock {\em IEEE SECURITY \& PRIVACY}, 4(5):0050--56, 2006.

\bibitem{essex2010aperio}
Aleks Essex, Jeremy Clark, and Carlisle Adams.
\newblock Aperio: High integrity elections for developing countries.
\newblock In {\em Towards Trustworthy Elections}, pages 388--401. Springer,
  2010.

\bibitem{essex2012eperio}
Aleksander Essex, Jeremy Clark, Urs Hengartner, and Carlisle Adams.
\newblock Eperio: Mitigating technical complexity in cryptographic election
  verification.
\newblock {\em IACR Cryptology ePrint Archive}, 2012:178, 2012.

\bibitem{essex2012hover}
Aleksander Essex and Urs Hengartner.
\newblock Hover: Trustworthy elections with hash-only verification.
\newblock {\em IEEE Security \& Privacy}, 10(5):18--24, 2012.

\bibitem{karlof2005cryptographic}
Chris Karlof, Naveen Sastry, and David Wagner.
\newblock Cryptographic voting protocols: A systems perspective.
\newblock In {\em USENIX Security}, volume~5, pages 33--50, 2005.

\bibitem{bruck2010modular}
Shuki Bruck, David Jefferson, and Ronald~L Rivest.
\newblock A modular voting architecture (``frog voting'').
\newblock In {\em Towards Trustworthy Elections}, pages 97--106. Springer,
  2010.

\bibitem{rivest2008notion}
Ronald~L Rivest.
\newblock On the notion of `software independence' in voting systems.
\newblock {\em Philosophical Transactions of The Royal Society A},
  366(1881):3759--3767, August 6, 2008.

\bibitem{popoveniuc2010performance}
Stefan Popoveniuc, John Kelsey, Andrew Regenscheid, and Poorvi Vora.
\newblock Performance requirements for end-to-end verifiable elections.
\newblock In {\em International conference on Electronic voting
  technology/workshop on trustworthy elections}, pages 1--16. USENIX
  Association, 2010.

\bibitem{paul2009trustworthy}
Nathanael Paul and Andrew~S Tanenbaum.
\newblock Trustworthy voting: From machine to system.
\newblock {\em IEEE Computer}, 42(5):23--29, 2009.

\bibitem{fink2009tpm}
Russell~A Fink, Alan~T Sherman, and Richard Carback.
\newblock {TPM} meets {DRE}: reducing the trust base for electronic voting
  using trusted platform modules.
\newblock {\em Information Forensics and Security, IEEE Transactions on},
  4(4):628--637, 2009.

\bibitem{smart2011true}
Matt Smart and Eike Ritter.
\newblock True trustworthy elections: remote electronic voting using trusted
  computing.
\newblock In {\em Autonomic and Trusted Computing}, pages 187--202. Springer,
  2011.

\bibitem{molnar2006tamper}
David Molnar, Tadayoshi Kohno, Naveen Sastry, and David Wagner.
\newblock Tamper-evident, history-independent, subliminal-free data structures
  on prom storage-or-how to store ballots on a voting machine.
\newblock In {\em IEEE Symposium on Security and Privacy}, pages 6--pp. IEEE,
  2006.

\bibitem{bethencourt2007cryptographic}
John Bethencourt, Dan Boneh, and Brent Waters.
\newblock Cryptographic methods for storing ballots on a voting machine.
\newblock In {\em NDSS}, 2007.

\bibitem{heather2009append}
James Heather and David Lundin.
\newblock The append-only web bulletin board.
\newblock In {\em Formal Aspects in Security and Trust}, pages 242--256.
  Springer, 2009.

\bibitem{culnane2014peered}
Chris Culnane and Steve Schneider.
\newblock A peered bulletin board for robust use in verifiable voting systems.
\newblock In {\em IEEE 27th Computer Security Foundations Symposium (CSF)},
  pages 169--183. IEEE, 2014.

\bibitem{haenni2013generic}
Rolf Haenni and Reto~E Koenig.
\newblock A generic approach to prevent board flooding attacks in
  coercion-resistant electronic voting schemes.
\newblock {\em Computers \& Security}, 33:59--69, 2013.

\bibitem{paul2003authentication}
Nathanael Paul, David Evans, Avi Rubin, and Dan Wallach.
\newblock Authentication for remote voting.
\newblock In {\em Workshop on Human-Computer Interaction and Security Systems,
  Fort Lauderdale}, 2003.

\bibitem{popoveniuc2008remote}
Stefan Popoveniuc and Poorvi~L Vora.
\newblock Remote ballot casting with captchas.
\newblock In {\em Workshop on information and System Security}, 2008.

\bibitem{popoveniuc2010speakup}
Stefan Popoveniuc.
\newblock Speakup: remote unsupervised voting.
\newblock {\em Industrial Track ACNS}, 2010.

\bibitem{yasinsac2013independent}
Alec Yasinsac.
\newblock Independent computations for safe remote electronic voting.
\newblock In {\em Security Protocols Workshop XXI}, pages 71--83. Springer,
  2013.

\bibitem{laskowski2004improving}
Sharon~J Laskowski, Marguerite Autry, John Cugini, William Killam, and James
  Yen.
\newblock Improving the usability and accessibility of voting systems and
  products.
\newblock {\em NIST Special Publication}, 2004.

\bibitem{schneider2011focus}
Steve Schneider, Morgan Llewellyn, Chris Culnane, James Heather, Sriramkrishnan
  Srinivasan, and Zhe Xia.
\newblock Focus group views on pr{\^e}t {\`a} voter 1.0.
\newblock In {\em Workshop on Requirements Engineering for Electronic Voting
  Systems}, pages 56--65, 2011.

\bibitem{delis2014pressing}
Alex Delis, Konstantina Gavatha, Aggelos Kiayias, Charalampos Koutalakis, Elias
  Nikolakopoulos, Lampros Paschos, Mema Rousopoulou, Georgios Sotirellis, Panos
  Stathopoulos, Pavlos Vasilopoulos, Thomas Zacharias, and Bingsheng Zhang.
\newblock Pressing the button for {E}uropean elections: verifiable e-voting and
  public attitudes toward internet voting in {G}reece.
\newblock In {\em International Conference on Electronic Voting: Verifying the
  Vote (EVOTE)}, pages 1--8. IEEE, 2014.

\bibitem{olembo2014voter}
M~Maina Olembo, Karen Renaud, Steffen Bartsch, and Melanie Volkamer.
\newblock Voter, what message will motivate you to verify your vote.
\newblock In {\em Workshop on Usable Security, USEC}, 2014.

\bibitem{olembo2013mental}
Maina~M Olembo, Steffen Bartsch, and Melanie Volkamer.
\newblock Mental models of verifiability in voting.
\newblock In {\em E-Voting and Identify}, pages 142--155. Springer, 2013.

\bibitem{nandi2010stamp}
Mridul Nandi, Stefan Popoveniuc, and Poorvi~L Vora.
\newblock Stamp-it: a method for enhancing the universal verifiability of e2e
  voting systems.
\newblock In {\em Information Systems Security}, pages 81--95. Springer, 2010.

\bibitem{alvarez2010electronic}
R~Michael Alvarez and Thad~E Hall.
\newblock {\em Electronic elections: The perils and promises of digital
  democracy}.
\newblock Princeton University Press, 2010.

\bibitem{loncke2004online}
Mieke Loncke and Jos Dumortier.
\newblock Online voting: a legal perspective.
\newblock {\em International Review of Law, Computers \& Technology},
  18(1):59--79, 2004.

\bibitem{vocrehab}
Title~29 of~the United States~Code.
\newblock Vocational rehabilitation act, 1973.
\newblock $\S$ 794d, Section 508.

\bibitem{schwartz2012establishing}
Bryan~P Schwartz and Dan Grice.
\newblock Establishing a legal framework for e-voting in canada.
\newblock {\em Man. LJ}, 36:301, 2012.

\bibitem{maaten2004towards}
Epp Maaten.
\newblock Towards remote e-voting: {E}stonian case.
\newblock {\em Electronic Voting in Europe}, 47:83--100, 2004.

\bibitem{osce2006netherlands}
Organization for Security and Cooperation in~Europe~(OSCE).
\newblock {Netherlands. Parliamentary Elections, 22 November 2006. Election
  Assessment Mission Report}, 2006.
\newblock \url{http://www.osce.org/odihr/elections/netherlands/24322}.

\bibitem{oostveen2004internet}
Anne-Marie Oostveen and Peter Van~den Besselaar.
\newblock Internet voting technologies and civic participation: the users'
  perspective.
\newblock {\em Javnost-the public}, 11(1):61--78, 2004.

\bibitem{hole2010toward}
Kjell Hole and Lars-Helge Netland.
\newblock Toward risk assessment of large-impact and rare events.
\newblock {\em IEEE Security \& Privacy}, (3):21--27, 2010.

\bibitem{pieters2005ethics}
W~Pieters and MJ~Becker.
\newblock Ethics of e-voting: An essay on requirements and values in internet
  elections.
\newblock {\em Ethics of New Information Technology: Proceedings of the Sixth
  International Conference of Computer Ethics: Philosophical Inquiry}, pages
  307--318, 2005.

\end{thebibliography}
\end{document}